\documentclass[12pt,a4paper]{article}
\usepackage{amssymb}

\usepackage{amsmath}

\newcounter{resultnum}[section]
\newtheorem{conclusion}{Conclusion}[section]

\newcounter{conclusionnum}[section]

\newcounter{conditionnum}[section]

\newcounter{conjecturenum}[section]
\newtheorem{example}{Example}[section]

\newcounter{examplenum}[section]

\newcounter{exercisenum}[section]
\newtheorem{lemma}{Lemma}[section]

\newcounter{lemmanum}[section]

\newcounter{notationnum}[section]
\newtheorem{theorem}{Theorem}[section]

\newcounter{theoremnum}[section]
\newtheorem{definition}{Definition}[section]

\newcounter{definitionnum}[section]
\newtheorem{corollary}{Corollary}[section]

\newcounter{corollarynum}[section]
\newtheorem{remark}{Remark}[section]

\newcounter{remarknum}[section]
\newtheorem{proposition}{Proposition}[section]

\newcounter{propositionnum}[section]

\newcounter{acknowledgementnum}[section]

\newcounter{algorithmnum}[section]

\newcounter{axiomnum}[section]

\newcounter{casenum}[section]

\newcounter{claimnum}[section]

\newcounter{summarynum}[section]

\newcounter{problemnum}[section]
\newenvironment{proof}[1][]{\textbf{Proof.} }{}

\begin{document}

\title{Curve Flows in Lagrange--Finsler Geometry, \\
Bi--Hamiltonian Structures and Solitons}
\author{ Stephen C. Anco\thanks{%
sanco@brocku.ca} \\
%EndAName
{\small \textit{Department of Mathematics, Brock University,}}\\
{\small \textit{St. Catharines, Ontario L2S 3A1, Canada}} \\
{\quad} \\
Sergiu I. Vacaru\thanks{Sergiu.Vacaru@gmail.com   }
\\
{\small \textit{Faculty of Mathematics, University "Al. I. Cuza" Ia\c si},} \\
{\small \textit{\ 700506, Ia\c si, Romania}}
}
\date{October 5, 2008}
\maketitle

\begin{abstract}
Methods in Riemann--Finsler geometry are applied to investigate
bi--Hamil\-ton\-ian structures and related mKdV hierarchies of soliton
equations derived geometrically from regular Lagrangians and flows of
non--stretching curves in tangent bundles. The total space geometry and
nonholonomic flows of curves are defined by Lagrangian semisprays inducing
canonical nonlinear connections (N--connections), Sa\-sa\-ki type metrics
and linear connections. The simplest examples of such geometries are given
by tangent bundles on Riemannian symmetric spaces $G/SO(n)$ provided with an
N--connection structure and an adapted metric, for which we elaborate a
complete classification, and by generalized Lagrange spaces with constant
Hessian. In this approach, bi--Hamiltonian structures are derived for
geometric mechanical models and (pseudo) Riemannian metrics in gravity. The
results yield horizontal/ vertical pairs of vector sine--Gordon equations
and vector mKdV equations, with the corresponding geometric curve flows in
the hierarchies described in an explicit form by nonholonomic wave maps and
mKdV analogs of nonholonomic Schr\"{o}dinger maps on a tangent bundle.


\end{abstract}

%\tableofcontents

%\newpage

\section{ Introduction}

Some interesting studies in geometric mechanics and field theory and theory
of partial differential equations are related to nonholonomic structures
characterizing integrability of certain nonlinear physical systems, their
global and local symmetries \cite{kern,ma1,ma2,deleon1,deleon2}.
In parallel, the differential geometry of plane and space curves has
received considerable attention in the theory of integrable nonlinear
partial differential equations and applications to modern physics
\cite{lam,gol,nak,lanper,chou1,chou2,mbsw}.
More particularly, it is well known
that both the modified Korteweg-de Vries (mKdV) equation and the sine-Gordon
(SG) equation can be encoded as flows of the curvature invariant of
plane curves in Euclidean plane geometry.
Similarly, curve flows in Riemannian manifolds of constant curvature give
rise to a vector generalization of the mKdV equation and encode its
bi-Hamiltonian structure in a natural geometric way \cite{saw,anc1}.
This approach provides an elegant geometric origin for previous results
on the Hamiltonian structure of multi-component mKdV equations
(see, e.g. Ref. \cite{ath1,ath2}).

In recent work \cite{anc2}, geometric flows of curves were studied in
Riemannian symmetric spaces $M=G/SO(n)$ which provide the simplest
generalization of $n$--dimensional constant-curvature Riemannian geometries.
The isometry groups $G$ of these spaces are exhausted by the Lie groups
$SO(n+1)$ and $SU(n)$, as known from Cartan's classification \cite{helag}.
The main results of \cite{anc2} were to show that,
firstly, the Cartan structure equations for torsion and
curvature of a moving parallel frame and its associated frame connection
1--form encode $O(n-1)$--invariant bi-Hamiltonian operators. Secondly, this
bi-Hamiltonian structure generates a hierarchy of integrable flows of curves
in which the frame components of the principal normal along the curve
(analogous to curvature invariants) satisfy $O(n-1)$--invariant
multi-component soliton equations that include vector mKdV equations and
vector sine-Gordon equations. The two groups $G=SO(n+1),SU(n)$ give
different soliton hierarchies and account precisely for the two known
integrable versions \cite{wang,sw,aw} of vector mKdV equations and vector
sine-Gordon equations. Thirdly, the curve flows corresponding to such vector
soliton equations were shown to be described geometrically by wave maps and
mKdV analogs of Schr\"{o}dinger maps into the curved manifolds $M=G/SO(n)$.

A crucial condition \cite{anc1,anc2} behind such constructions is the fact that
the frame curvature matrix is constant on these spaces $M=G/SO(n)$.
This approach can be developed into a geometric formalism for mapping arbitrary
(semi) Riemannian  metrics \cite{vhfrm} and regular Lagrange mechanical systems
into bi-Hamiltonian structures and related solitonic equations following
certain methods elaborated in the geometry of generalized Finsler and
Lagrange spaces \cite{ma1,ma2,bej} and nonholonomic manifolds with
applications in modern gravity \cite{bejf,vncg,vsgg}.

The first aim of this paper is to prove that solitonic hierarchies can be
generated by (semi) Riemannian metrics $g_{ij}$ on a manifold $V$ of
dimension $\dim V=n\geq 2$ if the geometrical objects are lifted into the
total space of the tangent bundle $TV$, or of a vector bundle $\mathcal{E}%
=(M,\pi ,E)$, $\dim E=m\geq n$, by a moving parallel frame formulation of
geometric curve flows when $V$ has constant matrix curvature as defined
canonically with respect to certain preferred frames. The second purpose is
to elaborate applications in geometric mechanics, in particular, that the
dynamics defined by any regular Lagrangian can be encoded in terms of
bi-Hamiltonian structures and related solitonic hierarchies. It will be
emphasized that, in a similar way, any solution of the Einstein equations
given by a generically off-diagonal metric
can be mapped into solitonic equations.

The paper is organized as follows:

In section 2 we outline the geometry of vector bundles equipped with a
nonlinear connection. We emphasize the possibility to define fundamental
geometric objects induced by a (semi) Riemannian metric on the base space
when the Riemannian curvature tensor has constant coefficients with respect
to a preferred nonholonomic framing.

In section 3 we consider curve flows on nonholonomic vector bundles. We
sketch an approach to classification of such spaces defined by conventional
horizontal and vertical symmetric (semi) Riemannian subspaces and equipped
with nonholonomic distributions defined by a nonlinear connection structure.
Bi-Hamiltonian operators are then derived for a canonical distinguished
connection, adapted to the nonlinear connection structure, for which the
distinguished curvature coefficients are constant.

Section 4 is devoted to the formalism of distinguished bi-Hamiltonian
operators and vector soliton equations for arbitrary (semi) Riemannian
spaces admitting nonholonomic deformations to symmetric Riemannian spaces.
We define the basic equations for nonholonomic curve flows. Then we consider
the properties of cosymplectic and symplectic operators adapted to the
nonlinear connection structure. Finally, we construct solitonic hierarchies
of bi-Hamiltonian anholonomic curve flows.

Section 5 contains some applications of the formalism in modern mechanics
and gravity.

We conclude with some further remarks in section 6. The Appendix contains
necessary definitions and formulas from the geometry of nonholonomic
manifolds.

\section{Nonholonomic Structures on Manifolds}

In this section, we prove that for any (semi) Riemannian metric $g_{ij}$ on
a manifold $V$ it is possible to define lifts to the tangent bundle $TV$
provided with canonical nonlinear connection (in brief, N--connection),
Sasaki type metric and canonical linear connection structure. The geometric
constructions will be elaborated in general form for vector bundles.

\subsection{N--connections induced by Riemannian metrics}

Let $\mathcal{E}=(E,\pi , F,M)$ be a (smooth) vector bundle of over base
manifold $M$, with dimensions $\dim M=n$ and $\dim E=(n+m)$, for $n\geq 2$,
and with $m\geq n$ being the dimension of typical fiber $F$. Here $\pi
:E\rightarrow M$ defines a surjective submersion. At any point $u\in E$, the
total space $E$ splits into ``horizontal'', $M_{u}$, and ``vertical'', $%
F_{u} $, subspaces. We denote the local coordinates in the form $u=(x,y)$,
or $u^{\alpha }=\left( x^{i},y^{a}\right) $, with horizontal indices $%
i,j,k,$ etc. $=1,2,\ldots,n$ and vertical indices $a,b,c,$ etc.
$=n+1,n+2,\ldots,n+m$.\footnote{%
In the particular case when we have a tangent bundle $E\mathbf{=}TM$, both
type of indices run over the same values since $n=m$ but it is convenient to
distinguish the horizontal and vertical ones by using different groups of
Latin indices.} The summation rule on repeated ``upper'' and ``lower'' indices
will be applied.

Let the base manifold $M$ be equipped with a (semi) Riemannian metric,
namely a second rank tensor of fixed signature,\footnote{%
In physics literature, one uses the term (pseudo) Riemannian/Euclidean space}
$h\underline{g}=\underline{g}_{ij}(x)dx^{i}\otimes dx^{j}$. It is possible
to introduce a vertical metric structure $v\underline{g}=\underline{g}%
_{ab}(x)dy^{a}\otimes dy^{b}$ by completing the matrix $\underline{g}%
_{ij}(x) $ diagonally with $\pm 1$ till any nondegenerate second rank tensor
$\underline{g}_{ab}(x)$ if $m>n$. This defines a metric structure $%
\underline{\mathbf{g}}=[h\underline{g},v\underline{g}]$ (we shall also use
the notation $\underline{g}_{\alpha \beta }=[\underline{g}_{ij},\underline{g}%
_{ab}])$ on $\mathcal{E}$. We can deform the metric structure, $\underline{g}%
_{\alpha \beta }\rightarrow g_{\alpha \beta }=[g_{ij},g_{ab}]$, by
considering a frame (vielbein) transformation,%
\begin{equation}
g_{\alpha \beta }(x,y)=e_{\alpha }^{~\underline{\alpha }}(x,y)~e_{\beta }^{~%
\underline{\beta }}(x,y)g_{\underline{\alpha }\underline{\beta }}(x),
\label{auxm}
\end{equation}%
with coefficients $\underline{g}_{\alpha \beta }(x)=g_{\underline{\alpha }%
\underline{\beta }}(x)$. The coefficients $e_{\alpha }^{~\underline{\alpha }%
}(x,y)$ will be defined below (see formula (\ref{aux4})) from the condition
of generating curvature tensors with constant coefficients with respect to
certain preferred frames.

For any $g_{ab}$ obtained from $g_{\alpha \beta }$, we consider a generating
function
\begin{equation*}
\mathcal{L}(x,y)=g_{ab}(x,y)y^{a}y^{b}
\end{equation*}%
inducing a vertical metric
\begin{equation}
\tilde{g}_{ab}=\frac{1}{2}\frac{\partial ^{2}\mathcal{L}}{\partial
y^{a}\partial y^{b}}  \label{ehes}
\end{equation}%
which is ``weakly'' regular if $\det |\tilde{g}_{ab}|\neq 0$. \footnote{%
Similar values, for $e_{\alpha }^{~\underline{\alpha }}=\delta _{\alpha }^{~%
\underline{\alpha }}$, where $\delta _{\alpha }^{~\underline{\alpha }}$ is
the Kronecker symbol, were introduced for the so--called generalized
Lagrange spaces when $\mathcal{L}$ was called the ``absolute energy'' \cite%
{ma1}.}

By straightforward calculations we can prove the following\footnote{%
See Refs. \cite{ma1,ma2} for details of a similar proof; here we note that
in our case, in general, \ $e_{\alpha }^{~\underline{\alpha }}\neq \delta
_{\alpha }^{~\underline{\alpha }}$}:

\begin{theorem}
\label{teleq}The Euler--Lagrange equations on $TM$,
\begin{equation*}
\frac{d}{d\tau }\left( \frac{\partial L}{\partial y^{i}}\right) -\frac{%
\partial L}{\partial x^{i}}=0,
\end{equation*}%
for the \ Lagrangian $L=\sqrt{|\mathcal{L}|}$, where $y^{i}=\frac{dx^{i}}{%
d\tau }$ for a path curve $x^{i}(\tau )$ on $M$, depending on parameter $%
\tau $, are equivalent to the ``nonlinear'' geodesic equations
\begin{equation*}
\frac{d^{2}x^{i}}{d\tau ^{2}}+2\widetilde{G}^{i}(x^{k},\frac{dx^{j}}{d\tau }%
)=0
\end{equation*}%
defining path curves of a canonical semispray $S=y^{i}\frac{\partial }{%
\partial x^{i}}-2\widetilde{G}^{i}(x,y)\frac{\partial }{\partial y^{i}}$,
where
\begin{equation*}
2\widetilde{G}^{i}(x,y)=\frac{1}{2}\ \tilde{g}^{ij}\left( \frac{\partial
^{2}L}{\partial y^{i}\partial x^{k}}y^{k}-\frac{\partial L}{\partial x^{i}}%
\right)
\end{equation*}%
with $\tilde{g}^{ij}$ being inverse to the vertical metric (\ref{ehes}).
\end{theorem}

This theorem has an important geometric mechanical interpretation.

\begin{corollary}
For any (semi) Riemannian metric $\underline{g}_{ij}(x)$ on $M$, we can
associate canonically an effective regular Lagrange mechanics on $TM$ with
the Euler--Lagrange equations transformed into nonlinear (semispray)
geodesic equations.
\end{corollary}

The differential of map $\pi:E\rightarrow M$ is defined by fiber preserving
morphisms of the tangent bundles $TE$ and $TM$. Its kernel is just the
vertical subspace $vE$ with a related inclusion mapping $i:vE\rightarrow TE$.

\begin{definition}
A nonlinear connection (N--connection) $\mathbf{N}$ on a vector bundle $%
\mathcal{E}$ \ is defined by the splitting on the left of an exact sequence
\begin{equation*}
0\rightarrow vE\overset{i}{\rightarrow }TE\rightarrow TE/vE\rightarrow 0,
\end{equation*}%
i.e. by a morphism of submanifolds $\mathbf{N:\ \ }TE\rightarrow vE$ such
that $\mathbf{N\circ }i$ is the identity map in $vE$.
\end{definition}

Equivalently, a N--connection is defined by a Whitney sum of conventional
horizontal (h) subspace, $\left( hE\right) $, and vertical (v) subspace, $%
\left( vE\right) $,
\begin{equation}
TE=hE\oplus vE.  \label{whitney}
\end{equation}%
This sum defines a nonholonomic (alternatively, anholonomic, or
non--integ\-rab\-le) distribution of horizontal (h) and vertical (v)
subspaces on $TE\mathbf{.}$ Locally, a N--connection is determined by its
coefficients $N_{i}^{a}(u)$,%
\begin{equation*}
\mathbf{N}=N_{i}^{a}(u)dx^{i}\otimes \frac{\partial }{\partial y^{a}}.
\end{equation*}%
The well known class of linear connections consists of a particular subclass
with the coefficients being linear in $y^{a}$, i.e., $N_{i}^{a}(u)=\Gamma
_{bj}^{a}(x)y^{b}$.

\begin{remark}
A manifold (or a bundle space) is called nonholonomic if it is provided with
a nonholonomic distribution (see historical details and summary of results
in \cite{bejf}). In a particular case, when the nonholonomic distribution is
of type (\ref{whitney}), such spaces are called N--anholonomic \cite{vsgg}.
\end{remark}

Any N--connection $\mathbf{N}=\left\{ N_{i}^{a}(u)\right\} $ may be
characterized by a N--adapted frame (vielbein) structure $\mathbf{e}_{\nu
}=(e_{i},e_{a})$, where
\begin{equation}
\mathbf{e}_{i}=\frac{\partial }{\partial x^{i}}-N_{i}^{a}(u)\frac{\partial }{%
\partial y^{a}}\mbox{ and
}e_{a}=\frac{\partial }{\partial y^{a}},  \label{dder}
\end{equation}%
and the dual frame (coframe) structure $\mathbf{e}^{\mu }=(e^{i},\mathbf{e}%
^{a})$, where
\begin{equation}
e^{i}=dx^{i}\mbox{ and }\mathbf{e}^{a}=dy^{a}+N_{i}^{a}(u)dx^{i}.
\label{ddif}
\end{equation}
We remark that $\mathbf{e}_{\nu }=(\mathbf{e}_{i},e_{a})$ and $\mathbf{e}%
^{\mu }=(e^{i},\mathbf{e}^{a})$ are, respectively, the former
``N--elongated'' partial derivatives $\delta _{\nu }=\delta /\partial u^{\nu
}=(\delta _{i},\partial _{a})$ and N--elongated differentials $\delta ^{\mu
}=\delta u^{\mu }=(d^{i},\delta ^{a})$ which emphasize that operators (\ref%
{dder}) and (\ref{ddif}) define, correspondingly, certain ``N--elongated''
partial derivatives and differentials which are more convenient for tensor
and integral calculations on such nonholonomic manifolds.\footnote{%
We shall use ``boldface'' symbols whenever necessary as emphasis for
any space and/or geometrical objects equipped with/adapted to a\ N--connection
structure, or for the coefficients computed with respect to N--adapted
frames.}

For any N--connection, we can introduce its N--connection curvature
\begin{equation*}
\mathbf{\Omega }=\frac{1}{2}\Omega _{ij}^{a}\ d^{i}\wedge d^{j}\otimes
\partial _{a},
\end{equation*}%
with the coefficients defined as the Nijenhuis tensor,%
\begin{equation}
\Omega _{ij}^{a}=\mathbf{e}_{[j}N_{i]}^{a}=\mathbf{e}_{j}N_{i}^{a}-\mathbf{e}%
_{i}N_{j}^{a}=\frac{\partial N_{i}^{a}}{\partial x^{j}}-\frac{\partial
N_{j}^{a}}{\partial x^{i}}+N_{i}^{b}\frac{\partial N_{j}^{a}}{\partial y^{b}}%
-N_{j}^{b}\frac{\partial N_{i}^{a}}{\partial y^{b}}.  \label{ncurv}
\end{equation}

The vielbeins (\ref{ddif}) satisfy the nonholonomy (equivalently,
anholonomy) relations
\begin{equation}
\lbrack \mathbf{e}_{\alpha },\mathbf{e}_{\beta }]=\mathbf{e}_{\alpha }%
\mathbf{e}_{\beta }-\mathbf{e}_{\beta }\mathbf{e}_{\alpha }=W_{\alpha \beta
}^{\gamma }\mathbf{e}_{\gamma }  \label{anhrel}
\end{equation}%
with (antisymmetric) nontrivial anholonomy coefficients $W_{ia}^{b}=\partial
_{a}N_{i}^{b}$ and $W_{ji}^{a}=\Omega _{ij}^{a}$.

These geometric objects can be defined in a form adapted to a
N--connec\-ti\-on structure via decompositions being invariant under
parallel transport preserving the splitting (\ref{whitney}). In this case we
call them ``distinguished'' (by the N--connection structure), i.e.
d--objects. For instance, a vector field $\mathbf{X}\in T\mathbf{V}$ \ is
expressed
\begin{equation*}
\mathbf{X}=(hX,\ vX),\mbox{ \ or \ }\mathbf{X}=X^{\alpha }\mathbf{e}_{\alpha
}=X^{i}\mathbf{e}_{i}+X^{a}e_{a},
\end{equation*}%
where $hX=X^{i}\mathbf{e}_{i}$ and $vX=X^{a}e_{a}$ state, respectively, the
N--adapted horizontal (h) and vertical (v) components of the vector (which
following Refs. \cite{ma1,ma2} is called a distinguished vector, in brief,
d--vector). In a similar fashion, the geometric objects on $\mathbf{V}$, for
instance, tensors, spinors, connections, etc. can be defined and called
respectively d--tensors, d--spinors, d--connections if they are adapted to
the N--connection splitting (\ref{whitney}).

\begin{theorem}
Any (semi) Riemannian metric $\underline{g}_{ij}(x)$ on $M$ induces a
canonical N--connection on $TM$.
\end{theorem}

\begin{proof}
We sketch a proof by defining the coefficients of N--connection
\begin{equation}
\tilde{N}_{\ j}^{i}(x,y)=\frac{\partial \tilde{G}^{i}}{\partial y^{j}}
\label{cnlce}
\end{equation}%
where
\begin{eqnarray}
\tilde{G}^{i} &=&\frac{1}{4}\tilde{g}^{ij}\left( \frac{\partial ^{2}\mathcal{%
L}}{\partial y^{i}\partial x^{k}}y^{k}-\frac{\partial \mathcal{L}}{\partial
x^{j}}\right) =\frac{1}{4}\tilde{g}^{ij}g_{jk}\gamma _{lm}^{k}y^{l}y^{m},
\label{aux2} \\
\gamma _{\ lm}^{i} &=&\frac{1}{2}g^{ih}(\partial _{m}g_{lh}+\partial
_{l}g_{mh}-\partial _{h}g_{lm}),\ \partial _{h}=\partial /\partial x^{h},
\notag
\end{eqnarray}%
with $g_{ah}$ and $\tilde{g}_{ij}$ defined respectively by formulas (\ref%
{auxm}) and (\ref{ehes}). $\square$
\end{proof}

The N--adapted operators (\ref{dder}) and (\ref{ddif}) defined by the
N--connection coefficients (\ref{cnlce}) are denoted respectively $\mathbf{%
\tilde{e}}_{\nu }=(\mathbf{\tilde{e}}_{i},e_{a})$ and $\mathbf{\tilde{e}}%
^{\mu }=(e^{i},\mathbf{\tilde{e}}^{a})$.

\subsection{Canonical linear connection and metric structures}

The constructions will be performed on a vector bundle $\mathbf{E}$ provided
with N--connection structure. We shall emphasize the special properties of a
tangent bundle $(TM,\!\pi ,\!M)$ when the linear connection and metric are
induced by a (semi) Riemannian metric on $M$.

\begin{definition}
A distinguished connection (i.e., d--connection) $\mathbf{D}=(h\mathbf{D}$, $%
v\mathbf{D})$ is a linear connection preserving under parallel transport the
nonholonomic decomposition (\ref{whitney}).
\end{definition}

The N--adapted components $\mathbf{\Gamma }_{\ \beta \gamma }^{\alpha }$ of
a d--connection $\mathbf{D}_{\alpha }=(\mathbf{e}_{\alpha }\rfloor \mathbf{D}%
)$ are defined by equations
\begin{equation}
\mathbf{D}_{\alpha }\mathbf{e}_{\beta }=\mathbf{\Gamma }_{\ \alpha \beta
}^{\gamma }\mathbf{e}_{\gamma },\mbox{\ or \ }\mathbf{\Gamma }_{\ \alpha
\beta }^{\gamma }\left( u\right) =\left( \mathbf{D}_{\alpha }\mathbf{e}%
_{\beta }\right) \rfloor \mathbf{e}^{\gamma }.  \label{dcon1}
\end{equation}%
The N--adapted splitting into h-- and v--covariant derivatives is stated by
\begin{equation*}
h\mathbf{D}=\{\mathbf{D}_{k}=\left( L_{jk}^{i},L_{bk\;}^{a}\right) \},%
\mbox{
and }\ v\mathbf{D}=\{\mathbf{D}_{c}=\left( C_{jk}^{i},C_{bc}^{a}\right) \},
\end{equation*}%
where, by definition, $L_{jk}^{i}=\left( \mathbf{D}_{k}\mathbf{e}_{j}\right)
\rfloor e^{i}$, $L_{bk}^{a}=\left( \mathbf{D}_{k}e_{b}\right) \rfloor
\mathbf{e}^{a}$, $C_{jc}^{i}=\left( \mathbf{D}_{c}\mathbf{e}_{j}\right)
\rfloor e^{i}$, $C_{bc}^{a}=\left( \mathbf{D}_{c}e_{b}\right) \rfloor
\mathbf{e}^{a}$. The components $\mathbf{\Gamma }_{\ \alpha \beta }^{\gamma
}=\left( L_{jk}^{i},L_{bk}^{a},C_{jc}^{i},C_{bc}^{a}\right) $ completely
define a d--connection $\mathbf{D}$ on $\mathbf{E}$.

The simplest way to perform N--adapted computations is to use differential
forms. For instance, starting with the d--connection 1--form,
\begin{equation}
\mathbf{\Gamma }_{\ \beta }^{\alpha }=\mathbf{\Gamma }_{\ \beta \gamma
}^{\alpha }\mathbf{e}^{\gamma },  \label{dconf}
\end{equation}%
with the coefficients defined with respect to N--elongated frames (\ref{ddif}%
) and (\ref{dder}), the torsion of a d--connection,
\begin{equation}
\mathcal{T}^{\alpha }\doteqdot \mathbf{De}^{\alpha }=d\mathbf{e}^{\alpha
}+\Gamma _{\ \beta }^{\alpha }\wedge \mathbf{e}^{\beta },  \label{tors}
\end{equation}
is characterized by (N--adapted) d--torsion components,
\begin{eqnarray}
T_{\ jk}^{i} &=&L_{\ jk}^{i}-L_{\ kj}^{i},\ T_{\ ja}^{i}=-T_{\ aj}^{i}=C_{\
ja}^{i},\ T_{\ ji}^{a}=\Omega _{\ ji}^{a},\   \notag \\
T_{\ bi}^{a} &=&-T_{\ ib}^{a}=\frac{\partial N_{i}^{a}}{\partial y^{b}}-L_{\
bi}^{a},\ T_{\ bc}^{a}=C_{\ bc}^{a}-C_{\ cb}^{a}.  \label{dtors}
\end{eqnarray}%
For d--connection structures on $TM$, we have to identify indices in the
form $i\leftrightarrows a,j\leftrightarrows b,..$. and the components of N--
and d--connections, for instance, $N_{i}^{a}\leftrightarrows N_{i}^{j}$ and $%
L_{\ jk}^{i}\leftrightarrows L_{\ bk}^{a},C_{\ ja}^{i}\leftrightarrows C_{\
ca}^{b}\leftrightarrows C_{\ jk}^{i}$.

\begin{definition}
A distinguished metric (i.e., d--metric) on a vector bundle $\mathbf{E}$ is
a non-degenerate second rank metric tensor $\mathbf{g=}g\mathbf{\oplus _{N}}%
h$, equivalently
\begin{equation}
\mathbf{g}=\ g_{ij}(x,y)\ e^{i}\otimes e^{j}+\ h_{ab}(x,y)\ \mathbf{e}%
^{a}\otimes \mathbf{e}^{b},  \label{m1}
\end{equation}%
adapted to the N--connection decomposition (\ref{whitney}).
\end{definition}

From the class of arbitrary d--connections $\mathbf{D}$ on $\mathbf{V,}$ one
distinguishes those which are metric compatible (metrical) satisfying the
condition%
\begin{equation}
\mathbf{Dg=0}  \label{metcomp}
\end{equation}%
including all h- and v-projections $D_{j}g_{kl}=0$, $D_{a}g_{kl}=0$, $%
D_{j}h_{ab}=0$, $D_{a}h_{bc}=0$. For d--metric structures on $\mathbf{%
V\simeq }TM$, with $g_{ij}=h_{ab}$, the condition of vanishing
``nonmetricity'' (\ref{metcomp}) transforms into%
\begin{equation}
h\mathbf{D(}g\mathbf{)=}0\mbox{\  and\ }v\mathbf{D(}h\mathbf{)=}0,
\label{metcompt}
\end{equation}%
i.e. $D_{j}g_{kl}=0$ and $D_{a}g_{kl}=0$.

For any metric structure $\mathbf{g}$ on a manifold, there is the unique
metric compatible and torsionless Levi Civita connection $\nabla $ for which
$\nabla \mathbf{g=0}$. This connection is not a d--connection because it
does not preserve under parallelism the N--connection splitting (\ref%
{whitney}). One has to consider less constrained cases, admitting nonzero
torsion coefficients, when a d--connection is constructed canonically for a
d--metric structure. A simple minimal metric compatible extension of $\nabla
$ is that of canonical d--connection with $T_{\ jk}^{i}=0$ and $T_{\
bc}^{a}=0$ but $T_{\ ja}^{i},T_{\ ji}^{a}$ and $T_{\ bi}^{a}$ are not zero,
see (\ref{dtors}). The coefficient formulas for such connections are given
in Appendix, see (\ref{candcon}) and related discussion.

\begin{lemma}
Any (semi) Riemannian metric $\underline{g}_{ij}(x)$ on a manifold $M$
induces a canonical d--metric structure on $TM$,
\begin{equation}
\mathbf{\tilde{g}}=\tilde{g}_{ij}(x,y)\ e^{i}\otimes e^{j}+\ \tilde{g}%
_{ij}(x,y)\ \mathbf{\tilde{e}}^{i}\otimes \mathbf{\tilde{e}}^{j},
\label{slme}
\end{equation}%
where $\mathbf{\tilde{e}}^{i}$ are elongated as in (\ref{ddif}), but with $%
\tilde{N}_{\ j}^{i}$ given by (\ref{cnlce}).
\end{lemma}

\begin{proof}
This construction is trivial by lifting to the so--called Sasaki metric \cite%
{yano} but in our case using the coefficients $\tilde{g}_{ij}$ (\ref{ehes})$.%
\square $
\end{proof}

\begin{proposition}
There exist canonical d--connections on $TM$ induced by a (semi) Riemannian
metric $\underline{g}_{ij}(x)$ on $M$.
\end{proposition}

\begin{proof}
We can construct an example in explicit form by introducing $\tilde{g}_{ij}$
and $\tilde{g}_{ab}$ in formulas (\ref{candcontm}), see Appendix, in order
to compute the coefficients $\tilde{\Gamma}_{\ \beta \gamma }^{\alpha }=(%
\tilde{L}_{\ jk}^{i},\tilde{C}_{bc}^{a}).\square $
\end{proof}

The above Lemma and Proposition establish the following result.

\begin{theorem}
\label{tisrg}Any (semi) Riemannian metric $\underline{g}_{ij}(x)$ on $M$
induces a nonholonomic (semi) Riemannian structure on $TM$.
\end{theorem}

We note that the induced Riemannian structure is nonholonomic because on $TM$
there is a non-integrable distribution (\ref{whitney}) defining $\tilde{N}%
_{\ j}^{i}$. The corresponding curvature curvature tensor $\widetilde{R}_{\
\beta \gamma \tau }^{\alpha }=\{\widetilde{R}_{\ hjk}^{i},\tilde{P}_{\
jka}^{i},\tilde{S}_{\ bcd}^{a}\}$ is computed by substituting $\tilde{g}%
_{ij},\tilde{N}_{\ j}^{i}$ and $\widetilde{\mathbf{e}}_{k}$ into formulas (%
\ref{dcurvtb}), from Appendix, for $\tilde{\Gamma}_{\ \beta \gamma }^{\alpha
}=(\tilde{L}_{\ jk}^{i},\tilde{C}_{bc}^{a})$. Here one should be noted that
the constructions on $TM$ depend on arbitrary vielbein coefficients $%
e_{\alpha }^{~\underline{\alpha }}(x,y)$ in (\ref{auxm}). We can restrict
such coefficients in order to generate various particular classes of (semi)
Riemannian geometries on $TM$, for instance, in order to generate symmetric
Riemannian spaces with constant curvature, see Refs. \cite{helag,kob,sharpe}.

\begin{corollary}
There are lifts of a (semi) Riemannian metric $\underline{g}_{ij}(x)$ on $M$,
$\dim M=n$, generating a Riemannian structure on $TM$ with the curvature
coefficients of the canonical d--connection coinciding (with respect to
N--adapted bases) with those for a Riemannian space of constant curvature of
dimension $n+n$.
\end{corollary}

\begin{proof}
For a given metric $\underline{g}_{ij}(x)$ on $M$, we chose such
coefficients $e_{\alpha }^{~\underline{\alpha }}(x,y)$ $=\left\{ e_{a}^{~%
\underline{a}}(x,y)\right\} $ in (\ref{auxm}) that
\begin{equation*}
g_{ab}(x,y)=e_{a}^{~\underline{a}}(x,y)~e_{b}^{~\underline{b}}(x,y)g_{%
\underline{a}\underline{b}}(x)
\end{equation*}%
produces a vertical metric (\ref{ehes}) of type%
\begin{equation}
\tilde{g}_{ef}=\frac{1}{2}\frac{\partial ^{2}\mathcal{L}}{\partial
y^{e}\partial y^{f}}=\frac{1}{2}\frac{\partial ^{2}(e_{a}^{~\underline{a}%
}~e_{b}^{~\underline{b}}y^{a}y^{b})}{\partial y^{e}\partial y^{f}}g_{%
\underline{a}\underline{b}}(x)=\ \mathring{g}_{ef},  \label{aux4}
\end{equation}%
where $\ \mathring{g}_{ab}$ is the metric of a symmetric Riemannian space
(of constant curvature). Considering a prescribed $\mathring{g}_{ab}$, we
have to integrate two times on $y^{e}$ in order to find any solution for $%
e_{a}^{~\underline{a}}$ defining a frame structure in the vertical subspace.
The next step is to construct the d--metric $\mathring{g}_{\alpha \beta }=[\
\mathring{g}_{ij},\ \mathring{g}_{ab}]$ of type (\ref{slme}), in our case,
with respect to a nonholonomic base elongated by $~\widetilde{\mathring{N}}{}%
_{\ j}^{i}$, generated by $\underline{g}_{ij}(x)$ and $\tilde{g}_{ef}=%
\mathring{g}_{ab}$, like in (\ref{cnlce}) and (\ref{aux2}). This defines a
constant curvature Riemannian space of dimension $n+n$. The coefficients of
the canonical d--connection, which in this case coincide with those for the
Levi Civita connection, and the coefficients of the Riemannian curvature can
be computed respectively by putting $\tilde{g}_{ef}=\ \mathring{g}_{ab}$ in
formulas (\ref{candcontm}) and (\ref{dcurvtb}), see Appendix. Finally, we
note that the induced symmetric Riemannian space contains additional
geometric structures like the N--connection and anholonomy coefficients $%
W_{\alpha \beta }^{\gamma }$, see (\ref{anhrel})$.\square $
\end{proof}

There are various possibilities to generate on $TM$ nonholonomic Riemannian
structures from a given metric $\underline{g}_{ij}(x)$ on $M$. They result
in different geometrical and physical models. In this work, we emphasize the
possibility of generating spaces with constant curvature because for such
symmetric spaces one can derive a bi-Hamiltonian hierarchy of curve flows
and associated solitonic equations.

\begin{example}
The simplest example where a Riemannian structure with constant matrix
curvature coefficients is generated on $TM$ comes from a d--metric induced
by $\tilde{g}_{ij}=\delta _{ij}$, i.e.%
\begin{equation}
\mathbf{\tilde{g}}_{[E]}=\delta _{ij}e^{i}\otimes e^{j}+\ \delta _{ij}\
\mathbf{\tilde{e}}^{i}\otimes \mathbf{\tilde{e}}^{j},  \label{clgs}
\end{equation}%
with $\mathbf{\tilde{e}}^{i}$ given by $\tilde{N}_{\ j}^{i}$ as defined by $%
\underline{g}_{ij}(x)$ on $M$.
\end{example}

It should be noted that the metric (\ref{clgs}) is generically off--diagonal
with respect to a coordinate basis because, in general, the anholonomy
coefficients from (\ref{anhrel}) are not zero. This way, we model on $TM$ a
nonholonomic Euclidean space with vanishing curvature coefficients of the
canonical d--connection (it can be verified by putting respectively the
constant coefficients of metric (\ref{clgs}) into formulas (\ref{candcontm})
and (\ref{dcurvtb})). We note that the conditions of Theorem \ref{teleq} are
not satisfied by the d--metric (\ref{clgs}) (the coefficients $\tilde{g}%
_{ij}=\delta _{ij}$ are not defined as in (\ref{ehes})), so we can not
associate a geometrical mechanics model for such constructions.

There is an important generalization:

\begin{example}
We can consider $\mathcal{L}$ as a hypersurface in $TM$ for which the matrix
$\partial ^{2}\mathcal{L}/\partial y^{a}\partial y^{b}$ (i.e. the Hessian,
following the analogy with Lagrange mechanics and field theory) is constant
and nondegenerate. This states that $\tilde{g}_{ij}=const$, which produces
vanishing curvature coefficients for the canonical d--connection induced by $%
\underline{g}_{ij}(x)$ on $M$.
\end{example}

Finally, we note that a number of geometric ideas and methods applied in
this section were considered in the approaches to the geometry of
nonholonomic spaces and generalized Finsler--Lagrange geometry elaborated by
the schools of G. Vranceanu and R. Miron and by A. Bejancu in Romania \cite%
{vr1,vr2,ma1,ma2,bej,bejf}, see also Section \ref{sec5}. In these approaches
it is possible to construct geometric models with metric compatible linear
connections, which is important for developments connected with modern
(non)commutative gravity and string theory \cite{vncg,vsgg}. For Finsler
spaces with nontrivial nonmetricity, for instance, those defined by the the
Berwald and Chern connections \cite{bcs}, the resulting physical theories
with local anisotropy fall outside of the class of standard models.

\section{Curve Flows and Anholonomic Constraints}

We now formulate the geometry of curve flows adapted to the nonlinear
connection structure.

\subsection{N--adapted curve flows}

Let us consider a vector bundle $\mathcal{E}=(E,\pi ,F,M)$, $\dim E=$ $n+m$
(in the particular case, $E=TM$, we have $m=n)$ equipped with a d--metric $%
\mathbf{g}=[g,h]$ (\ref{m1}) and N--connection $N_{i}^{a}$ (\ref{whitney})
structures. A non--stretching curve $\gamma (\tau ,l)$ on $\mathbf{V,}$
where $\tau $ is a parameter and $l$ is the arclength of the curve on $%
\mathbf{V,}$ is defined by an evolution d--vector $\mathbf{Y}=\gamma _{\tau
} $ and tangent d--vector $\mathbf{X}=\gamma _{l}$ such that $\mathbf{g(X,X)}%
=1 $. Such curves $\gamma (\tau ,l)$ sweep out a two--dimensional surface in
$T_{\gamma (\tau ,l)}\mathbf{V}\subset T\mathbf{V}$.

We shall work with N--adapted bases (\ref{dder}) and (\ref{ddif}) and the
connection 1--form $\mathbf{\Gamma }_{\ \beta }^{\alpha }=\mathbf{\Gamma }%
_{\ \beta \gamma }^{\alpha }\mathbf{e}^{\gamma }$ with the coefficients $%
\mathbf{\Gamma }_{\ \beta \gamma }^{\alpha }$ for the canonical
d--connection operator $\mathbf{D}$ (\ref{candcon}) (see Appendix), acting
in the form%
\begin{equation}
\mathbf{D}_{\mathbf{X}}\mathbf{e}_{\alpha }=(\mathbf{X\rfloor \Gamma }%
_{\alpha \ }^{\ \gamma })\mathbf{e}_{\gamma }\mbox{ and }\mathbf{D}_{\mathbf{%
Y}}\mathbf{e}_{\alpha }=(\mathbf{Y\rfloor \Gamma }_{\alpha \ }^{\ \gamma })%
\mathbf{e}_{\gamma },  \label{part01}
\end{equation}%
where ``$\mathbf{\rfloor}$'' denotes the interior product and the indices
are lowered and raised respectively by the d--metric $\mathbf{g}_{\alpha
\beta }=[g_{ij},h_{ab}]$ and its inverse $\mathbf{g}^{\alpha \beta
}=[g^{ij},h^{ab}]$. We note that $\mathbf{D}_{\mathbf{X}}=\mathbf{X}^{\alpha
}\mathbf{D}_{\alpha }$ is the covariant derivation operator along curve $%
\gamma (\tau ,l)$. It is convenient to fix the N--adapted frame to be
parallel to curve $\gamma (l)$ adapted in the form
\begin{eqnarray}
e^{1} &\doteqdot &h\mathbf{X,}\mbox{ for }i=1,\mbox{ and }e^{\widehat{i}},%
\mbox{ where }h\mathbf{g(}h\mathbf{X,}e^{\widehat{i}}\mathbf{)=}0,
\label{curvframe} \\
\mathbf{e}^{n+1} &\doteqdot &v\mathbf{X,}\mbox{ for }a=n+1,\mbox{ and }%
\mathbf{e}^{\widehat{a}},\mbox{ where }v\mathbf{g(}v\mathbf{X,\mathbf{e}}^{%
\widehat{a}}\mathbf{)=}0,  \notag
\end{eqnarray}%
for $\widehat{i}=2,3,\ldots,n$ and $\widehat{a}=n+2,n+3,\ldots,n+m$. For such
frames, the covariant derivative of each ``normal'' d--vector $\mathbf{e}^{%
\widehat{\alpha }}$ is parallel to the d--vectors adapted to $\gamma (\tau
,l)$,
\begin{eqnarray}
\mathbf{D}_{\mathbf{X}}e^{\widehat{i}} &\mathbf{=}&\mathbf{-}\rho ^{\widehat{%
i}}\mathbf{(}u\mathbf{)\ X}\mbox{ and }\mathbf{D}_{h\mathbf{X}}h\mathbf{X}%
=\rho ^{\widehat{i}}\mathbf{(}u\mathbf{)\ \mathbf{e}}_{\widehat{i}},
\label{part02} \\
\mathbf{D}_{\mathbf{X}}\mathbf{\mathbf{e}}^{\widehat{a}} &\mathbf{=}&\mathbf{%
-}\rho ^{\widehat{a}}\mathbf{(}u\mathbf{)\ X}\mbox{ and }\mathbf{D}_{v%
\mathbf{X}}v\mathbf{X}=\rho ^{\widehat{a}}\mathbf{(}u\mathbf{)\ }e_{\widehat{%
a}},  \notag
\end{eqnarray}%
in terms of some coefficient functions $\rho ^{\widehat{i}}\mathbf{(}u%
\mathbf{)}$ and $\rho ^{\widehat{a}}\mathbf{(}u\mathbf{).}$ The formulas (%
\ref{part01}) and (\ref{part02}) are distinguished into h-- and
v--components for $\mathbf{X=}h\mathbf{X}+v\mathbf{X}$ and $\mathbf{D=(}h%
\mathbf{D},v\mathbf{D)}$ for $\mathbf{D=\{\Gamma }_{\ \alpha \beta }^{\gamma
}\},h\mathbf{D}=\{L_{jk}^{i},L_{bk}^{a}\}$ and $v\mathbf{D=\{}%
C_{jc}^{i},C_{bc}^{a}\}$. Along $\gamma (l)$, we can pull back differential
forms in a parallel N--adapted form. For instance, $\mathbf{\Gamma }_{\
\mathbf{X}}^{\alpha \beta }\doteqdot \mathbf{X\rfloor \Gamma }_{\ }^{\alpha
\beta }$.

An algebraic characterization of parallel frames can be obtained if we
perform a frame transformation preserving the decomposition (\ref{whitney})
to an orthonormal basis $\mathbf{e}_{\alpha ^{\prime }}$,
\begin{equation}
\mathbf{e}_{\alpha }\rightarrow A_{\alpha }^{\ \alpha ^{\prime }}(u)\
\mathbf{e}_{\alpha ^{\prime }},  \label{orthbas}
\end{equation}%
called an orthonormal d--basis, where the coefficients of the d--metric (\ref%
{m1}) are transformed into the Euclidean ones $\mathbf{g}_{\alpha ^{\prime
}\beta ^{\prime }}=\delta _{\alpha ^{\prime }\beta ^{\prime }}$. In
distinguished form, we obtain two skew matrices%
\begin{equation*}
\mathbf{\Gamma }_{h\mathbf{X}}^{i^{\prime }j^{\prime }}\doteqdot h\mathbf{%
X\rfloor \Gamma }_{\ }^{i^{\prime }j^{\prime }}=2\ e_{h\mathbf{X}%
}^{[i^{\prime }}\ \rho ^{j^{\prime }]}\mbox{ and }\mathbf{\Gamma }_{v\mathbf{%
X}}^{a^{\prime }b^{\prime }}\doteqdot v\mathbf{X\rfloor \Gamma }_{\
}^{a^{\prime }b^{\prime }}=2\mathbf{\ e}_{v\mathbf{X}}^{[a^{\prime }}\ \rho
^{b^{\prime }]}
\end{equation*}%
where
\begin{equation*}
\ e_{h\mathbf{X}}^{i^{\prime }}\doteqdot g(h\mathbf{X,}e^{i^{\prime }})=[1,%
\underbrace{0,\ldots ,0}_{n-1}]\mbox{ and }\ e_{v\mathbf{X}}^{a^{\prime
}}\doteqdot h(v\mathbf{X,}e^{a^{\prime }})=[1,\underbrace{0,\ldots ,0}_{m-1}]
\end{equation*}%
and
\begin{equation*}
\mathbf{\Gamma }_{h\mathbf{X\,}i^{\prime }}^{\qquad j^{\prime }}=\left[
\begin{array}{cc}
0 & \rho ^{j^{\prime }} \\
-\rho _{i^{\prime }} & \mathbf{0}_{[h]}%
\end{array}%
\right] \mbox{ and }\mathbf{\Gamma }_{v\mathbf{X\,}a^{\prime }}^{\qquad
b^{\prime }}=\left[
\begin{array}{cc}
0 & \rho ^{b^{\prime }} \\
-\rho _{a^{\prime }} & \mathbf{0}_{[v]}%
\end{array}%
\right]
\end{equation*}%
with $\mathbf{0}_{[h]}$ and $\mathbf{0}_{[v]}$ being respectively $%
(n-1)\times (n-1)$ and $(m-1)\times (m-1)$ matrices. The above presented
row--matrices and skew--matrices show that locally a N--anholonomic manifold
$\mathbf{V}$ of dimension $n+m$, with respect to distinguished orthonormal
frames, are characterized algebraically by pairs of unit vectors in $\mathbb{%
R}^{n}$ and $\mathbb{R}^{m}$ preserved respectively by the $SO(n-1)$ and $%
SO(m-1)$ rotation subgroups of the local N--adapted frame structure group $%
SO(n)\oplus SO(m)$. The connection matrices $\mathbf{\Gamma }_{h\mathbf{X\,}%
i^{\prime }}^{\qquad j^{\prime }}$ and $\mathbf{\Gamma }_{v\mathbf{X\,}%
a^{\prime }}^{\qquad b^{\prime }}$ belong to the orthogonal complements of
the corresponding Lie subalgebras and algebras, $\mathfrak{so}(n-1)\subset
\mathfrak{so}(n)$ and $\mathfrak{so}(m-1)\subset \mathfrak{so}(m)$. The
torsion (\ref{tors}) and curvature (\ref{curv}) (see Appendix) tensors in
orthonormal component form with respect to (\ref{curvframe}) can be mapped
into a distinguished orthonormal dual frame (\ref{orthbas}),%
\begin{equation}
\mathcal{T}^{\alpha ^{\prime }}\doteqdot \mathbf{D}_{\mathbf{X}}\mathbf{e}_{%
\mathbf{Y}}^{\alpha ^{\prime }}-\mathbf{D}_{\mathbf{Y}}\mathbf{e}_{\mathbf{X}%
}^{\alpha ^{\prime }}+\mathbf{e}_{\mathbf{Y}}^{\beta ^{\prime }}\Gamma _{%
\mathbf{X}\beta ^{\prime }}^{\quad \alpha ^{\prime }}-\mathbf{e}_{\mathbf{X}%
}^{\beta ^{\prime }}\Gamma _{\mathbf{Y}\beta ^{\prime }}^{\quad \alpha
^{\prime }},  \label{mtors}
\end{equation}%
and
\begin{equation}
\mathcal{R}_{\beta ^{\prime }}^{\;\alpha ^{\prime }}(\mathbf{X,Y})=\mathbf{D}%
_{\mathbf{Y}}\Gamma _{\mathbf{X}\beta ^{\prime }}^{\quad \alpha ^{\prime }}-%
\mathbf{D}_{\mathbf{X}}\Gamma _{\mathbf{Y}\beta ^{\prime }}^{\quad \alpha
^{\prime }}+\Gamma _{\mathbf{Y}\beta ^{\prime }}^{\quad \gamma ^{\prime
}}\Gamma _{\mathbf{X}\gamma ^{\prime }}^{\quad \alpha ^{\prime }}-\Gamma _{%
\mathbf{X}\beta ^{\prime }}^{\quad \gamma ^{\prime }}\Gamma _{\mathbf{Y}%
\gamma ^{\prime }}^{\quad \alpha ^{\prime }},  \label{mcurv}
\end{equation}%
where $\mathbf{e}_{\mathbf{Y}}^{\alpha ^{\prime }}\doteqdot \mathbf{g}(%
\mathbf{Y},\mathbf{e}^{\alpha ^{\prime }})$ and $\Gamma _{\mathbf{Y}\beta
^{\prime }}^{\quad \alpha ^{\prime }}\doteqdot \mathbf{Y\rfloor }\Gamma
_{\beta ^{\prime }}^{\;\alpha ^{\prime }}=\mathbf{g}(\mathbf{e}^{\alpha
^{\prime }},\mathbf{D}_{\mathbf{Y}}\mathbf{e}_{\beta ^{\prime }})$ define
respectively the N--adapted orthonormal frame row--matrix and the canonical
d--connection skew--matrix in the flow directions, where $\mathcal{R}_{\beta
^{\prime }}^{\;\alpha ^{\prime }}(\mathbf{X,Y})\doteqdot \mathbf{g}(\mathbf{e%
}^{\alpha ^{\prime }},[\mathbf{D}_{\mathbf{X}}$, $\mathbf{D}_{\mathbf{Y}}]%
\mathbf{e}_{\beta ^{\prime }})$ is the curvature matrix. Both torsion and
curvature components can be distinguished into h-- and v--components like (%
\ref{dtors}) and (\ref{dcurv}), by considering N--adapted decompositions of
type
\begin{equation*}
\mathbf{g}=[g,h],\mathbf{e}_{\beta ^{\prime }}=(\mathbf{e}_{j^{\prime
}},e_{b^{\prime }}),\mathbf{e}^{\alpha ^{\prime }}=(e^{i^{\prime
}},e^{a^{\prime }}),\mathbf{X=}h\mathbf{X}+v\mathbf{X,D=(}h\mathbf{D},v%
\mathbf{D).}
\end{equation*}%
Finally, we note that the matrices for torsion (\ref{mtors}) and curvature (%
\ref{mcurv}) can be computed for any metric compatible linear connection
like the Levi Civita and the canonical d--connection. For our purposes, we
are interested to define a frame in which the curvature tensor has constant
coefficients and the torsion tensor vanishes.

\subsection{Anholonomic bundles with constant matrix curvature}

For vanishing N--connection curvature and torsion, we get a holonomic
Riemannian manifold with constant curvature, wherein the equations (\ref%
{mtors}) and (\ref{mcurv}) directly encode a bi-Hamiltonian structure, see
details in Refs. \cite{saw,anc1}. A larger class of Riemannian manifolds for
which the frame curvature matrix is constant consists of the symmetric
spaces $M=G/H$ for compact semisimple Lie groups $G\supset H$ (where $H$ is
required to be invariant under an involutive automorphism of $G$). A complete
classification and summary of main results for such spaces is given in Refs. %
\cite{helag,kob}. Constancy of the frame curvature matrix is a consequence
of the fact that the Riemannian curvature and the metric tensors on the
curved manifold $M=G/H$ are covariantly constant and $G$--invariant.
In Ref. \cite{anc2}, a bi-Hamiltonian structure was shown to be encoded
in the frame equations analogous to (\ref{mtors}) and (\ref{mcurv})
for the symmetric spaces $M=G/SO(n)$ with $H=SO(n)\supset O(n-1)$.
All such spaces are exhausted by $G=SO(n+1),SU(n)$. The derivation of the
bi-Hamiltonian structure exploited the canonical soldering of the spaces
$G/SO(n)$ onto Klein geometries \cite{sharpe}.
A similar derivation \cite{ancima} holds for the Lie groups $G=SO(n+1),SU(n)$
themselves when they are viewed as symmetric spaces
in the standard manner \cite{kob}.
A broad generalization of this derivation to arbitrary
Riemannian symmetric spaces, including arbitrary compact semisimple Lie groups,
has been obtained in Ref. \cite{ancjgp}.

\subsubsection{Symmetric nonholonomic tangent bundles}

We suppose that the base manifold is a constant-curvature symmetric space of
the form $M=hG/SO(n)$ with the isotropy subgroup $hH=SO(n)\supset O(n)$
and that the typical fiber space is likewise of the constant-curvature
symmetric form $F=vG/SO(m)$
with the isotropy subgroup $vH=SO(m)\supset O(m)$. This means
(according to Cartan's classification of symmetric spaces \cite{helag}) that
we have $hG=SO(n+1)$ and $vG=SO(m+1)$,
which is general enough for a study of real holonomic and nonholonomic
manifolds and geometric mechanics models.
\footnote{%
It is necessary to consider $hG=SU(n)$ and $vG=SU(m)$ for geometric
models involving spinor and gauge fields} \

Our aim is to solder in a canonical way (like in the N--connection geometry)
the horizontal and vertical symmetric Riemannian spaces of dimension $n$ and
$m$ with a (total) symmetric Riemannian space $V$ of dimension $n+m$, when $%
V=G/SO(n+m)$ with the isotropy group $H=SO(n+m)\supset O(n+m)$ and $%
G=SO(n+m+1)$. First, we note that the just mentioned horizontal, vertical
and total symmetric Riemannian spaces can be identified with respective
Klein geometries. For instance, the metric tensor $hg=\{\mathring{g}_{ij}\}$
on $M$ is defined by the Cartan--Killing inner product $<\cdot ,\cdot >_{h}$
on $T_{x}hG\simeq h\mathfrak{g}$ restricted to the Lie algebra quotient
spaces $h\mathfrak{p=}h\mathfrak{g/}h\mathfrak{h,}$ with $T_{x}hH\simeq h%
\mathfrak{h,}$ where $h\mathfrak{g=}h\mathfrak{h}\oplus h\mathfrak{p}$ is
stated such that there is an involutive automorphism of $hG$ leaving $hH$
fixed, i.e. $[h\mathfrak{h,}h\mathfrak{p]}\subseteq $ $h\mathfrak{p}$ and $[h%
\mathfrak{p,}h\mathfrak{p]}\subseteq h\mathfrak{h.}$ In a similar form, we
can define the inner products and\ Lie algebras: $vg =\{\mathring{h}_{ab}\}$
is given by restriction of $<\cdot ,\cdot>_{v}$ on $T_{y}vG\simeq v\mathfrak{%
g}$ to $v\mathfrak{p=}v\mathfrak{g/}v\mathfrak{h}$ with $T_{y}vH \simeq v%
\mathfrak{h}$, $v\mathfrak{g=}v\mathfrak{h}\oplus v\mathfrak{p}$, where
\begin{equation}
[v\mathfrak{h,}v\mathfrak{p]}\subseteq v\mathfrak{p} ,\qquad [v\mathfrak{p,}v%
\mathfrak{p]}\subseteq v\mathfrak{h} ;  \label{algstr}
\end{equation}
likewise $\mathbf{g} =\{\mathring{g}_{\alpha \beta }\}$ is given by
restriction of $<\cdot ,\cdot>_{\mathbf{g}}$ on $T_{(x,y)}G\simeq \mathfrak{g%
}$ to $\mathfrak{p=g/h}$, with $T_{(x,y)}H \simeq \mathfrak{h,g=h}\oplus
\mathfrak{p}$ where
\begin{equation}
[\mathfrak{h,p]}\subseteq \mathfrak{p} ,\quad [\mathfrak{p,p]} \subseteq
\mathfrak{h} .  \notag
\end{equation}
We parametrize the metric structure with constant coefficients on $%
V=G/SO(n+m)$ in the form%
\begin{equation*}
\mathring{g}=\mathring{g}_{\alpha \beta }du^{\alpha }\otimes du^{\beta },
\end{equation*}%
where $u^{\alpha }$ are local coordinates and
\begin{equation}
\mathring{g}_{\alpha \beta }=\left[
\begin{array}{cc}
\mathring{g}_{ij}+\mathring{N}_{i}^{a}N_{j}^{b}\mathring{h}_{ab} & \mathring{%
N}_{j}^{e}\mathring{h}_{ae} \\
\mathring{N}_{i}^{e}\mathring{h}_{be} & \mathring{h}_{ab}%
\end{array}%
\right]  \label{constans}
\end{equation}%
where trivial, constant, N--connection coefficients $\mathring{N}_{j}^{e}=%
\mathring{h}^{eb}\mathring{g}_{jb}$ are defined in terms of any given
coefficients $\mathring{h}^{eb}$ and $\mathring{g}_{jb}$, i.e. from the
inverse metrics coefficients defined respectively by the metric on $%
hG=SO(n+1)$ and by  the metric coefficients out of $(n\times n)$-- and $%
(m\times m)$-- terms of the metric $\mathring{g}_{\alpha \beta }$. As a
result, we define an equivalent d--metric structure of type (\ref{m1})
\begin{eqnarray}
\mathbf{\mathring{g}} &=&\ \mathring{g}_{ij}\ e^{i}\otimes e^{j}+\ \mathring{%
h}_{ab}\ \mathbf{\mathring{e}}^{a}\otimes \mathbf{\mathring{e}}^{b},
\label{m1const} \\
e^{i} &=&dx^{i},\ \;\mathbf{\mathring{e}}^{a}=dy^{a}+\mathring{N}%
_{i}^{e}dx^{i} ,  \notag
\end{eqnarray}%
defining a trivial $(n+m)$--splitting $\mathbf{\mathring{g}=}\mathring{g}%
\mathbf{\oplus _{\mathring{N}}}\mathring{h}\mathbf{\ }$because all
nonholonomy coefficients $\mathring{W}_{\alpha \beta }^{\gamma }$ and
N--connection curvature coefficients $\mathring{\Omega}_{ij}^{a}$ are zero.
In a more general form, we can consider any covariant coordinate transforms
of (\ref{m1const}) preserving the\ $(n+m)$--splitting resulting in any $%
W_{\alpha \beta }^{\gamma }=0$ (\ref{anhrel}) and $\Omega _{ij}^{a}=0$ (\ref%
{ncurv}). We denote such trivial N--anholonomic Riemannian spaces as $%
\mathbf{\mathring{V}}=[hG=SO(n+1)$, $vG=SO(m+1),\;\mathring{N}_{i}^{e}]$. It
can be considered that such trivially N--anholonomic group spaces possess a
Lie d--algebra symmetry $\mathfrak{so}_{\mathring{N}}(n+m)\doteqdot
\mathfrak{so}(n)\oplus \mathfrak{so}(m)$.

The simplest generalization for a vector bundle $\mathbf{\mathring{E}}$ is
to consider nonholonomic distributions on $V=G/SO(n+m)$ defined locally by
arbitrary N--connection coefficients $N_{i}^{a}(x,y)$ with nonvanishing $%
W_{\alpha \beta }^{\gamma }$ and $\Omega _{ij}^{a}$ but with constant
d--metric coefficients when
\begin{eqnarray}
\mathbf{g} &=&\ \mathring{g}_{ij}\ e^{i}\otimes e^{j}+\ \mathring{h}_{ab}\
\mathbf{e}^{a}\otimes \mathbf{e}^{b},  \label{m1b} \\
e^{i} &=&dx^{i},\ \mathbf{e}^{a}=dy^{a}+N_{i}^{a}(x,y)dx^{i}.  \notag
\end{eqnarray}%
This metric is very similar to (\ref{clgs}) but with the coefficients $\
\mathring{g}_{ij}\ $\ and $\ \mathring{h}_{ab}$ induced by the corresponding
Lie d--algebra structure $\mathfrak{so}_{\mathring{N}}(n+m)$. Such spaces
transform into N--anholonomic Riemann--Cartan manifolds $\mathbf{\mathring{V}%
}_{\mathbf{N}}=[hG=SO(n+1)$, $vG=SO(m+1),\;N_{i}^{e}]$ with nontrivial
N--connection curvature and induced d--torsion coefficients of the canonical
d--connection (see formulas (\ref{dtors}) computed for constant d--metric
coefficients and the canonical d--connection coefficients in (\ref{candcon}%
)). One has zero curvature for the canonical d--connection (in general, such
spaces are curved ones with generically off--diagonal metric (\ref{m1b}) and
nonzero curvature tensor for the Levi Civita connection).\footnote{%
With constant values for the d--metric coefficients we get zero coefficients
for the canonical d--connection which then yields vanishing curvature
coefficients (\ref{dcurv}).} This allows us to classify the N--anholonomic
manifolds (and vector bundles) as having the same group and algebraic
properties as pairs of symmetric Riemannian spaces of dimension $n$ and $m$
but nonholonomically soldered to the symmetric Riemannian space of dimension
$n+m$. With respect to N--adapted orthonormal bases (\ref{orthbas}), with
distinguished h-- and v--subspaces, we obtain the same inner products and
group and Lie algebra spaces as in (\ref{algstr}).

The classification of N--anholonomic vector bundles is almost similar to
that for symmetric Riemannian spaces if we consider that $n=m$ and try to
model tangent bundles of such spaces, provided with N--connection structure.
For instance, we can take a (semi) Riemannian structure with the
N--connection induced by a absolute energy structure like in (\ref{cnlce})
and with the canonical d--connection structure (\ref{candcon}), for $\tilde{g%
}_{ef}=\ \mathring{g}_{ab}$ (\ref{aux4}). A straightforward computation of
the canonical d--connection coefficients\footnote{%
On tangent bundles, such d--connections can be defined to be torsionless}
and of d--curvatures for $\;^{\circ }\tilde{g}_{ij}$ and $\;^{\circ }%
\tilde{N}_{\ j}^{i}$ proves that the nonholonomic Riemannian manifold $%
\left( M=SO(n+1)/SO(n),\;^{\circ }\mathcal{L}\right) $ possess constant both
zero canonical d--con\-nec\-ti\-on curvature and torsion but with induced
nontrivial N--connection curvature $\;^{\circ }\tilde{\Omega}_{jk}^{i}$.
Such spaces, being tangent to symmetric Riemannian spaces, are classified
similarly to the Riemannian ones with constant matrix curvature, see (\ref%
{algstr}) for $n=m$, but possessing a nonholonomic structure induced by
generating function $\;^{\circ }\mathcal{L}$.

\subsubsection{N--anholonomic Klein spaces}

The bi-Hamiltonian and solitonic constructions
in Refs. \cite{ancjgp,anc2,anc1,aw}
are based on soldering Riemannian symmetric--space geometries onto Klein
geometries \cite{sharpe}. For the N--anholonomic spaces of dimension $n+n$,
with constant d--curvatures, similar constructions will hold but we have to
adapt them to the N--connection structure.

There are two Hamiltonian variables given by the principal normals $%
\;^{h}\nu $ and $\;^{v}\nu $, respectively, in the horizontal and vertical
subspaces, defined by the canonical d--connection $\mathbf{D}=(h\mathbf{D},v%
\mathbf{D})$, see formulas (\ref{curvframe}) and (\ref{part02}),
\begin{equation*}
\;^{h}\nu \doteqdot \mathbf{D}_{h\mathbf{X}}h\mathbf{X}=\nu ^{\widehat{i}}%
\mathbf{\mathbf{e}}_{\widehat{i}}\mbox{\ and \ }\;^{v}\nu \doteqdot \mathbf{D%
}_{v\mathbf{X}}v\mathbf{X}=\nu ^{\widehat{a}}e_{\widehat{a}}.
\end{equation*}%
This normal d--vector $\mathbf{v}=(\;^{h}\nu $, $\;^{v}\nu )$, with
components of type $\mathbf{\nu }^{\alpha }=(\nu ^{i}$, $\;\nu ^{a})=(\nu
^{1}$, $\nu ^{\widehat{i}},\nu ^{n+1},\nu ^{\widehat{a}})$, is defined with
respect to the tangent direction of curve $\gamma $. There is also the
principal normal d--vector $\mathbf{\varpi }=(\;^{h}\varpi ,\;^{v}\varpi )$
with components of type $\mathbf{\varpi }^{\alpha }=(\varpi ^{i}$, $\;\varpi
^{a})=(\varpi ^{1},\varpi ^{\widehat{i}},\varpi ^{n+1},\varpi ^{\widehat{a}})
$ defined with respect to the flow direction, with
\begin{equation*}
\;^{h}\varpi \doteqdot \mathbf{D}_{h\mathbf{Y}}h\mathbf{X=}\varpi ^{\widehat{%
i}}\mathbf{\mathbf{e}}_{\widehat{i}},\;^{v}\varpi \doteqdot \mathbf{D}_{v%
\mathbf{Y}}v\mathbf{X}=\varpi ^{\widehat{a}}e_{\widehat{a}},
\end{equation*}%
representing a Hamiltonian d--covector field. We can consider that the
normal part of the flow d--vector
\begin{equation*}
\mathbf{h}_{\perp }\doteqdot \mathbf{Y}_{\perp }=h^{\widehat{i}}\mathbf{%
\mathbf{e}}_{\widehat{i}}+h^{\widehat{a}}e_{\widehat{a}}
\end{equation*}%
represents a Hamiltonian d--vector field. For such configurations, we can
consider parallel N--adapted frames $\mathbf{e}_{\alpha ^{\prime }}=(\mathbf{%
e}_{i^{\prime }},e_{a^{\prime }})$ when the h--variables $\nu ^{\widehat{%
i^{\prime }}}$, $\varpi ^{\widehat{i^{\prime }}},h^{\widehat{i^{\prime }}}$
are respectively encoded in the top row of the horizontal canonical
d--connection matrices $\mathbf{\Gamma }_{h\mathbf{X\,}i^{\prime }}^{\qquad
j^{\prime }}$ and $\mathbf{\Gamma }_{h\mathbf{Y\,}i^{\prime }}^{\qquad
j^{\prime }}$ and in the row matrix $\left( \mathbf{e}_{\mathbf{Y}%
}^{i^{\prime }}\right) _{\perp }\doteqdot \mathbf{e}_{\mathbf{Y}}^{i^{\prime
}}-g_{\parallel }\;\mathbf{e}_{\mathbf{X}}^{i^{\prime }}$ where $%
g_{\parallel }\doteqdot g(h\mathbf{Y,}h\mathbf{X})$ is the tangential
h--part of the flow d--vector. A similar encoding holds for v--variables $%
\nu ^{\widehat{a^{\prime }}},\varpi ^{\widehat{a^{\prime }}},h^{\widehat{%
a^{\prime }}}$ in the top row of the vertical canonical d--connection
matrices \ $\mathbf{\Gamma }_{v\mathbf{X\,}a^{\prime }}^{\qquad b^{\prime }}$
and $\mathbf{\Gamma }_{v\mathbf{Y\,}a^{\prime }}^{\qquad b^{\prime }}$ and
in the row matrix $\left( \mathbf{e}_{\mathbf{Y}}^{a^{\prime }}\right)
_{\perp }\doteqdot \mathbf{e}_{\mathbf{Y}}^{a^{\prime }}-h_{\parallel }\;%
\mathbf{e}_{\mathbf{X}}^{a^{\prime }}$ where $h_{\parallel }\doteqdot h(v%
\mathbf{Y,}v\mathbf{X})$ is the tangential v--part of the flow d--vector. In
an abbreviated notation, we shall write $\mathbf{v}^{\alpha ^{\prime }}$ and
$\mathbf{\varpi }^{\alpha ^{\prime }}$ where the primed Greek indices $%
\alpha ^{\prime },\beta ^{\prime },..$. will denote both N--adapted and then
orthonormalized components of geometric objects (like d--vectors,
d--covectors, d--tensors, d--groups, d--algebras, d--matrices) admitting
further decompositions into h-- and v--components defined as non-integrable
distributions of such objects.

With respect to N--adapted orthonormal frames, the geometry of
N--anholonomic manifolds is defined algebraically, on their tangent bundles,
by pairs of horizontal and vertical Klein spaces (see \cite{sharpe} for a
summary of Klein geometry, and for bi-Hamiltonian soliton constructions,
see \cite{ancjgp,anc1}).
The N--connection structure induces a N--anholonomic Klein geometry
stated by two left--invariant $h\mathfrak{g}$-- and $v\mathfrak{g}$--valued
Maurer--Cartan forms on the Lie d--group $\mathbf{G}=(h\mathbf{G},v\mathbf{G}%
)$ that are identified with the zero--curvature canonical d--connection
1--form $\;^{\mathbf{G}}\mathbf{\Gamma }=\{\;^{\mathbf{G}}\mathbf{\Gamma }%
_{\ \beta ^{\prime }}^{\alpha ^{\prime }}\}$, via
\begin{equation*}
\;^{\mathbf{G}}\mathbf{\Gamma }_{\ \beta ^{\prime }}^{\alpha ^{\prime }}=\;^{%
\mathbf{G}}\mathbf{\Gamma }_{\ \beta ^{\prime }\gamma ^{\prime }}^{\alpha
^{\prime }}\mathbf{e}^{\gamma ^{\prime }}=\;^{h\mathbf{G}}L_{\;j^{\prime
}k^{\prime }}^{i^{\prime }}\mathbf{e}^{k^{\prime }}+\;^{v\mathbf{G}%
}C_{\;j^{\prime }k^{\prime }}^{i^{\prime }}e^{k^{\prime }}.
\end{equation*}%
For the case of a trivial N--connection structure in vector bundles with the
base and typical fiber spaces being symmetric Riemannian spaces, we identify
$\;^{h\mathbf{G}}L_{\;j^{\prime }k^{\prime }}^{i^{\prime }}$ and $\;^{v%
\mathbf{G}}C_{\;j^{\prime }k^{\prime }}^{i^{\prime }}$ with the coefficients
of the Cartan connections $\;^{h\mathbf{G}}L$ and $\;^{v\mathbf{G}}C$,
respectively, for the $h\mathbf{G}$ and $v\mathbf{G,}$ both with vanishing
curvatures. We have
\begin{equation*}
d\;^{\mathbf{G}}\mathbf{\Gamma +}\frac{1}{2}\mathbf{[\;^{\mathbf{G}}\mathbf{%
\Gamma ,}\;^{\mathbf{G}}\mathbf{\Gamma }]=0}
\end{equation*}%
and, respectively, for h-- and v--components, $d\;^{h\mathbf{G}}\mathbf{L}+%
\frac{1}{2}\mathbf{[\;^{h\mathbf{G}}L,\;^{h\mathbf{G}}L]}=0$ and $d\;^{v%
\mathbf{G}}\mathbf{C}+\frac{1}{2}[\;^{v\mathbf{G}}\mathbf{C}$, $\;^{v\mathbf{%
G}}\mathbf{C}]=0, $ where $d$ denotes the total derivative on the d--group
manifold $\mathbf{G}=h\mathbf{G}\oplus v\mathbf{G}$ or its restrictions on $h%
\mathbf{G}$ or $v\mathbf{G.}$ We can consider that $\;^{\mathbf{G}}\mathbf{%
\Gamma }$ defines the so--called Cartan d--connection for non-integrable
N--connection structures (see details and supersymmetric/ noncommutative
developments in \cite{vncg,vsgg}).

The Cartan d--connection determines a N--anholonomic Riemannian geometric
structure on the nonholonomic bundle
\begin{equation*}
\mathbf{\mathring{E}}=[hG=SO(n+1), vG=SO(m+1),\;N_{i}^{e}],
\end{equation*}
derived through the following decomposition of the Lie d--algebra $\mathfrak{%
g}=h\mathfrak{g}\oplus v\mathfrak{g}$.  For the horizontal splitting, we
have $h\mathfrak{g}=\mathfrak{so}(n)\oplus h\mathfrak{p,}$ with $[h\mathfrak{%
p},h\mathfrak{p}]\subset \mathfrak{so}(n)$ and, for the vertical splitting,
we have $[\mathfrak{so}(n),h\mathfrak{p}]\subset h\mathfrak{p;} $ for the
vertical splitting, $v\mathfrak{g}=\mathfrak{so}(m)\oplus v\mathfrak{p,}$
with $[v\mathfrak{p},v\mathfrak{p}]\subset \mathfrak{so}(m)$ and $[\mathfrak{%
so}(m),v\mathfrak{p}]\subset v\mathfrak{p}$. When $n=m$, any canonical
d--objects (N--connection, d--metric, d--connection, etc.) derived from (\ref%
{m1b})  using the Cartan d--connection agree with ones determined by the
canonical d--connection (\ref{candcontm}) on a tangent bundle.

It is useful to consider a quotient space with distinguished structure group
$\mathbf{V}_{\mathbf{N}}=\mathbf{G}/SO(n)\oplus $ $SO(m)$ regarding $\mathbf{%
G}$ as a principal $\left( SO(n)\oplus SO(m)\right) $--bundle over $\mathbf{%
\mathring{E}}$, which is a N--anholonomic bundle. In this case, we can
always fix a local section of this bundle and pull--back $\;^{\mathbf{G}}%
\mathbf{\Gamma }$ to give a $\left( h\mathfrak{g}\oplus v\mathfrak{g}\right)
$--valued 1--form $^{\mathfrak{g}}\mathbf{\Gamma }$ in a point $u\in \mathbf{%
\mathring{E}}$. Any change of local sections define $SO(n)\oplus $ $SO(m)$
gauge transformations of the canonical d--connection $^{\mathfrak{g}}\mathbf{%
\Gamma ,}$ all preserving the nonholonomic decomposition (\ref{whitney}).

There are involutive automorphisms $h\sigma =\pm 1$ and $v\sigma =\pm 1$,
respectively, of $h\mathfrak{g}$ and $v\mathfrak{g,}$ defined that $%
\mathfrak{so}(n)$ (or $\mathfrak{so}(m)$) is the eigenspace $h\sigma =+1$
(or $v\sigma =+1)$ and $h\mathfrak{p}$ (or $v\mathfrak{p}$) is the
eigenspace $h\sigma =-1$ (or $v\sigma =-1)$. By means of an N--adapted
decomposition relative to these eigenspaces, the symmetric part of the
canonical d--connection
\begin{equation*}
\mathbf{\Gamma \doteqdot }\frac{1}{2}\left( ^{\mathfrak{g}}\mathbf{\Gamma +}%
\sigma \left( ^{\mathfrak{g}}\mathbf{\Gamma }\right) \right) ,
\end{equation*}%
defines a $\left( \mathfrak{so}(n)\oplus \mathfrak{so}(m)\right) $--valued
d--connection 1--form, with respective h- and v--splitting $\mathbf{%
L\doteqdot }\frac{1}{2}\left( ^{h\mathfrak{g}}\mathbf{L+}h\sigma \left( ^{h%
\mathfrak{g}}\mathbf{L}\right) \right)$ and $\mathbf{C\doteqdot }\frac{1}{2}%
( ^{v\mathfrak{g}}\mathbf{C} +h\sigma ( ^{v \mathfrak{g}}\mathbf{C}))$.
Likewise the antisymmetric part
\begin{equation*}
\mathbf{e\doteqdot }\frac{1}{2}\left( ^{\mathfrak{g}}\mathbf{\Gamma -}\sigma
\left( ^{\mathfrak{g}}\mathbf{\Gamma }\right) \right) ,
\end{equation*}%
defines a $\left( h\mathfrak{p}\oplus v\mathfrak{p}\right) $--valued
N--adapted coframe for the Cartan--Killing inner product $<\cdot ,\cdot >_{%
\mathfrak{p}}$ on $T_{u}\mathbf{G}\simeq h\mathfrak{g}\oplus v\mathfrak{g}$
restricted to $T_{u}\mathbf{V}_{\mathbf{N}}\simeq \mathfrak{p}$, with
respective h- and v--splitting $h\mathbf{e\doteqdot }\frac{1}{2}\left( ^{h%
\mathfrak{g}}\mathbf{L-}h\sigma \left( ^{h\mathfrak{g}}\mathbf{L}\right)
\right)$ and $v\mathbf{e\doteqdot }\frac{1}{2} ( ^{v\mathfrak{g}}\mathbf{C}
- h\sigma ( ^{v\mathfrak{g}}\mathbf{C}))$. This inner product, distinguished
into h- and v--components, provides a d--metric structure of type $\mathbf{g}%
=[g,h]$ (\ref{m1}), where $g=<h\mathbf{e\otimes }h\mathbf{e}>_{h\mathfrak{p}}
$ and $h=<v\mathbf{e\otimes }v\mathbf{e}>_{v\mathfrak{p}}$ on $\mathbf{V}_{%
\mathbf{N}}=\mathbf{G}/SO(n)\oplus $ $SO(m)$.

We generate a $\mathbf{G}$--invariant d--derivative $\mathbf{D}$ whose
restriction to the tangent space $T\mathbf{V}_{\mathbf{N}}$ for any
N--anholonomic curve flow $\gamma (\tau ,{l})$ in $\mathbf{V}_{\mathbf{N}}=%
\mathbf{G}/SO(n)$ $\oplus $ $SO(m)$ is defined via%
\begin{equation}
\mathbf{D}_{\mathbf{X}}\mathbf{e=}\left[ \mathbf{e}, \gamma _l\rfloor
\mathbf{\Gamma }\right] \mbox{\ and \ }\mathbf{D}_{\mathbf{Y}}\mathbf{e=}%
\left[ \mathbf{e},\gamma _{\mathbf{\tau }}\rfloor \mathbf{\Gamma }\right] ,
\label{aux33}
\end{equation}%
which have h- and v--decompositions. The derivatives $\mathbf{D}_{\mathbf{X}%
} $ and $\mathbf{D}_{\mathbf{Y}}$ are equivalent to those considered in (\ref%
{part01}) and obey the Cartan structure equations (\ref{mtors}) and (\ref%
{mcurv}). For canonical d--connections, a large class of N--anholonomic
spaces of dimension $n=m$, the d--torsions are zero and the d--curvatures
have constant coefficients.

Let $\mathbf{e}^{\alpha ^{\prime }}=(e^{i^{\prime }},\mathbf{e}^{a^{\prime
}})$ be a N--adapted orthonormal coframe being identified with the $\left( h%
\mathfrak{p}\oplus v\mathfrak{p}\right) $--valued coframe $\mathbf{e} $ in a
fixed orthonormal basis for $\mathfrak{p=}h\mathfrak{p}\oplus v\mathfrak{%
p\subset }h\mathfrak{g}\oplus v\mathfrak{g.}$ Considering the kernel/
cokernel of Lie algebra multiplications in the h- and v--subspaces,
respectively, $\left[ \mathbf{e}_{h\mathbf{X}},\cdot \right] _{h\mathfrak{g}%
} $ and $\left[ \mathbf{e}_{v\mathbf{X}},\cdot \right] _{v\mathfrak{g}}$, we
can decompose the coframes into parallel and perpendicular parts with
respect to $\mathbf{e}_{\mathbf{X}}$. We write
\begin{equation*}
\mathbf{e=(e}_{C}=h\mathbf{e}_{C}+v\mathbf{e}_{C},\mathbf{e}_{C^{\perp }}=h%
\mathbf{e}_{C^{\perp }}+v\mathbf{e}_{C^{\perp }}\mathbf{),}
\end{equation*}%
for $\mathfrak{p(}=h\mathfrak{p}\oplus v\mathfrak{p)}$--valued mutually
orthogonal d--vectors $\mathbf{e}_{C}$ and $\mathbf{e}_{C^{\perp }}$,
satisfying the conditions $\left[ \mathbf{e}_{\mathbf{X}},\mathbf{e}_{C}%
\right] _{\mathfrak{g}}=0$ and $\left[ \mathbf{e}_{\mathbf{X}},\mathbf{e}%
_{C^{\perp }}\right] _{\mathfrak{g}}\neq 0;$ such conditions can be stated
in h- and v--component form, respectively, $\left[ h\mathbf{e}_{\mathbf{X}},h%
\mathbf{e}_{C}\right] _{h\mathfrak{g}}=0$, $\left[ h\mathbf{e}_{\mathbf{X}},h%
\mathbf{e}_{C^{\perp }}\right] _{h\mathfrak{g}}\neq 0$ and $\left[ v\mathbf{e%
}_{\mathbf{X}},v\mathbf{e}_{C}\right] _{v\mathfrak{g}}=0$, $\left[ v\mathbf{e%
}_{\mathbf{X}},v\mathbf{e}_{C^{\perp }}\right] _{v\mathfrak{g}}\neq 0$. Then
we have the algebraic decompositions
\begin{equation*}
T_{u}\mathbf{V}_{\mathbf{N}}\simeq \mathfrak{p=}h\mathfrak{p}\oplus v%
\mathfrak{p}=\mathfrak{g=}h\mathfrak{g}\oplus v\mathfrak{g}/\mathfrak{so}%
(n)\oplus \mathfrak{so}(m)
\end{equation*}%
and
\begin{equation*}
\mathfrak{p=p}_{C}\oplus \mathfrak{p}_{C^{\perp }}=\left( h\mathfrak{p}%
_{C}\oplus v\mathfrak{p}_{C}\right) \oplus \left( h\mathfrak{p}_{C^{\perp
}}\oplus v\mathfrak{p}_{C^{\perp }}\right) ,
\end{equation*}%
with $\mathfrak{p}_{\parallel }\subseteq \mathfrak{p}_{C}$ and $\mathfrak{p}%
_{C^{\perp }}\subseteq \mathfrak{p}_{\perp }$, where $\left[ \mathfrak{p}%
_{\parallel },\mathfrak{p}_{C}\right] =0$, $<\mathfrak{p}_{C^{\perp }},%
\mathfrak{p}_{C}>=0$, but $\left[ \mathfrak{p}_{\parallel },\mathfrak{p}%
_{C^{\perp }}\right] \neq 0$ (i.e. $\mathfrak{p}_{C}$ is the centralizer of $%
\mathbf{e}_{\mathbf{X}}$ in $\mathfrak{p=}h\mathfrak{p}\oplus v\mathfrak{%
p\subset }h\mathfrak{g}\oplus v\mathfrak{g);}$ in h- \ and v--components,
thus $h\mathfrak{p}_{\parallel }\subseteq h\mathfrak{p}_{C}$ and $h\mathfrak{%
p}_{C^{\perp }}\subseteq h\mathfrak{p}_{\perp }$, where $\left[ h\mathfrak{p}%
_{\parallel },h\mathfrak{p}_{C}\right] =0$, $<h\mathfrak{p}_{C^{\perp }},h%
\mathfrak{p}_{C}>=0$, but $\left[ h\mathfrak{p}_{\parallel },h\mathfrak{p}%
_{C^{\perp }}\right] \neq 0$ (i.e. $h\mathfrak{p}_{C}$ is the centralizer of
$\mathbf{e}_{h\mathbf{X}}$ in $h\mathfrak{p\subset }h\mathfrak{g)}$ and $v%
\mathfrak{p}_{\parallel }\subseteq v\mathfrak{p}_{C}$ and $v\mathfrak{p}%
_{C^{\perp }}\subseteq v\mathfrak{p}_{\perp }$, where $\left[ v\mathfrak{p}%
_{\parallel },v\mathfrak{p}_{C}\right] =0$, $<v\mathfrak{p}_{C^{\perp }},v%
\mathfrak{p}_{C}>=0$, but $\left[ v\mathfrak{p}_{\parallel },v\mathfrak{p}%
_{C^{\perp }}\right] \neq 0$ (i.e. $v\mathfrak{p}_{C}$ is the centralizer of
$\mathbf{e}_{v\mathbf{X}}$ in $v\mathfrak{p\subset }v\mathfrak{g).}$ Using
the canonical d--connection derivative $\mathbf{D}_{\mathbf{X}}$ of a
d--covector algebraically perpendicular (or parallel) to $\mathbf{e}_{%
\mathbf{X}}$, we get a new d--vector which is algebraically parallel (or
perpendicular) to $\mathbf{e}_{\mathbf{X}}$, i.e. $\mathbf{D}_{\mathbf{X}}%
\mathbf{e}_{C}\in \mathfrak{p}_{C^{\perp }}$ (or $\mathbf{D}_{\mathbf{X}}%
\mathbf{e}_{C^{\perp }}\in \mathfrak{p}_{C});$ in h- \ and v--components
such formulas are written $\mathbf{D}_{h\mathbf{X}}h\mathbf{e}_{C}\in h%
\mathfrak{p}_{C^{\perp }}$ (or $\mathbf{D}_{h\mathbf{X}}h\mathbf{e}%
_{C^{\perp }}\in h\mathfrak{p}_{C})$ and $\mathbf{D}_{v\mathbf{X}}v\mathbf{e}%
_{C}\in v\mathfrak{p}_{C^{\perp }}$ (or $\mathbf{D}_{v\mathbf{X}}v\mathbf{e}%
_{C^{\perp }}\in v\mathfrak{p}_{C})$. All such d--algebraic relations can be
written in N--anholonomic manifolds and canonical d--connection settings,
for instance, using certain relations of type
\begin{equation*}
\mathbf{D}_{\mathbf{X}}(\mathbf{e}^{\alpha ^{\prime }})_{C}=\mathbf{v}%
_{~\beta ^{\prime }}^{\alpha ^{\prime }}(\mathbf{e}^{\beta ^{\prime
}})_{C^{\perp }}\mbox{ \ and \ }\mathbf{D}_{\mathbf{X}}(\mathbf{e}^{\alpha
^{\prime }})_{C^{\perp }}=-\mathbf{v}_{~\beta ^{\prime }}^{\alpha ^{\prime
}}(\mathbf{e}^{\beta ^{\prime }})_{C},
\end{equation*}%
for some antisymmetric d--tensors $\mathbf{v}^{\alpha ^{\prime }\beta
^{\prime }}=-\mathbf{v}^{\beta ^{\prime }\alpha ^{\prime }}$. Note we will
get a N--adapted $\left( SO(n)\oplus SO(m)\right) $--parallel frame defining
a generalization of the concept of Riemannian parallel frame \cite{anc1} on
N--adapted manifolds whenever $\mathfrak{p}_{C}$ is larger than $\mathfrak{p}%
_{\parallel }$. Substituting $\mathbf{e}^{\alpha ^{\prime }}=(e^{i^{\prime
}},\mathbf{e}^{a^{\prime }})$ into the last formulas and considering h- and
v--components, we define $SO(n)$--parallel and $SO(m)$--parallel frames (for
simplicity we omit these formulas when the Greek letter indices are split
into Latin letter h- and v--indices).

The final conclusion of this section is that the Cartan structure equations
on hypersurfaces swept out by nonholonomic curve flows on N--anholonomic
spaces with constant matrix curvature for the canonical d--connection
geometrically encode separate $O(n-1)$-- and $O(m-1)$--invariant
(respectively, horizontal and vertical) bi-Hamiltonian operators, whenever
the N--connection freedom of the d--group action $SO(n)\oplus SO(m)$ on $%
\mathbf{e}$ and $\mathbf{\Gamma }$ is used to fix them to be a N--adapted
parallel coframe and its associated canonical d--connection 1--form, related
to the canonical covariant derivative on N--anholonomic manifolds.

\section{Anholonomic bi-Hamiltonian Structures \newline
and Vector Soliton Equations}

Introducing N--adapted orthonormal bases, for N--anholonomic spaces of
dimension $n+n$, with constant curvatures of the canonical d--connection,
we will be able to derive bi-Hamiltonian and vector soliton structures
similarly to the case of symmetric Riemannian spaces
(see Ref. \cite{aw,anc2,ancjgp}).

\subsection{Basic equations for N--anholonomic curve flows}

In this section, we shall prove the results for the h--components of certain
N--anholonomic manifolds with constant d--curvature and then duplicate the
formulas for the v--components omitting similar details.

There is an isomorphism between the real space $\mathfrak{so}(n)$ and the
Lie algebra of $n\times n$ skew--symmetric matrices. This yields an
isomorphism between $h\mathfrak{p}$ $\simeq \mathbb{R}^{n}$ and the tangent
spaces $T_{x}M=\mathfrak{so}(n+1)/$ $\mathfrak{so}(n)$ of the Riemannian
manifold $M=SO(n+1)/$ $SO(n)$ as described by the following canonical
decomposition
\begin{equation*}
h\mathfrak{g}=\mathfrak{so}(n+1)\supset h\mathfrak{p\in }\left[
\begin{array}{cc}
0 & h\mathbf{p} \\
-h\mathbf{p}^{T} & h\mathbf{0}%
\end{array}%
\right] \mbox{\ for\ }h\mathbf{0\in }h\mathfrak{h=so}(n)
\end{equation*}%
with $h\mathbf{p=\{}p^{i^{\prime }}\mathbf{\}\in }\mathbb{R}^{n}$ being the
h--component of the d--vector $\mathbf{p=(}p^{i^{\prime }}\mathbf{,}%
p^{a^{\prime }}\mathbf{)}$ and $h\mathbf{p}^{T}$ denoting the transposition
of the row $h\mathbf{p.}$ The Cartan--Killing inner product on $h\mathfrak{p}
$ is given by
\begin{eqnarray*}
h\mathbf{p\cdot }h\mathbf{p} &\mathbf{=}&\left\langle \left[
\begin{array}{cc}
0 & h\mathbf{p} \\
-h\mathbf{p}^{T} & h\mathbf{0}%
\end{array}%
\right] ,\left[
\begin{array}{cc}
0 & h\mathbf{p} \\
-h\mathbf{p}^{T} & h\mathbf{0}%
\end{array}%
\right] \right\rangle \\
&\mathbf{\doteqdot }&\frac{1}{2}tr\left\{ \left[
\begin{array}{cc}
0 & h\mathbf{p} \\
-h\mathbf{p}^{T} & h\mathbf{0}%
\end{array}%
\right] ^{T}\left[
\begin{array}{cc}
0 & h\mathbf{p} \\
-h\mathbf{p}^{T} & h\mathbf{0}%
\end{array}%
\right] \right\} ,
\end{eqnarray*}%
where $tr$ denotes the trace of the corresponding product of matrices. This
product identifies canonically $h\mathfrak{p}$ $\simeq \mathbb{R}^{n}$ with
its dual $h\mathfrak{p}^{\ast }$ $\simeq \mathbb{R}^{n}$. In a similar
fashion, we can define the Cartan--Killing inner product $v\mathbf{p\cdot }v%
\mathbf{p}$ where
\begin{equation*}
v\mathfrak{g}=\mathfrak{so}(m+1)\supset v\mathfrak{p\in }\left[
\begin{array}{cc}
0 & v\mathbf{p} \\
-v\mathbf{p}^{T} & v\mathbf{0}%
\end{array}%
\right] \mbox{\ for\ }v\mathbf{0\in }v\mathfrak{h=so}(m)
\end{equation*}%
with $v\mathbf{p=\{}p^{a^{\prime }}\mathbf{\}\in }\mathbb{R}^{m}$ being the
v--component of the d--vector $\mathbf{p=(}p^{i^{\prime }}p^{a^{\prime }}%
\mathbf{)}$. In general, in the tangent bundle of a N--anholonomic manifold,
we can consider the Cartan--Killing N--adapted inner product $\mathbf{p\cdot
p=}h\mathbf{p\cdot }h\mathbf{p+}v\mathbf{p\cdot }v\mathbf{p.}$

Following the introduced Cartan--Killing parametrizations, we analyze the
flow $\gamma (\tau ,l)$ of a non--stretching curve in $\mathbf{V}_{\mathbf{N}%
}=\mathbf{G}/SO(n)\oplus $ $SO(m)$. Let us introduce a coframe $\mathbf{e}%
\in T_{\gamma }^{\ast }\mathbf{V}_{\mathbf{N}}\otimes (h\mathfrak{p\oplus }v%
\mathfrak{p})$, which is a N--adapted $\left( SO(n)\mathfrak{\oplus }%
SO(m)\right) $--parallel covector frame along $\gamma $, and its associated
canonical d--con\-nec\-tion 1--form $\mathbf{\Gamma }\in T_{\gamma }^{\ast }%
\mathbf{V}_{\mathbf{N}}\otimes (\mathfrak{so}(n)\mathfrak{\oplus so}(m))$.
Such d--objects are respectively parametrized:%
\begin{equation*}
\mathbf{e}_{\mathbf{X}}=\mathbf{e}_{h\mathbf{X}}+\mathbf{e}_{v\mathbf{X}},
\end{equation*}%
for
\begin{equation*}
\mathbf{e}_{h\mathbf{X}}=\gamma _{h\mathbf{X}}\rfloor h\mathbf{e=}\left[
\begin{array}{cc}
0 & (1,\overrightarrow{0}) \\
-(1,\overrightarrow{0})^{T} & h\mathbf{0}%
\end{array}%
\right]
\end{equation*}%
and
\begin{equation*}
\mathbf{e}_{v\mathbf{X}}=\gamma _{v\mathbf{X}}\rfloor v\mathbf{e=}\left[
\begin{array}{cc}
0 & (1,\overleftarrow{0}) \\
-(1,\overleftarrow{0})^{T} & v\mathbf{0}%
\end{array}%
\right] ,
\end{equation*}%
where we write $(1,\overrightarrow{0})\in \mathbb{R}^{n},\overrightarrow{0}%
\in \mathbb{R}^{n-1}$ and $(1,\overleftarrow{0})\in \mathbb{R}^{m},%
\overleftarrow{0}\in \mathbb{R}^{m-1};$%
\begin{equation*}
\mathbf{\Gamma =}\left[ \mathbf{\Gamma }_{h\mathbf{X}},\mathbf{\Gamma }_{v%
\mathbf{X}}\right] ,
\end{equation*}%
with
\begin{equation*}
\mathbf{\Gamma }_{h\mathbf{X}}\mathbf{=}\gamma _{h\mathbf{X}}\rfloor \mathbf{%
L=}\left[
\begin{array}{cc}
0 & (0,\overrightarrow{0}) \\
-(0,\overrightarrow{0})^{T} & \mathbf{L}%
\end{array}%
\right] \in \mathfrak{so}(n+1),
\end{equation*}%
where
\begin{equation*}
\mathbf{L=}\left[
\begin{array}{cc}
0 & \overrightarrow{v} \\
-\overrightarrow{v}^{T} & h\mathbf{0}%
\end{array}%
\right] \in \mathfrak{so}(n),~\overrightarrow{v}\in \mathbb{R}^{n-1},~h%
\mathbf{0\in }\mathfrak{so}(n-1),
\end{equation*}%
and also
\begin{equation*}
\mathbf{\Gamma }_{v\mathbf{X}}\mathbf{=}\gamma _{v\mathbf{X}}\rfloor \mathbf{%
C=}\left[
\begin{array}{cc}
0 & (0,\overleftarrow{0}) \\
-(0,\overleftarrow{0})^{T} & \mathbf{C}%
\end{array}%
\right] \in \mathfrak{so}(m+1),
\end{equation*}%
where
\begin{equation*}
\mathbf{C=}\left[
\begin{array}{cc}
0 & \overleftarrow{v} \\
-\overleftarrow{v}^{T} & v\mathbf{0}%
\end{array}%
\right] \in \mathfrak{so}(m),~\overleftarrow{v}\in \mathbb{R}^{m-1},~v%
\mathbf{0\in }\mathfrak{so}(m-1).
\end{equation*}%
The above parametrizations involve no loss of generality, because they can
be achieved by the available gauge freedom using $SO(n)$ and $SO(m)$
rotations on the N--adapted coframe and canonical d--connection 1--form,
distinguished in h- and v--components.

We introduce decompositions of horizontal $SO(n+1)/$ $SO(n)$ matrices
\begin{eqnarray*}
h\mathfrak{p} \mathfrak{\ni }\left[
\begin{array}{cc}
0 & h\mathbf{p} \\
-h\mathbf{p}^{T} & h\mathbf{0}%
\end{array}%
\right] = &&\left[
\begin{array}{cc}
0 & \left( h\mathbf{p}_{\parallel },\overrightarrow{0}\right) \\
-\left( h\mathbf{p}_{\parallel },\overrightarrow{0}\right) ^{T} & h\mathbf{0}%
\end{array}%
\right] \\
&&+\left[
\begin{array}{cc}
0 & \left( 0,h\overrightarrow{\mathbf{p}}_{\perp }\right) \\
-\left( 0,h\overrightarrow{\mathbf{p}}_{\perp }\right) ^{T} & h\mathbf{0}%
\end{array}%
\right] ,
\end{eqnarray*}%
into tangential and normal parts relative to $\mathbf{e}_{h\mathbf{X}}$ via
corresponding decompositions of h--vectors $h\mathbf{p=(}h\mathbf{\mathbf{p}%
_{\parallel },}h\mathbf{\overrightarrow{\mathbf{p}}_{\perp })\in }\mathbb{R}%
^{n}$ relative to $\left( 1,\overrightarrow{0}\right) $, when $h\mathbf{%
\mathbf{p}_{\parallel }}$ is identified with $h\mathfrak{p}_{C}$ and $h%
\mathbf{\overrightarrow{\mathbf{p}}_{\perp }}$ is identified with $h%
\mathfrak{p}_{\perp }=h\mathfrak{p}_{C^{\perp }}$. In a similar form, we
decompose vertical $SO(m+1)/$ $SO(m)$ matrices
\begin{eqnarray*}
v\mathfrak{p} \mathfrak{\ni }\left[
\begin{array}{cc}
0 & v\mathbf{p} \\
-v\mathbf{p}^{T} & v\mathbf{0}%
\end{array}%
\right] = &&\left[
\begin{array}{cc}
0 & \left( v\mathbf{p}_{\parallel },\overleftarrow{0}\right) \\
-\left( v\mathbf{p}_{\parallel },\overleftarrow{0}\right) ^{T} & v\mathbf{0}%
\end{array}%
\right] \\
&&+\left[
\begin{array}{cc}
0 & \left( 0,v\overleftarrow{\mathbf{p}}_{\perp }\right) \\
-\left( 0,v\overleftarrow{\mathbf{p}}_{\perp }\right) ^{T} & v\mathbf{0}%
\end{array}%
\right] ,
\end{eqnarray*}%
into tangential and normal parts relative to $\mathbf{e}_{v\mathbf{X}}$ via
corresponding decompositions of h--vectors $v\mathbf{p=(}v\mathbf{\mathbf{p}%
_{\parallel },}v\overleftarrow{\mathbf{\mathbf{p}}}\mathbf{_{\perp })\in }%
\mathbb{R}^{m}$ relative to $\left( 1,\overleftarrow{0}\right) $, when $v%
\mathbf{\mathbf{p}_{\parallel }}$ is identified with $v\mathfrak{p}_{C}$ and
$v\overleftarrow{\mathbf{\mathbf{p}}}\mathbf{_{\perp }}$ is identified with $%
v\mathfrak{p}_{\perp }=v\mathfrak{p}_{C^{\perp }}$.

There are analogous matrix decompositions relative to the flow direction. In
the h--direction, we parametrize%
\begin{equation*}
\mathbf{e}_{h\mathbf{Y}}=\gamma _{\tau }\rfloor h\mathbf{e=}\left[
\begin{array}{cc}
0 & \left( h\mathbf{e}_{\parallel },h\overrightarrow{\mathbf{e}}_{\perp
}\right) \\
-\left( h\mathbf{e}_{\parallel },h\overrightarrow{\mathbf{e}}_{\perp
}\right) ^{T} & h\mathbf{0}%
\end{array}%
\right] ,
\end{equation*}%
when $\mathbf{e}_{h\mathbf{Y}}\in h\mathfrak{p,}\left( h\mathbf{e}%
_{\parallel },h\overrightarrow{\mathbf{e}}_{\perp }\right) \in \mathbb{R}%
^{n} $ and $h\overrightarrow{\mathbf{e}}_{\perp }\in \mathbb{R}^{n-1}$, and
\begin{equation}
\mathbf{\Gamma }_{h\mathbf{Y}}\mathbf{=}\gamma _{h\mathbf{Y}}\rfloor \mathbf{%
L=}\left[
\begin{array}{cc}
0 & (0,\overrightarrow{0}) \\
-(0,\overrightarrow{0})^{T} & h\mathbf{\varpi }_{\tau }%
\end{array}%
\right] \in \mathfrak{so}(n+1),  \label{auxaaa}
\end{equation}%
where
\begin{equation*}
h\mathbf{\varpi }_{\tau }\mathbf{=}\left[
\begin{array}{cc}
0 & \overrightarrow{\varpi } \\
-\overrightarrow{\varpi }^{T} & h\mathbf{\Theta }%
\end{array}%
\right] \in \mathfrak{so}(n),~\overrightarrow{\varpi }\in \mathbb{R}^{n-1},~h%
\mathbf{\Theta \in }\mathfrak{so}(n-1).
\end{equation*}%
In the v--direction, we parametrize
\begin{equation*}
\mathbf{e}_{v\mathbf{Y}}=\gamma _{\tau }\rfloor v\mathbf{e=}\left[
\begin{array}{cc}
0 & \left( v\mathbf{e}_{\parallel },v\overleftarrow{\mathbf{e}}_{\perp
}\right) \\
-\left( v\mathbf{e}_{\parallel },v\overleftarrow{\mathbf{e}}_{\perp }\right)
^{T} & v\mathbf{0}%
\end{array}%
\right] ,
\end{equation*}%
when $\mathbf{e}_{v\mathbf{Y}}\in v\mathfrak{p,}\left( v\mathbf{e}%
_{\parallel },v\overleftarrow{\mathbf{e}}_{\perp }\right) \in \mathbb{R}^{m}$
and $v\overleftarrow{\mathbf{e}}_{\perp }\in \mathbb{R}^{m-1}$, and
\begin{equation*}
\mathbf{\Gamma }_{v\mathbf{Y}}\mathbf{=}\gamma _{v\mathbf{Y}}\rfloor \mathbf{%
C=}\left[
\begin{array}{cc}
0 & (0,\overleftarrow{0}) \\
-(0,\overleftarrow{0})^{T} & v\mathbf{\varpi }_{\tau }%
\end{array}%
\right] \in \mathfrak{so}(m+1),
\end{equation*}%
where
\begin{equation*}
v\mathbf{\varpi }_{\tau }\mathbf{=}\left[
\begin{array}{cc}
0 & \overleftarrow{\varpi } \\
-\overleftarrow{\varpi }^{T} & v\mathbf{\Theta }%
\end{array}%
\right] \in \mathfrak{so}(m),~\overleftarrow{\varpi }\in \mathbb{R}^{m-1},~v%
\mathbf{\Theta \in }\mathfrak{so}(m-1).
\end{equation*}%
The components $h\mathbf{e}_{\parallel }$ and $h\overrightarrow{\mathbf{e}}%
_{\perp }$ correspond to the decomposition
\begin{equation*}
\mathbf{e}_{h\mathbf{Y}}=h\mathbf{g(\gamma }_{\tau },\mathbf{\gamma }_{l}%
\mathbf{)e}_{h\mathbf{X}}+\mathbf{(\gamma }_{\tau })_{\perp }\rfloor h%
\mathbf{e}_{\perp }
\end{equation*}%
into tangential and normal parts relative to $\mathbf{e}_{h\mathbf{X}}$. In
a similar form, one considers $v\mathbf{e}_{\parallel }$ and $v%
\overleftarrow{\mathbf{e}}_{\perp }$ corresponding to the decomposition
\begin{equation*}
\mathbf{e}_{v\mathbf{Y}}=v\mathbf{g(\gamma }_{\tau },\mathbf{\gamma }_{l}%
\mathbf{)e}_{v\mathbf{X}}+\mathbf{(\gamma }_{\tau })_{\perp }\rfloor v%
\mathbf{e}_{\perp }.
\end{equation*}

Using the above stated matrix parametrizations, we get%
\begin{eqnarray}
\left[ \mathbf{e}_{h\mathbf{X}},\mathbf{e}_{h\mathbf{Y}}\right] &=&-\left[
\begin{array}{cc}
0 & 0 \\
0 & h\mathbf{e}_{\perp }%
\end{array}%
\right] \in \mathfrak{so}(n+1), \qquad\mbox{with}  \label{aux41} \\
h\mathbf{e}_{\perp } &=&\left[
\begin{array}{cc}
0 & h\overrightarrow{\mathbf{e}}_{\perp } \\
-(h\overrightarrow{\mathbf{e}}_{\perp })^{T} & h\mathbf{0}%
\end{array}%
\right] \in \mathfrak{so}(n);  \notag \\
\left[ \mathbf{\Gamma }_{h\mathbf{Y}},\mathbf{e}_{h\mathbf{Y}}\right] &=&-%
\left[
\begin{array}{cc}
0 & \left( 0,\overrightarrow{\varpi }\right) \\
-\left( 0,\overrightarrow{\varpi }\right) ^{T} & 0%
\end{array}%
\right] \in h\mathfrak{p}_{\perp };  \notag \\
\left[ \mathbf{\Gamma }_{h\mathbf{X}},\mathbf{e}_{h\mathbf{Y}}\right] &=&-%
\left[
\begin{array}{cc}
0 & \left( -\overrightarrow{v}\cdot h\overrightarrow{\mathbf{e}}_{\perp },h%
\mathbf{e}_{\parallel }\overrightarrow{v}\right) \\
-\left( -\overrightarrow{v}\cdot h\overrightarrow{\mathbf{e}}_{\perp },h%
\mathbf{e}_{\parallel }\overrightarrow{v}\right) ^{T} & h\mathbf{0}%
\end{array}%
\right] \in h\mathfrak{p};  \notag
\end{eqnarray}%
and
\begin{eqnarray}
\left[ \mathbf{e}_{v\mathbf{X}},\mathbf{e}_{v\mathbf{Y}}\right] &=&-\left[
\begin{array}{cc}
0 & 0 \\
0 & v\mathbf{e}_{\perp }%
\end{array}%
\right] \in \mathfrak{so}(m+1), \qquad\mbox{with}  \label{aux41a} \\
v\mathbf{e}_{\perp } &=&\left[
\begin{array}{cc}
0 & v\overrightarrow{\mathbf{e}}_{\perp } \\
-(v\overrightarrow{\mathbf{e}}_{\perp })^{T} & v\mathbf{0}%
\end{array}%
\right] \in \mathfrak{so}(m);  \notag \\
\left[ \mathbf{\Gamma }_{v\mathbf{Y}},\mathbf{e}_{v\mathbf{Y}}\right] &=&-%
\left[
\begin{array}{cc}
0 & \left( 0,\overleftarrow{\varpi }\right) \\
-\left( 0,\overleftarrow{\varpi }\right) ^{T} & 0%
\end{array}%
\right] \in v\mathfrak{p}_{\perp };  \notag \\
\left[ \mathbf{\Gamma }_{v\mathbf{X}},\mathbf{e}_{v\mathbf{Y}}\right] &=&-%
\left[
\begin{array}{cc}
0 & \left( -\overleftarrow{v}\cdot v\overleftarrow{\mathbf{e}}_{\perp },v%
\mathbf{e}_{\parallel }\overleftarrow{v}\right) \\
-\left( -\overleftarrow{v}\cdot v\overleftarrow{\mathbf{e}}_{\perp },v%
\mathbf{e}_{\parallel }\overleftarrow{v}\right) ^{T} & v\mathbf{0}%
\end{array}%
\right] \in v\mathfrak{p}.  \notag
\end{eqnarray}

We can use formulas (\ref{aux41}) and (\ref{aux41a}) in order to write the
structure equations (\ref{mtors}) and (\ref{mcurv}) in terms of N--adapted
frames and connection 1--forms soldered to the Klein geometry of
N--anholonomic spaces using the relations (\ref{aux33}). One obtains
respectively the $\mathbf{G}$--invariant N--adapted torsion and curvature
generated by the canonical d--connection,
\begin{equation}
\mathbf{T}(\gamma _{\tau },\gamma _{l})=\left( \mathbf{D}_{\mathbf{X}}\gamma
_{\tau }-\mathbf{D}_{\mathbf{Y}}\gamma _{l}\right) \rfloor \mathbf{e=D}_{%
\mathbf{X}}\mathbf{e}_{\mathbf{Y}}-\mathbf{D}_{\mathbf{Y}}\mathbf{e}_{%
\mathbf{X}}+\left[ \mathbf{\Gamma }_{\mathbf{X}},\mathbf{e}_{\mathbf{Y}}%
\right] -\left[ \mathbf{\Gamma }_{\mathbf{Y}},\mathbf{e}_{\mathbf{X}}\right]
\label{torscf}
\end{equation}%
and
\begin{equation}
\mathbf{R}(\gamma _{\tau },\gamma _{l})\mathbf{e=}\left[ \mathbf{D}_{\mathbf{%
X}},\mathbf{D}_{\mathbf{Y}}\right] \mathbf{e=D}_{\mathbf{X}}\mathbf{\Gamma }%
_{\mathbf{Y}}-\mathbf{D}_{\mathbf{Y}}\mathbf{\Gamma }_{\mathbf{X}}+\left[
\mathbf{\Gamma }_{\mathbf{X}},\mathbf{\Gamma }_{\mathbf{Y}}\right]
\label{curvcf}
\end{equation}%
where $\mathbf{e}_{\mathbf{X}}\doteqdot \gamma _{l}\rfloor \mathbf{e,}$ $%
\mathbf{e}_{\mathbf{Y}}\doteqdot \gamma _{\mathbf{\tau }}\rfloor \mathbf{e,}$
$\mathbf{\Gamma }_{\mathbf{X}}\doteqdot \gamma _{l}\rfloor \mathbf{\Gamma }$
and $\mathbf{\Gamma }_{\mathbf{Y}}\doteqdot \gamma _{\mathbf{\tau }}\rfloor
\mathbf{\Gamma .}$ The formulas (\ref{torscf}) and (\ref{curvcf}) are
respectively equivalent to (\ref{dtors}) and (\ref{dcurv}). In general, $%
\mathbf{T}(\gamma _{\tau },\gamma _{l})\neq 0$ and $\mathbf{R}(\gamma _{\tau
},\gamma _{l})\mathbf{e}$ cannot be defined to have constant matrix
coefficients with respect to a N--adapted basis. For N--anholonomic spaces
with dimensions $n=m$, we have $\mathbf{T}(\gamma _{\tau },\gamma _{l})=0$
and $\mathbf{R}(\gamma _{\tau },\gamma _{l})\mathbf{e}$ defined by constant,
or vanishing, d--curvature coefficients (see discussions related to formulas
(\ref{dcurvtb}) and (\ref{candcontm})). For such cases, we can consider the
h-- and v--components of (\ref{torscf}) and (\ref{curvcf}) in a similar
manner as for symmetric Riemannian spaces but with the canonical
d--connection instead of the Levi Civita one. One obtains, respectively,%
\begin{eqnarray}
0 &=&\left( \mathbf{D}_{h\mathbf{X}}\gamma _{\tau }-\mathbf{D}_{h\mathbf{Y}%
}\gamma _{l}\right) \rfloor h\mathbf{e}  \label{torseq} \\
&\mathbf{=}&\mathbf{D}_{h\mathbf{X}}\mathbf{e}_{h\mathbf{Y}}-\mathbf{D}_{h%
\mathbf{Y}}\mathbf{e}_{h\mathbf{X}}+\left[ \mathbf{L}_{h\mathbf{X}},\mathbf{e%
}_{h\mathbf{Y}}\right] -\left[ \mathbf{L}_{h\mathbf{Y}},\mathbf{e}_{h\mathbf{%
X}}\right] ;  \notag \\
0 &=&\left( \mathbf{D}_{v\mathbf{X}}\gamma _{\tau }-\mathbf{D}_{v\mathbf{Y}%
}\gamma _{l}\right) \rfloor v\mathbf{e}  \notag \\
&\mathbf{=}&\mathbf{D}_{v\mathbf{X}}\mathbf{e}_{v\mathbf{Y}}-\mathbf{D}_{v%
\mathbf{Y}}\mathbf{e}_{v\mathbf{X}}+\left[ \mathbf{C}_{v\mathbf{X}},\mathbf{e%
}_{v\mathbf{Y}}\right] -\left[ \mathbf{C}_{v\mathbf{Y}},\mathbf{e}_{v\mathbf{%
X}}\right] ,  \notag
\end{eqnarray}%
and%
\begin{eqnarray}
h\mathbf{R}(\gamma _{\tau },\gamma _{l})h\mathbf{e} &\mathbf{=}&\left[
\mathbf{D}_{h\mathbf{X}},\mathbf{D}_{h\mathbf{Y}}\right] h\mathbf{e=D}_{h%
\mathbf{X}}\mathbf{L}_{h\mathbf{Y}}-\mathbf{D}_{h\mathbf{Y}}\mathbf{L}_{h%
\mathbf{X}}+\left[ \mathbf{L}_{h\mathbf{X}},\mathbf{L}_{h\mathbf{Y}}\right]
\label{curveq} \\
v\mathbf{R}(\gamma _{\tau },\gamma _{l})v\mathbf{e} &\mathbf{=}&\left[
\mathbf{D}_{v\mathbf{X}},\mathbf{D}_{v\mathbf{Y}}\right] v\mathbf{e=D}_{v%
\mathbf{X}}\mathbf{C}_{v\mathbf{Y}}-\mathbf{D}_{v\mathbf{Y}}\mathbf{C}_{v%
\mathbf{X}}+\left[ \mathbf{C}_{v\mathbf{X}},\mathbf{C}_{v\mathbf{Y}}\right] .
\notag
\end{eqnarray}

Following N--adapted curve flow parametrizations (\ref{aux41}) and (\ref%
{aux41a}), the equations (\ref{torseq}) and (\ref{curveq}) are written
\begin{eqnarray}
0 &=&\mathbf{D}_{h\mathbf{X}}h\mathbf{e}_{\parallel }+\overrightarrow{v}%
\cdot h\overrightarrow{\mathbf{e}}_{\perp },~0=\mathbf{D}_{v\mathbf{X}}v%
\mathbf{e}_{\parallel }+\overleftarrow{v}\cdot v\overleftarrow{\mathbf{e}}%
_{\perp },;  \label{torseqd} \\
~0 &=&\overrightarrow{\varpi }-h\mathbf{e}_{\parallel }\overrightarrow{v}+%
\mathbf{D}_{h\mathbf{X}}h\overrightarrow{\mathbf{e}}_{\perp },~0=%
\overleftarrow{\varpi }-v\mathbf{e}_{\parallel }\overleftarrow{v}+\mathbf{D}%
_{v\mathbf{X}}v\overleftarrow{\mathbf{e}}_{\perp };  \notag
\end{eqnarray}%
and%
\begin{eqnarray}
\mathbf{D}_{h\mathbf{X}}\overrightarrow{\varpi }-\mathbf{D}_{h\mathbf{Y}}%
\overrightarrow{v}+\overrightarrow{v}\rfloor h\mathbf{\Theta } &\mathbf{=}&h%
\overrightarrow{\mathbf{e}}_{\perp },~\mathbf{D}_{v\mathbf{X}}\overleftarrow{%
\varpi }-\mathbf{D}_{v\mathbf{Y}}\overleftarrow{v}+\overleftarrow{v}\rfloor v%
\mathbf{\Theta =}v\overleftarrow{\mathbf{e}}_{\perp };  \notag \\
\mathbf{D}_{h\mathbf{X}}h\mathbf{\Theta -}\overrightarrow{v}\otimes
\overrightarrow{\varpi }+\overrightarrow{\varpi }\otimes \overrightarrow{v}
&=&0,~\mathbf{D}_{v\mathbf{X}}v\mathbf{\Theta -}\overleftarrow{v}\otimes
\overleftarrow{\varpi }+\overleftarrow{\varpi }\otimes \overleftarrow{v}=0.
\label{curveqd}
\end{eqnarray}%
The tensor products and interior products are defined in the form: for the
h--components, $\otimes $ denotes the outer product of pairs of vectors ($%
1\times n$ row matrices), producing $n\times n$ matrices $\overrightarrow{A}%
\otimes \overrightarrow{B}=\overrightarrow{A}^{T}\overrightarrow{B}$, and $%
\rfloor $ denotes multiplication of $n\times n$ matrices on vectors ($%
1\times n$ row matrices), such that $\overrightarrow{A}\rfloor \left(
\overrightarrow{B}\otimes \overrightarrow{C}\right) =\left( \overrightarrow{A%
}\cdot \overrightarrow{B}\right) \overrightarrow{C}$ which is the transpose
of the standard matrix product on column vectors. Likewise, for the
v--components, we just change $n\rightarrow m$ and $\overrightarrow{A}%
\rightarrow \overleftarrow{A}$. For the sequel, $\wedge$ will denote the
skew product of vectors $\overrightarrow{A}\wedge \overrightarrow{B}=%
\overrightarrow{A}\otimes \overrightarrow{B}-\overrightarrow{B}\otimes
\overrightarrow{A}$.

The variables $\mathbf{e}_{\parallel }$ and $\mathbf{\Theta ,}$ written in
h-- and v--components, can be expressed in terms of the variables $%
\overrightarrow{v},\overrightarrow{\varpi },h\overrightarrow{\mathbf{e}}%
_{\perp }$ and $\overleftarrow{v},\overleftarrow{\varpi },v\overleftarrow{%
\mathbf{e}}_{\perp }$ (see respectively the first two equations in (\ref%
{torseqd}) and the last two equations in (\ref{curveqd})) by
\begin{equation*}
h\mathbf{e}_{\parallel }=-\mathbf{D}_{h\mathbf{X}}^{-1}(\overrightarrow{v}%
\cdot h\overrightarrow{\mathbf{e}}_{\perp }),~v\mathbf{e}_{\parallel }=-%
\mathbf{D}_{v\mathbf{X}}^{-1}(\overleftarrow{v}\cdot v\overleftarrow{\mathbf{%
e}}_{\perp }),
\end{equation*}%
and%
\begin{equation*}
h\mathbf{\Theta =D}_{h\mathbf{X}}^{-1}\left( \overrightarrow{v}\otimes
\overrightarrow{\varpi }-\overrightarrow{\varpi }\otimes \overrightarrow{v}%
\right) ,~v\mathbf{\Theta =D}_{v\mathbf{X}}^{-1}\left( \overleftarrow{v}%
\otimes \overleftarrow{\varpi }-\overleftarrow{\varpi }\otimes
\overleftarrow{v}\right) .
\end{equation*}%
Substituting these expressions, correspondingly, in the last two equations
in (\ref{torseqd}) and in the first two equations in (\ref{curveqd}), we
express
\begin{equation*}
\overrightarrow{\varpi }=-\mathbf{D}_{h\mathbf{X}}h\overrightarrow{\mathbf{e}%
}_{\perp }-\mathbf{D}_{h\mathbf{X}}^{-1}(\overrightarrow{v}\cdot h%
\overrightarrow{\mathbf{e}}_{\perp })\overrightarrow{v},~\overleftarrow{%
\varpi }=-\mathbf{D}_{v\mathbf{X}}v\overleftarrow{\mathbf{e}}_{\perp }-%
\mathbf{D}_{v\mathbf{X}}^{-1}(\overleftarrow{v}\cdot v\overleftarrow{\mathbf{%
e}}_{\perp })\overleftarrow{v},
\end{equation*}%
contained in the h-- and v--flow equations respectively on $\overrightarrow{v%
}$ and $\overleftarrow{v}$, considered as evolution equations via $\mathbf{D}%
_{h\mathbf{Y}}\overrightarrow{v}=\overrightarrow{v}_{\tau }$ and $\mathbf{D}%
_{h\mathbf{Y}}\overleftarrow{v}=\overleftarrow{v}_{\tau }$,
\begin{eqnarray}
\overrightarrow{v}_{\tau } &=&\mathbf{D}_{h\mathbf{X}}\overrightarrow{\varpi
}-\overrightarrow{v}\rfloor \mathbf{D}_{h\mathbf{X}}^{-1}\left(
\overrightarrow{v}\otimes \overrightarrow{\varpi }-\overrightarrow{\varpi }%
\otimes \overrightarrow{v}\right) -\overrightarrow{R}h\overrightarrow{%
\mathbf{e}}_{\perp },  \label{floweq} \\
\overleftarrow{v}_{\tau } &=&\mathbf{D}_{v\mathbf{X}}\overleftarrow{\varpi }-%
\overleftarrow{v}\rfloor \mathbf{D}_{v\mathbf{X}}^{-1}\left( \overleftarrow{v%
}\otimes \overleftarrow{\varpi }-\overleftarrow{\varpi }\otimes
\overleftarrow{v}\right) -\overleftarrow{S}v\overleftarrow{\mathbf{e}}%
_{\perp },  \notag
\end{eqnarray}%
where the scalar curvatures of the canonical d--connection, $\overrightarrow{%
R}$ and $\overleftarrow{S}$ are defined by formulas (\ref{sdccurv}) in
Appendix. For symmetric Riemannian spaces like $SO(n+1)/SO(n)\simeq S^{n}$, $%
\overrightarrow{R}$ is just the scalar curvature $\chi =1$, \ see \cite{anc2}%
. On N--anholonomic manifolds under consideration, the values $%
\overrightarrow{R}$ and $\overleftarrow{S}$ are constrained  by be certain
zero or nonzero constants.

The above presented considerations lead to a proof of the following main
results.

\begin{lemma}
On N--anholonomic spaces with constant curvature matrix coefficients for the
canonical d--connection, there are N--adapted Hamiltonian symplectic
operators,
\begin{equation}
h\mathcal{J}=\mathbf{D}_{h\mathbf{X}}+\mathbf{D}_{h\mathbf{X}}^{-1}\left(
\overrightarrow{v}\cdot \right) \overrightarrow{v}\mbox{ \ and \ }v\mathcal{J%
}=\mathbf{D}_{v\mathbf{X}}+\mathbf{D}_{v\mathbf{X}}^{-1}\left(
\overleftarrow{v}\cdot \right) \overleftarrow{v},  \label{sop}
\end{equation}%
and cosymplectic operators%
\begin{equation}
h\mathcal{H}\doteqdot \mathbf{D}_{h\mathbf{X}}+\overrightarrow{v}\rfloor
\mathbf{D}_{h\mathbf{X}}^{-1}\left( \overrightarrow{v}\wedge \right)
\mbox{
\ and \ }v\mathcal{H}\doteqdot \mathbf{D}_{v\mathbf{X}}+\overleftarrow{v}%
\rfloor \mathbf{D}_{v\mathbf{X}}^{-1}\left( \overleftarrow{v}\wedge \right) .
\label{csop}
\end{equation}
\end{lemma}

The properties of operators (\ref{sop}) and (\ref{csop}) are defined by

\begin{theorem}
\label{mr1}The d--operators $\mathcal{J=}\left( h\mathcal{J},v\mathcal{J}%
\right) $ and $\mathcal{H=}\left( h\mathcal{H},v\mathcal{H}\right) $ are
res\-pec\-tively $\left( O(n-1),O(m-1)\right) $--invariant Hamiltonian
symplectic and cosymplectic operators with respect to the flow
variables $\left( \overrightarrow{v},\overleftarrow{v}\right)$. Consequently
the curve flow equations on N--anholonomic manifolds with constant
d--connection curvature have a Hamiltonian form: the h--flows are given by%
\begin{eqnarray}
\overrightarrow{v}_{\tau } &=&h\mathcal{H}\left( \overrightarrow{\varpi }%
\right) -\overrightarrow{R}~h\overrightarrow{\mathbf{e}}_{\perp }=h\mathfrak{%
R}\left( h\overrightarrow{\mathbf{e}}_{\perp }\right) -\overrightarrow{R}~h%
\overrightarrow{\mathbf{e}}_{\perp },  \notag \\
\overrightarrow{\varpi } &=&h\mathcal{J}\left( h\overrightarrow{\mathbf{e}}%
_{\perp }\right) ;  \label{hhfeq1}
\end{eqnarray}%
the v--flows are given by
\begin{eqnarray}
\overleftarrow{v}_{\tau } &=&v\mathcal{H}\left( \overleftarrow{\varpi }%
\right) -\overleftarrow{S}~v\overleftarrow{\mathbf{e}}_{\perp }=v\mathfrak{R}%
\left( v\overleftarrow{\mathbf{e}}_{\perp }\right) -\overleftarrow{S}~v%
\overleftarrow{\mathbf{e}}_{\perp },  \notag \\
\overleftarrow{\varpi } &=&v\mathcal{J}\left( v\overleftarrow{\mathbf{e}}%
_{\perp }\right) ,  \label{vhfeq1}
\end{eqnarray}%
where the hereditary recursion d--operator has the respective h-- and
v--compo\-nents
\begin{equation}
h\mathfrak{R}=h\mathcal{H}\circ h\mathcal{J}\mbox{ \ and \ }v\mathfrak{R}=v%
\mathcal{H}\circ v\mathcal{J}.  \label{reqop}
\end{equation}
\end{theorem}

\begin{proof}
Details in the case of holonomic structures are given in Ref. \cite{saw} and
generalized to arbitrary gauge groups in \cite{ancjgp}.
For the present case, the main additional considerations consist of soldering
certain classes of generalized Lagrange spaces with
$\left(O(n-1),O(m-1)\right)$--gauge symmetry onto the Klein geometry of
N--anholonomic spaces. $\square$
\end{proof}

\subsection{bi-Hamiltonian anholonomic curve flows and solitonic hierarchies}

Following the usual solitonic techniques,
see details in Refs. \cite{anc2,serg}%
, the recursion h--operator from (\ref{reqop}),%
\begin{eqnarray}
h\mathfrak{R} &=&\mathbf{D}_{h\mathbf{X}}\left( \mathbf{D}_{h\mathbf{X}}+%
\mathbf{D}_{h\mathbf{X}}^{-1}\left( \overrightarrow{v}\cdot \right)
\overrightarrow{v}\right) +\overrightarrow{v}\rfloor \mathbf{D}_{h\mathbf{X}%
}^{-1}\left( \overrightarrow{v}\wedge \mathbf{D}_{h\mathbf{X}}\right)
\label{reqoph} \\
&=&\mathbf{D}_{h\mathbf{X}}^{2}+|\mathbf{D}_{h\mathbf{X}}|^{2}+\mathbf{D}_{h%
\mathbf{X}}^{-1}\left( \overrightarrow{v}\cdot \right) \overrightarrow{v}%
_{l}-\overrightarrow{v}\rfloor \mathbf{D}_{h\mathbf{X}}^{-1}(\overrightarrow{%
v}_{l}\wedge ),  \notag
\end{eqnarray}%
generates a horizontal hierarchy of commuting Hamiltonian vector fields $h%
\overrightarrow{\mathbf{e}}_{\perp }^{(k)}$ starting from $h\overrightarrow{%
\mathbf{e}}_{\perp }^{(0)}=\overrightarrow{v}_{l}$ given by the
infinitesimal generator of $l$--translations in terms of arclength ${l}$
along the curve.

A vertical hierarchy of commuting vector fields $v\overleftarrow{\mathbf{e}}%
_{\perp }^{(k)}$ starting from $v\overleftarrow{\mathbf{e}}_{\perp }^{(0)}$ $%
=\overleftarrow{v}_{l}$ is generated by the recursion v--operator%
\begin{eqnarray}
v\mathfrak{R} &=&\mathbf{D}_{v\mathbf{X}}\left( \mathbf{D}_{v\mathbf{X}}+%
\mathbf{D}_{v\mathbf{X}}^{-1}\left( \overleftarrow{v}\cdot \right)
\overleftarrow{v}\right) +\overleftarrow{v}\rfloor \mathbf{D}_{v\mathbf{X}%
}^{-1}\left( \overleftarrow{v}\wedge \mathbf{D}_{v\mathbf{X}}\right)
\label{reqopv} \\
&=&\mathbf{D}_{v\mathbf{X}}^{2}+|\mathbf{D}_{v\mathbf{X}}|^{2}+\mathbf{D}_{v%
\mathbf{X}}^{-1}\left( \overleftarrow{v}\cdot \right) \overleftarrow{v}_{l}-%
\overleftarrow{v}\rfloor \mathbf{D}_{v\mathbf{X}}^{-1}(\overleftarrow{v}%
_{l}\wedge ).  \notag
\end{eqnarray}%
There are related hierarchies, generated by adjoint operators $\mathfrak{R}%
^{\ast }=(h\mathfrak{R}^{\ast }$, $v\mathfrak{R}^{\ast })$, of involutive
variational h--covector fields $\overrightarrow{\varpi }^{(k)}=\delta \left(
hH^{(k)}\right) /\delta \overrightarrow{v}$ in terms of Hamiltonians $%
hH=hH^{(k)}(\overrightarrow{v},\overrightarrow{v}_{l},\overrightarrow{v}%
_{2l},\ldots )$ starting from $\overrightarrow{\varpi }^{(0)}=\overrightarrow{v}%
,hH^{(0)}=\frac{1}{2}|\overrightarrow{v}|^{2}$ and of involutive variational
v--covector fields $\overleftarrow{\varpi }^{(k)}=\delta \left(
vH^{(k)}\right) /$ $\delta \overleftarrow{v}$ in terms of Hamiltonians $%
vH=vH^{(k)}(\overleftarrow{v},\overleftarrow{v}_{l},\overleftarrow{v}%
_{2l},\ldots )$ starting from $\overleftarrow{\varpi }^{(0)}=\overleftarrow{v}%
,vH^{(0)}=\frac{1}{2}|\overleftarrow{v}|^{2}$. The relations between
hierarchies are established correspondingly by formulas%
\begin{equation*}
h\overrightarrow{\mathbf{e}}_{\perp }^{(k)}=h\mathcal{H}\left(
\overrightarrow{\varpi }^{(k)},\overrightarrow{\varpi }^{(k+1)}\right) =h%
\mathcal{J}\left( h\overrightarrow{\mathbf{e}}_{\perp }^{(k)}\right)
\end{equation*}%
and
\begin{equation*}
v\overleftarrow{\mathbf{e}}_{\perp }^{(k)}=v\mathcal{H}\left( \overleftarrow{%
\varpi }^{(k)},\overleftarrow{\varpi }^{(k+1)}\right) =v\mathcal{J}\left( v%
\overleftarrow{\mathbf{e}}_{\perp }^{(k)}\right) ,
\end{equation*}%
where $k=0,1,2,\ldots $. All hierarchies (horizontal, vertical and their adjoint
ones) have a typical mKdV scaling symmetry, for instance, $l\mathbf{%
\rightarrow \lambda }l$ and $\overrightarrow{v}\rightarrow \mathbf{\lambda }%
^{-1}\overrightarrow{v}$ under which the values $h\overrightarrow{\mathbf{e}}%
_{\perp }^{(k)}$ and $hH^{(k)}$ have scaling weight $2+2k$, while $%
\overrightarrow{\varpi }^{(k)}$ has scaling weight $1+2k$.

The above presented considerations prove

\begin{corollary}
\label{c2} The recursion d--operator (\ref{reqop}) gives rise to
N--adapted hierarchies of distinguished horizontal and vertical
commuting bi-Hamiltonian flows, respectively, on
$\overrightarrow{v}$ and $\overleftarrow{v}$ that are given by
$O(n-1)$-- and $O(m-1)$ --invariant d--vector evolution equations,%
\begin{eqnarray*}
\overrightarrow{v}_{\tau } &=&h\overrightarrow{\mathbf{e}}_{\perp }^{(k+1)}-%
\overrightarrow{R}~h\overrightarrow{\mathbf{e}}_{\perp }^{(k)}=h\mathcal{H}%
\left( \delta \left( hH^{(k,\overrightarrow{R})}\right) /\delta
\overrightarrow{v}\right) \\
&=&\left( h\mathcal{J}\right) ^{-1}\left( \delta \left( hH^{(k+1,%
\overrightarrow{R})}\right) /\delta \overrightarrow{v}\right)
\end{eqnarray*}%
with horizontal Hamiltonians $hH^{(k+1,\overrightarrow{R})}=hH^{(k+1,%
\overrightarrow{R})}-\overrightarrow{R}~hH^{(k,\overrightarrow{R})}$ and
\begin{eqnarray*}
\overleftarrow{v}_{\tau } &=&v\overleftarrow{\mathbf{e}}_{\perp }^{(k+1)}-%
\overleftarrow{S}~v\overleftarrow{\mathbf{e}}_{\perp }^{(k)}=v\mathcal{H}%
\left( \delta \left( vH^{(k,\overleftarrow{S})}\right) /\delta
\overleftarrow{v}\right) \\
&=&\left( v\mathcal{J}\right) ^{-1}\left( \delta \left( vH^{(k+1,%
\overleftarrow{S})}\right) /\delta \overleftarrow{v}\right)
\end{eqnarray*}%
with vertical Hamiltonians $vH^{(k+1,\overleftarrow{S})}=vH^{(k+1,%
\overleftarrow{S})}-\overleftarrow{S}~vH^{(k,\overleftarrow{S})}$, for $%
k=0,1,2,\ldots$. The d--operators $\mathcal{H}$ and $\mathcal{J}$ are
N--adapted and mutually compatible, from which one can construct an
alternative (explicit) Hamiltonian d--operator $\mathcal{Q=H\circ J}$ $\circ
\mathcal{H=}\mathfrak{R\circ }\mathcal{H}$.
\end{corollary}

\subsubsection{Formulation of the main theorem}

The main goal of this paper is to prove that for any regular Lagrange system
we can define naturally a N--adapted bi-Hamiltonian hierarchy of flows
inducing anholonomic solitonic configurations.

\begin{theorem}
\label{mt} For any anholonomic vector bundle with prescribed d--metric
structure, there is a hierarchy of bi-Hamiltonian N--adapted flows of curves
$\gamma (\tau ,l)=[h\gamma (\tau ,l),v\gamma (\tau ,l)]$ described by
geometric nonholonomic map equations. $0$ flows are defined as convective
(traveling wave) maps%
\begin{equation}
\left( h\gamma \right) _{\tau }=\left( h\gamma \right) _{h\mathbf{X}}%
\mbox{\
and \ }\left( v\gamma \right) _{\tau }=\left( v\gamma \right) _{v\mathbf{X}}.
\label{trmap}
\end{equation}%
The $+1$ flows are defined by non--stretching mKdV maps%
\begin{eqnarray}
-\left( h\gamma \right) _{\tau } &=&\mathbf{D}_{h\mathbf{X}}^{2}\left(
h\gamma \right) _{h\mathbf{X}}+\frac{3}{2}\left| \mathbf{D}_{h\mathbf{X}%
}\left( h\gamma \right) _{h\mathbf{X}}\right| _{h\mathbf{g}}^{2}~\left(
h\gamma \right) _{h\mathbf{X}},  \label{1map} \\
-\left( v\gamma \right) _{\tau } &=&\mathbf{D}_{v\mathbf{X}}^{2}\left(
v\gamma \right) _{v\mathbf{X}}+\frac{3}{2}\left| \mathbf{D}_{v\mathbf{X}%
}\left( v\gamma \right) _{v\mathbf{X}}\right| _{v\mathbf{g}}^{2}~\left(
v\gamma \right) _{v\mathbf{X}},  \notag
\end{eqnarray}%
and the $+2,\ldots $ flows as higher order analogs. Finally, there are $-1$
flows are defined by the kernels of recursion operators (\ref{reqoph}) and (%
\ref{reqopv}) inducing non--stretching wave maps%
\begin{equation}
\mathbf{D}_{h\mathbf{Y}}\left( h\gamma \right) _{h\mathbf{X}}=0%
\mbox{\ and \
}\mathbf{D}_{v\mathbf{Y}}\left( v\gamma \right) _{v\mathbf{X}}=0.
\label{-1map}
\end{equation}
\end{theorem}

We carry out the proof in the next section \ref{ssp}.

\begin{remark}
\label{rema} Counterparts of N--adapted bi-Hamiltonian hierarchies and
related solitonic equations in Theorem~\ref{mt} can be constructed for $%
SU(n)\oplus SU(m)$ / $SO(n)\oplus SO(m)$ as done in Ref. \cite{anc1} for
Riemannian symmetric spaces $SU(n)/SO(n)$. Such results may be very relevant
in modern quantum / (non)commutative gravity.
\end{remark}

Indeed, constructions similar to Theorem~\ref{mt} can be carried out
in gravity models with nontrivial torsion and nonholonomic
structure and related geometry of noncommutative/ superspaces and
anholonomic spinors.
Finally, it should be emphasized that a number of exact solutions in gravity
\cite{vncg,vsgg}
can be nonholonomically deformed in order to generate nonholonomic
hierarchies of gravitational solitons of type (\ref{trmap}), (\ref{1map}) or
(\ref{-1map}), which will be considered in further publications.

\subsubsection{Proof of the main theorem}

\label{ssp}We provide a proof of Theorem \ref{mt} for the horizontal flows.
The approach is based on the method provided in Ref. \cite{anc1,anc2},
but in this proof
the Levi Civita connection on symmetric Riemannian spaces is replaced
by the horizontal components of the canonical d--connection in a generalized
Lagrange space with constant d--curvature coefficients. The vertical
constructions are similar.

One obtains a vector mKdV equation up to a convective term (which can be
absorbed by redefinition of coordinates) defining the +1 flow for $h%
\overrightarrow{\mathbf{e}}_{\perp }=\overrightarrow{v}_{l}$,%
\begin{equation*}
\overrightarrow{v}_{\tau }=\overrightarrow{v}_{3l}+\frac{3}{2}|%
\overrightarrow{v}|^{2}-\overrightarrow{R}~\overrightarrow{v}_{l},
\end{equation*}%
when the $+(k+1)$ flow gives a vector mKdV equation of higher order $3+2k$
on $\overrightarrow{v}$ and there is a $0$ h--flow $\overrightarrow{v}_{\tau
}=\overrightarrow{v}_{l}$ arising from $h\overrightarrow{\mathbf{e}}_{\perp
}=0$ and $h\overrightarrow{\mathbf{e}}_{\parallel }=1$ belonging outside the
hierarchy generated by $h\mathfrak{R.}$ Such flows correspond to N--adapted
horizontal motions of the curve $\gamma (\tau ,l)=[h\gamma (\tau ,l),v\gamma
(\tau ,l)]$ given by
\begin{equation*}
\left( h\gamma \right) _{\tau }=f\left( \left( h\gamma \right) _{h\mathbf{X}%
},\mathbf{D}_{h\mathbf{X}}\left( h\gamma \right) _{h\mathbf{X}},\mathbf{D}_{h%
\mathbf{X}}^{2}\left( h\gamma \right) _{h\mathbf{X}},\ldots\right)
\end{equation*}%
subject to the non--stretching condition $|\left( h\gamma \right) _{h\mathbf{%
X}}|_{h\mathbf{g}}=1$, where the equation of motion is derived from the
identifications
\begin{equation*}
\left( h\gamma \right) _{\tau }\longleftrightarrow \mathbf{e}_{h\mathbf{Y}},%
\mathbf{D}_{h\mathbf{X}}\left( h\gamma \right) _{h\mathbf{X}%
}\longleftrightarrow \mathcal{D}_{h\mathbf{X}}\mathbf{e}_{h\mathbf{X}}=\left[
\mathbf{L}_{h\mathbf{X}},\mathbf{e}_{h\mathbf{X}}\right]
\end{equation*}%
and so on, which maps the constructions from the tangent space of the curve
to the space $h\mathfrak{p}$. For such identifications we have
\begin{eqnarray*}
\left[ \mathbf{L}_{h\mathbf{X}},\mathbf{e}_{h\mathbf{X}}\right] &=&-\left[
\begin{array}{cc}
0 & \left( 0,\overrightarrow{v}\right) \\
-\left( 0,\overrightarrow{v}\right) ^{T} & h\mathbf{0}%
\end{array}%
\right] \in h\mathfrak{p}, \\
\left[ \mathbf{L}_{h\mathbf{X}},\left[ \mathbf{L}_{h\mathbf{X}},\mathbf{e}_{h%
\mathbf{X}}\right] \right] &=&-\left[
\begin{array}{cc}
0 & \left( |\overrightarrow{v}|^{2},\overrightarrow{0}\right) \\
-\left( |\overrightarrow{v}|^{2},\overrightarrow{0}\right) ^{T} & h\mathbf{0}%
\end{array}%
\right]
\end{eqnarray*}%
and so on, see similar calculus in (\ref{aux41}). At the next step, noting
for the +1 h--flow that
\begin{equation*}
h\overrightarrow{\mathbf{e}}_{\perp }=\overrightarrow{v}_{l}\mbox{
and }h\overrightarrow{\mathbf{e}}_{\parallel }=-\mathbf{D}_{h\mathbf{X}%
}^{-1}\left( \overrightarrow{v}\cdot \overrightarrow{v}_{l}\right) =-\frac{1%
}{2}|\overrightarrow{v}|^{2},
\end{equation*}%
we compute
\begin{eqnarray*}
\mathbf{e}_{h\mathbf{Y}} &=&\left[
\begin{array}{cc}
0 & \left( h\mathbf{e}_{\parallel },h\overrightarrow{\mathbf{e}}_{\perp
}\right) \\
-\left( h\mathbf{e}_{\parallel },h\overrightarrow{\mathbf{e}}_{\perp
}\right) ^{T} & h\mathbf{0}%
\end{array}%
\right] \\
&=&-\frac{1}{2}|\overrightarrow{v}|^{2}\left[
\begin{array}{cc}
0 & \left( 1,\overrightarrow{\mathbf{0}}\right) \\
-\left( 0,\overrightarrow{\mathbf{0}}\right) ^{T} & h\mathbf{0}%
\end{array}%
\right] +\left[
\begin{array}{cc}
0 & \left( 0,\overrightarrow{v}_{h\mathbf{X}}\right) \\
-\left( 0,\overrightarrow{v}_{h\mathbf{X}}\right) ^{T} & h\mathbf{0}%
\end{array}%
\right] \\
&=&\mathbf{D}_{h\mathbf{X}}\left[ \mathbf{L}_{h\mathbf{X}},\mathbf{e}_{h%
\mathbf{X}}\right] +\frac{1}{2}\left[ \mathbf{L}_{h\mathbf{X}},\left[
\mathbf{L}_{h\mathbf{X}},\mathbf{e}_{h\mathbf{X}}\right] \right] \\
&=&-\mathcal{D}_{h\mathbf{X}}\left[ \mathbf{L}_{h\mathbf{X}},\mathbf{e}_{h%
\mathbf{X}}\right] -\frac{3}{2}|\overrightarrow{v}|^{2}\mathbf{e}_{h\mathbf{X%
}}.
\end{eqnarray*}%
 From the above presented identifications related to the first and second
terms, it follows that
\begin{eqnarray*}
|\overrightarrow{v}|^{2} &=&<\left[ \mathbf{L}_{h\mathbf{X}},\mathbf{e}_{h%
\mathbf{X}}\right] ,\left[ \mathbf{L}_{h\mathbf{X}},\mathbf{e}_{h\mathbf{X}}%
\right] >_{h\mathfrak{p}}\longleftrightarrow h\mathbf{g}\left( \mathbf{D}_{h%
\mathbf{X}}\left( h\gamma \right) _{h\mathbf{X}},\mathbf{D}_{h\mathbf{X}%
}\left( h\gamma \right) _{h\mathbf{X}}\right) \\
&=&\left| \mathbf{D}_{h\mathbf{X}}\left( h\gamma \right) _{h\mathbf{X}%
}\right| _{h\mathbf{g}}^{2},
\end{eqnarray*}%
and we can identify $\mathcal{D}_{h\mathbf{X}}\left[ \mathbf{L}_{h\mathbf{X}%
},\mathbf{e}_{h\mathbf{X}}\right] $ with $\mathbf{D}_{h\mathbf{X}}^{2}\left(
h\gamma \right) _{h\mathbf{X}}$ and write
\begin{equation*}
-\mathbf{e}_{h\mathbf{Y}}\longleftrightarrow \mathbf{D}_{h\mathbf{X}%
}^{2}\left( h\gamma \right) _{h\mathbf{X}}+\frac{3}{2}\left| \mathbf{D}_{h%
\mathbf{X}}\left( h\gamma \right) _{h\mathbf{X}}\right| _{h\mathbf{g}%
}^{2}~\left( h\gamma \right) _{h\mathbf{X}}
\end{equation*}%
which is just the first equation (\ref{1map}) in the Theorem \ref{mt}
defining a non--stretching mKdV map h--equation induced by the h--part of
the canonical d--connection.

Using the adjoint representation $ad\left( \cdot \right) $ acting in the Lie
algebra $h\mathfrak{g}=h\mathfrak{p}\oplus \mathfrak{so}(n)$, with
\begin{equation*}
ad\left( \left[ \mathbf{L}_{h\mathbf{X}},\mathbf{e}_{h\mathbf{X}}\right]
\right) \mathbf{e}_{h\mathbf{X}}=\left[
\begin{array}{cc}
0 & \left( 0,\overrightarrow{\mathbf{0}}\right) \\
-\left( 0,\overrightarrow{\mathbf{0}}\right) ^{T} & \overrightarrow{\mathbf{v%
}}%
\end{array}%
\right] \in \mathfrak{so}(n+1),
\end{equation*}%
where%
\begin{equation*}
\overrightarrow{\mathbf{v}}=-\left[
\begin{array}{cc}
0 & \overrightarrow{v} \\
-\overrightarrow{v}^{T} & h\mathbf{0}%
\end{array}%
\in \mathfrak{so}(n)\right] ,
\end{equation*}%
we note (applying $ad\left( \left[ \mathbf{L}_{h\mathbf{X}},\mathbf{e}_{h%
\mathbf{X}}\right] \right) $ again )
\begin{equation*}
ad\left( \left[ \mathbf{L}_{h\mathbf{X}},\mathbf{e}_{h\mathbf{X}}\right]
\right) ^{2}\mathbf{e}_{h\mathbf{X}}=-|\overrightarrow{v}|^{2}\left[
\begin{array}{cc}
0 & \left( 1,\overrightarrow{\mathbf{0}}\right) \\
-\left( 1,\overrightarrow{\mathbf{0}}\right) ^{T} & \mathbf{0}%
\end{array}%
\right] =-|\overrightarrow{v}|^{2}\mathbf{e}_{h\mathbf{X}} .
\end{equation*}%
Hence equation (\ref{1map}) can be represented in alternative form
\begin{equation*}
-\left( h\gamma \right) _{\tau }=\mathbf{D}_{h\mathbf{X}}^{2}\left( h\gamma
\right) _{h\mathbf{X}}-\frac{3}{2}\overrightarrow{R}^{-1}ad\left( \mathbf{D}%
_{h\mathbf{X}}\left( h\gamma \right) _{h\mathbf{X}}\right) ^{2}~\left(
h\gamma \right) _{h\mathbf{X}},
\end{equation*}%
which is more convenient for analysis of higher order flows on $%
\overrightarrow{v}$ subjected to higher--order geometric partial
differential equations. Here we note that the $0$ flow on $\overrightarrow{v}
$ corresponds to just a convective (linear traveling h--wave but subjected
to certain nonholonomic constraints ) map equation (\ref{trmap}).

Now we consider a $-1$ flow contained in the h--hierarchy derived from the
property that $h\overrightarrow{\mathbf{e}}_{\perp }$ is annihilated by the
h--operator $h\mathcal{J}$ and mapped into $h\mathfrak{R}(h\overrightarrow{%
\mathbf{e}}_{\perp })=0$. This mean that $h\mathcal{J}(h\overrightarrow{%
\mathbf{e}}_{\perp })=\overrightarrow{\varpi }=0$. Such properties together
with (\ref{auxaaa}) and equations (\ref{floweq}) imply $\mathbf{L}_{\tau }=0$
and hence $h\mathcal{D}_{\tau }\mathbf{e}_{h\mathbf{X}}=[\mathbf{L}_{\tau },%
\mathbf{e}_{h\mathbf{X}}]=0$ for $h\mathcal{D}_{\tau }=h\mathbf{D}_{\tau }+[%
\mathbf{L}_{\tau },\cdot ]$. We obtain the equation of motion for the
h--component of curve, $h\gamma (\tau ,l)$, following the correspondences $%
\mathbf{D}_{h\mathbf{Y}}\longleftrightarrow h\mathcal{D}_{\tau }$ and $%
h\gamma _{l}\longleftrightarrow \mathbf{e}_{h\mathbf{X}}$,%
\begin{equation*}
\mathbf{D}_{h\mathbf{Y}}\left( h\gamma (\tau ,l)\right) =0,
\end{equation*}%
which is just the first equation in (\ref{-1map}).

Finally, we note that the formulas for the v--components, stated by Theorem %
\ref{mt} can be derived in a similar form by respective substitution in
the above proof of the h--operators and h--variables into v--ones, for
instance, $h\gamma \rightarrow v\gamma $, $h\overrightarrow{\mathbf{e}}%
_{\perp }\rightarrow v\overleftarrow{\mathbf{e}}_{\perp }$, $\overrightarrow{%
v}\rightarrow \overleftarrow{v},\overrightarrow{\varpi }\rightarrow
\overleftarrow{\varpi },\mathbf{D}_{h\mathbf{X}}\rightarrow \mathbf{D}_{v%
\mathbf{X}}$, $\mathbf{D}_{h\mathbf{Y}}\rightarrow \mathbf{D}_{v\mathbf{Y}},%
\mathbf{L\rightarrow C,}\overrightarrow{R}\rightarrow \overleftarrow{S},h%
\mathcal{D\rightarrow }v\mathcal{D}$, $h\mathfrak{R\rightarrow }v\mathfrak{R,%
}h\mathcal{J\rightarrow }v\mathcal{J}$.

\subsection{Nonholonomic mKdV and SG hierarchies}

We present explicit constructions when solitonic hierarchies are derived
following the conditions of Theorem \ref{mt}.

The h--flow and v--flow equations resulting from (\ref{-1map}) are%
\begin{equation}
\overrightarrow{v}_{\tau }=-\overrightarrow{R}h\overrightarrow{\mathbf{e}}%
_{\perp }\mbox{ \ and \ }\overleftarrow{v}_{\tau }=-\overleftarrow{S}v%
\overleftarrow{\mathbf{e}}_{\perp },  \label{deveq}
\end{equation}%
when, respectively,%
\begin{equation*}
0=\overrightarrow{\varpi }=-\mathbf{D}_{h\mathbf{X}}h\overrightarrow{\mathbf{%
e}}_{\perp }+h\mathbf{e}_{\parallel }\overrightarrow{v},~\mathbf{D}_{h%
\mathbf{X}}h\mathbf{e}_{\parallel }=h\overrightarrow{\mathbf{e}}_{\perp
}\cdot \overrightarrow{v}
\end{equation*}%
and
\begin{equation*}
0=\overleftarrow{\varpi }=-\mathbf{D}_{v\mathbf{X}}v\overleftarrow{\mathbf{e}%
}_{\perp }+v\mathbf{e}_{\parallel }\overleftarrow{v},~\mathbf{D}_{v\mathbf{X}%
}v\mathbf{e}_{\parallel }=v\overleftarrow{\mathbf{e}}_{\perp }\cdot
\overleftarrow{v}.
\end{equation*}%
The d--flow equations possess horizontal and vertical conservation laws%
\begin{equation*}
\mathbf{D}_{h\mathbf{X}}\left( (h\mathbf{e}_{\parallel })^{2}+|h%
\overrightarrow{\mathbf{e}}_{\perp }|^{2}\right) =0,
\end{equation*}%
for $(h\mathbf{e}_{\parallel })^{2}+|h\overrightarrow{\mathbf{e}}_{\perp
}|^{2}=<h\mathbf{e}_{\tau },h\mathbf{e}_{\tau }>_{h\mathfrak{p}}=|\left(
h\gamma \right) _{\tau }|_{h\mathbf{g}}^{2}$, and
\begin{equation*}
\mathbf{D}_{v\mathbf{Y}}\left( (v\mathbf{e}_{\parallel })^{2}+|v%
\overleftarrow{\mathbf{e}}_{\perp }|^{2}\right) =0,
\end{equation*}%
for $(v\mathbf{e}_{\parallel })^{2}+|v\overleftarrow{\mathbf{e}}_{\perp
}|^{2}=<v\mathbf{e}_{\tau },v\mathbf{e}_{\tau }>_{v\mathfrak{p}}=|\left(
v\gamma \right) _{\tau }|_{v\mathbf{g}}^{2}$. This corresponds to
\begin{equation*}
\mathbf{D}_{h\mathbf{X}}|\left( h\gamma \right) _{\tau }|_{h\mathbf{g}}^{2}=0%
\mbox{ \ and \ }\mathbf{D}_{v\mathbf{X}}|\left( v\gamma \right) _{\tau }|_{v%
\mathbf{g}}^{2}=0.
\end{equation*}%
(The problem of formulating conservation laws on N--anholonomic spaces ---
in particular, on nonholonomic vector bundles --- is analyzed in Ref. \cite%
{vsgg}. In general, such laws are more sophisticated than those on (semi)
Riemannian spaces because of nonholonomic constraints resulting in
non--symmetric Ricci tensors and different types of identities. But for the
geometries modeled for dimensions $n=m$ with canonical d--connections, we
get similar h-- and v--components of the conservation law equations as on
symmetric Riemannian spaces.)

Without loss of generality we can rescale conformally the variable $\tau $
in order to get $|\left( h\gamma \right) _{\tau }|_{h\mathbf{g}}^{2}$ $=1$
and (perhaps by a different rescaling)
$|\left( v\gamma \right) _{\tau }|_{v\mathbf{g}}^{2}=1$,
so that
\begin{equation*}
(h\mathbf{e}_{\parallel })^{2}+|h\overrightarrow{\mathbf{e}}_{\perp }|^{2}=1%
\mbox{ \ and \ }(v\mathbf{e}_{\parallel })^{2}+|v\overleftarrow{\mathbf{e}}%
_{\perp }|^{2}=1.
\end{equation*}%
In this case, we can express $h\mathbf{e}_{\parallel }$ and $h%
\overrightarrow{\mathbf{e}}_{\perp }$ in terms of $\overrightarrow{v}$ and
its derivatives and, similarly, we can express $v\mathbf{e}_{\parallel }$
and $v\overleftarrow{\mathbf{e}}_{\perp }$ in terms of $\overleftarrow{v}$
and its derivatives, which follows from (\ref{deveq}). The N--adapted wave
map equations describing the $-1$ flows reduce to a system of two
independent nonlocal evolution equations for the h-- and v--components,%
\begin{equation*}
\overrightarrow{v}_{\tau }=-\mathbf{D}_{h\mathbf{X}}^{-1}\left( \sqrt{%
\overrightarrow{R}^{2}-|\overrightarrow{v}_{\tau }|^{2}}~\overrightarrow{v}%
\right) \mbox{ \ and \ }\overleftarrow{v}_{\tau }=-\mathbf{D}_{v\mathbf{X}%
}^{-1}\left( \sqrt{\overleftarrow{S}^{2}-|\overleftarrow{v}_{\tau }|^{2}}~%
\overleftarrow{v}\right) .
\end{equation*}%
For N--anholonomic spaces of constant scalar d--curvatures, we can rescale
the equations on $\tau $ to the case when the terms $\overrightarrow{R}^{2},%
\overleftarrow{S}^{2}=1$, and the evolution equations transform into a
system of hyperbolic d--vector equations,%
\begin{equation}
\mathbf{D}_{h\mathbf{X}}(\overrightarrow{v}_{\tau })=-\sqrt{1-|%
\overrightarrow{v}_{\tau }|^{2}}~\overrightarrow{v}\mbox{ \ and \ }\mathbf{D}%
_{v\mathbf{X}}(\overleftarrow{v}_{\tau })=-\sqrt{1-|\overleftarrow{v}_{\tau
}|^{2}}~\overleftarrow{v},  \label{heq}
\end{equation}%
where $\mathbf{D}_{h\mathbf{X}}=\partial _{hl}$ and $\mathbf{D}_{v\mathbf{X}%
}=\partial _{vl}$ are usual partial derivatives, respectively, along $hl$
and $vl$ with $\overrightarrow{v}_{\tau }$ and $\overleftarrow{v}_{\tau }$
considered as scalar functions for the covariant derivatives $\mathbf{D}_{h%
\mathbf{X}}$ and $\mathbf{D}_{v\mathbf{X}}$ defined by the canonical
d--connection. It also follows that $h\overrightarrow{\mathbf{e}}_{\perp }$
and $v\overleftarrow{\mathbf{e}}_{\perp }$ obey corresponding vector
sine--Gordon (SG) equations%
\begin{equation}
\left( \sqrt{(1-|h\overrightarrow{\mathbf{e}}_{\perp }|^{2})^{-1}}~\partial
_{h\mathbf{l}}(h\overrightarrow{\mathbf{e}}_{\perp })\right) _{\tau }=-h%
\overrightarrow{\mathbf{e}}_{\perp }  \label{sgeh}
\end{equation}%
and
\begin{equation}
\left( \sqrt{(1-|v\overleftarrow{\mathbf{e}}_{\perp }|^{2})^{-1}}~\partial
_{v\mathbf{l}}(v\overleftarrow{\mathbf{e}}_{\perp })\right) _{\tau }=-v%
\overleftarrow{\mathbf{e}}_{\perp }.  \label{sgev}
\end{equation}

The above presented formulas and Corollary \ref{c2} imply

\begin{theorem}%{conclusion}
The recursion d--operator $\mathfrak{R}=(h\mathfrak{R,}h\mathfrak{R})$ (\ref%
{reqop}) generates two hierarchies of vector mKdV symmetries: the first one
is horizontal (see (\ref{reqoph})),
\begin{eqnarray}
\overrightarrow{v}_{\tau }^{(0)} &=&\overrightarrow{v}_{hl},~\overrightarrow{%
v}_{\tau }^{(1)}=h\mathfrak{R}(\overrightarrow{v}_{hl})=\overrightarrow{v}%
_{3hl}+\frac{3}{2}|\overrightarrow{v}|^{2}~\overrightarrow{v}_{hl},
\label{mkdv1a} \\
\overrightarrow{v}_{\tau }^{(2)} &=&h\mathfrak{R}^{2}(\overrightarrow{v}%
_{hl})=\overrightarrow{v}_{5hl}+\frac{5}{2}\left( |\overrightarrow{v}|^{2}~%
\overrightarrow{v}_{2hl}\right) _{hl}  \notag \\
&&+\frac{5}{2}\left( (|\overrightarrow{v}|^{2})_{hl\mathbf{~}hl}+|%
\overrightarrow{v}_{h\mathbf{l}}|^{2}+\frac{3}{4}|\overrightarrow{v}%
|^{4}\right) ~\overrightarrow{v}_{hl}-\frac{1}{2}|\overrightarrow{v}%
_{hl}|^{2}~\overrightarrow{v},  \notag \\
&&\text{ and so on},  \notag
\end{eqnarray}%
with all such flows commuting with the $-1$ flow
\begin{equation}
(\overrightarrow{v}_{\tau })^{-1}=h\overrightarrow{\mathbf{e}}_{\perp }
\label{mkdv1b}
\end{equation}%
associated to the vector SG equation (\ref{sgeh}); the second one is
vertical (see (\ref{reqopv})),
\begin{eqnarray}
\overleftarrow{v}_{\tau }^{(0)} &=&\overleftarrow{v}_{v\mathbf{l}},~%
\overleftarrow{v}_{\tau }^{(1)}=v\mathfrak{R}(\overleftarrow{v}_{vl})=%
\overleftarrow{v}_{3vl}+\frac{3}{2}|\overleftarrow{v}|^{2}~\overleftarrow{v}%
_{vl},  \label{mkdv2a} \\
\overleftarrow{v}_{\tau }^{(2)} &=&v\mathfrak{R}^{2}(\overleftarrow{v}_{vl})=%
\overleftarrow{v}_{5vl}+\frac{5}{2}\left( |\overleftarrow{v}|^{2}~%
\overleftarrow{v}_{2vl}\right) _{vl}  \notag \\
&&+\frac{5}{2}\left( (|\overleftarrow{v}|^{2})_{vl\mathbf{~}vl}+|%
\overleftarrow{v}_{vl}|^{2}+\frac{3}{4}|\overleftarrow{v}|^{4}\right) ~%
\overleftarrow{v}_{v\mathbf{l}}-\frac{1}{2}|\overleftarrow{v}_{vl}|^{2}~%
\overleftarrow{v},  \notag \\
&&\text{ and so on},  \notag
\end{eqnarray}%
with all such flows commuting with the $-1$ flow
\begin{equation}
(\overleftarrow{v}_{\tau })^{-1}=v\overleftarrow{\mathbf{e}}_{\perp }
\label{mkdv2b}
\end{equation}%
associated to the vector SG equation (\ref{sgev}).
\end{theorem}

Furthermore, the adjoint d--operator $\mathfrak{R}^{\ast }=\mathcal{J\circ H}$
generates a horizontal hierarchy of Hamiltonians,%
\begin{eqnarray}
hH^{(0)} &=&\frac{1}{2}|\overrightarrow{v}|^{2},~hH^{(1)}=-\frac{1}{2}|%
\overrightarrow{v}_{hl}|^{2}+\frac{1}{8}|\overrightarrow{v}|^{4},
\label{hhh} \\
hH^{(2)} &=&\frac{1}{2}|\overrightarrow{v}_{2hl}|^{2}-\frac{3}{4}|%
\overrightarrow{v}|^{2}~|\overrightarrow{v}_{hl}|^{2}-\frac{1}{2}\left(
\overrightarrow{v}\cdot \overrightarrow{v}_{hl}\right) +\frac{1}{16}|%
\overrightarrow{v}|^{6},  \notag \\
&&\text{ and so on},  \notag
\end{eqnarray}%
and vertical hierarchy of Hamiltonians%
\begin{eqnarray}
vH^{(0)} &=&\frac{1}{2}|\overleftarrow{v}|^{2},~vH^{(1)}=-\frac{1}{2}|%
\overleftarrow{v}_{vl}|^{2}+\frac{1}{8}|\overleftarrow{v}|^{4},  \label{hhv}
\\
vH^{(2)} &=&\frac{1}{2}|\overleftarrow{v}_{2vl}|^{2}-\frac{3}{4}|%
\overleftarrow{v}|^{2}~|\overleftarrow{v}_{vl}|^{2}-\frac{1}{2}\left(
\overleftarrow{v}\cdot \overleftarrow{v}_{vl}\right) +\frac{1}{16}|%
\overleftarrow{v}|^{6},  \notag \\
&&\text{ and so on},  \notag
\end{eqnarray}%
all of which are conserved densities for respective horizontal and vertical $%
-1$ flows and determining higher conservation laws for the corresponding
hyperbolic equations (\ref{sgeh}) and (\ref{sgev}).

The above presented horizontal equations (\ref{sgeh}), (\ref{mkdv1a}), (\ref%
{mkdv1b}) and (\ref{hhh}) and of vertical equations (\ref{sgev}), (\ref%
{mkdv2a}), (\ref{mkdv2b}) and (\ref{hhv}) have similar mKdV scaling
symmetries but with distinct parameters $\lambda _{h}$ and $\lambda _{v}$
because, in general, there are two independent scalar curvatures $%
\overrightarrow{R}$ and $\overleftarrow{S}$, see (\ref{sdccurv}). The
horizontal scaling symmetries are $hl\mathbf{\rightarrow }\lambda _{h}hl%
\mathbf{,}\overrightarrow{v}\rightarrow \left( \lambda _{h}\right) ^{-1}%
\overrightarrow{v}$ and $\tau \rightarrow \left( \lambda _{h}\right)
^{1+2k}, $ for $k=-1,0,1,2,..$., and likewise for the vertical scaling
symmetries. %, one has $vl%
%\mathbf{\rightarrow }\lambda _{v}vl\mathbf{,}\overleftarrow{v}\rightarrow
%\left( \lambda _{v}\right) ^{-1}\overleftarrow{v}$ and $\tau \rightarrow
%\left( \lambda _{v}\right) ^{1+2k}$, for $k=-1,0,1,2,..$.

Finally, we consider again the Remark \ref{rema} stating that similar
results (proved in Section 4) can be derived for unitary groups with complex
variables. The generated bi-Hamiltonian horizontal and vertical hierarchies
and solitonic equations will be different from those defined for real
orthogonal groups; for holonomic spaces this is demonstrated in Ref. \cite%
{anc1}. This distinguishes substantially the models of gauge gravity with
structure groups like the unitary one from those with orthogonal groups.

\section{Applications in Geometric Mechanics, Fin\-sler Geometry and Gravity}

\label{sec5}Here we consider some interesting examples when the data
defining fundamental geometric structures in mechanics and Finsler geometry,
or exact solutions in gravity, can be transformed into solitonic
hierarchies. The possibility to describe gravitational and electromagnetic
interactions by mechanical models, and vice versa, will be investigated.

\subsection{Geometric mechanics and Finsler geometry}

A differentiable Lagrangian $L(x,y)$, i.e. a fundamental Lagrange function,
is defined by a map $L:(x,y)\in TM\rightarrow L(x,y)\in \mathbb{R}$ of class
$\mathcal{C}^{\infty }$ on $\widetilde{TM}=TM\backprime \{0\}$ and
continuous on the null section $0:\ M\rightarrow TM$ of $\pi $. A regular
Lagrangian has non-degenerate Hessian
\begin{equation}
\ ^{(L)}g_{ij}(x,y)=\frac{1}{2}\frac{\partial ^{2}L(x,y)}{\partial
y^{i}\partial y^{j}}  \label{lqf}
\end{equation}%
when $rank\left| g_{ij}\right| =n$ on $\widetilde{TM}$.

\begin{definition}
A Lagrange space is a pair $L^{n}=\left[ M,L(x,y)\right] $ with $\
^{(L)}g_{ij}$ being of fixed signature over $\mathbf{V}=\widetilde{TM}$.
\end{definition}

The notion of Lagrange space was introduced by J. Kern \cite{kern} and
elaborated in details by R. Miron's school, see Refs. \cite{ma1,ma2,mh}, as
a natural extension of Finsler geometry \cite{fcartan,bej,matsumoto,bej}
(see also Refs. \cite{vsgg,vmon1}, on Lagrange--Finsler super/noncommutative
geometry). Straightforward calculations (on nonholonomic manifolds they are
reviewed in Refs. \cite{vsgg}) establish the following results:

\begin{enumerate}
\item The Euler--Lagrange equations%
\begin{equation*}
\frac{d}{d\tau }\left( \frac{\partial L}{\partial y^{i}}\right) -\frac{%
\partial L}{\partial x^{i}}=0
\end{equation*}%
where $y^{i}=\frac{dx^{i}}{d\tau }$ for $x^{i}(\tau )$ depending on
parameter $\tau $, are equivalent to the ``nonlinear'' geodesic equations
\begin{equation*}
\frac{d^{2}x^{i}}{d\tau ^{2}}+2G^{i}(x^{k},\frac{dx^{j}}{d\tau })=0
\end{equation*}%
defining paths of a canonical semispray%
\begin{equation*}
S=y^{i}\frac{\partial }{\partial x^{i}}-2G^{i}(x,y)\frac{\partial }{\partial
y^{i}}
\end{equation*}%
where
\begin{equation*}
2G^{i}(x,y)=\frac{1}{2}\ ^{(L)}g^{ij}\left( \frac{\partial ^{2}L}{\partial
y^{i}\partial x^{k}}y^{k}-\frac{\partial L}{\partial x^{i}}\right)
\end{equation*}%
with $^{(L)}g^{ij}$ being inverse to (\ref{lqf}).

\item There exists on $\mathbf{V\simeq }$ $\widetilde{TM}$ a canonical
N--connection $\ $%
\begin{equation}
^{(L)}N_{j}^{i}=\frac{\partial G^{i}(x,y)}{\partial y^{i}}  \label{cncl}
\end{equation}%
defined by the fundamental Lagrange function $L(x,y)$, which prescribes
nonholonomic frame structures of type (\ref{dder}) and (\ref{ddif}), $^{(L)}%
\mathbf{e}_{\nu }=(\mathbf{e}_{i},e_{a})$ and $^{(L)}\mathbf{e}^{\mu
}=(e^{i},\mathbf{e}^{a})$.

\item There is a canonical metric structure%
\begin{equation}
\ ^{(L)}\mathbf{g}=\ g_{ij}(x,y)\ e^{i}\otimes e^{j}+\ g_{ij}(x,y)\ \mathbf{e%
}^{i}\otimes \mathbf{e}^{j}  \label{slm}
\end{equation}%
constructed as a Sasaki type lift from $M$ for $g_{ij}(x,y)$.

\item There is a unique metrical and, in this case, torsionless canonical
d--connection$\ ^{(L)}\mathbf{D}=(hD,vD)$ with the nontrivial coefficients
with respect to $^{(L)}\mathbf{e}_{\nu }$ and $^{(L)}\mathbf{e}^{\mu }$
paramet\-riz\-ed respectively $\Gamma _{\ \beta \gamma }^{\alpha }=(L_{\
jk}^{i},C_{bc}^{a})$, for
\begin{eqnarray}
L_{\ jk}^{i} &=&\frac{1}{2}g^{ih}(\mathbf{e}_{k}g_{jh}+\mathbf{e}_{j}g_{kh}-%
\mathbf{e}_{h}g_{jk}),  \label{cdctb} \\
C_{\ jk}^{i} &=&\frac{1}{2}g^{ih}(e_{k}g_{jh}+e_{b}g_{kh}-e_{e}g_{bc})
\notag
\end{eqnarray}%
defining the generalized Christoffel symbols, where (for simplicity, we
omitted the left up labels $(L)$ for N--adapted bases). The connection $%
^{(L)}\mathbf{D}$ is metric compatible and torsionless, see an outline of
main definitions and formulas in the next subsection and Appendix.
\end{enumerate}

We conclude that any regular Lagrange mechanics can be geometrized as a
nonholonomic Riemann manifold $\mathbf{V}$ equipped with canonical
N--connection (\ref{cncl}) and adapted d--connection (\ref{cdctb}) and
d--metric structures (\ref{slm}) all induced by a $L(x,y)$. In some
approaches to Finsler geometry and generalizations \cite{bcs}, one consider
nontrivial non-metric structures. For instance, the so called Chern
d--connection $\ ^{Ch}\mathbf{D}$ is also a minimal extension of the Levi
Civita connection when $h\ ^{Ch}\mathbf{D(}g\mathbf{)=}0$ but $v\ ^{Ch}%
\mathbf{D(}h\mathbf{)\neq }0$, see formulas (\ref{metcompt}). Such
generalized Riemann--Finsler spaces can be modeled on nonholonomic
metric--affine manifolds, with nonmetricity, see detailed discussions in %
\cite{vsgg,ma1,ma2,bejf}. We note that the N--connection is induced by the
semispray configurations subjected to generalized nonlinear geodesic
equations equivalent to the Euler--Lagrange equations. A N--connection
structure transforms a Riemannian space into a nonholonomic one with
preferred honholonomic frame structure of type (\ref{dder}) and (\ref{ddif}).

\begin{remark}
Any Finsler geometry with a fundamental Finsler function $F(x,y)$, being
homogeneous of type $F(x,\lambda y)=\lambda F(x,y)$, for nonzero $\lambda
\in \mathbb{R}$, may be considered as a particular case of Lagrange geometry
when $L=F^{2}$. We shall apply the methods of Finsler geometry in this work.%
\footnote{%
In another direction, there is a proof \cite{bcs} that any Lagrange
fundamental function $L$ can be modeled as a singular case in a certain
class of Finsler geometries of extra dimension. Nevertheless the concept of
Lagrangian is a very important geometrical and physical one and we shall
distinguish the cases when we model a Lagrange or a Finsler geometry:\ A
physical or mechanical model with a Lagrangian is not only a ``singular''
case for a Finsler geometry but reflects a proper set of geometric objects
and structures with possible new concepts in physical theories.}
\end{remark}

For applications in optics of nonhomogeneous media and gravity (see, for
instance, Refs. \cite{ma2,vsgg,esv,vncg}) one considers metric forms of type
$g_{ij}\sim e^{\lambda (x,y)}\ ^{(L)}g_{ij}(x,y)\ $\ which can not be
derived explicitly from a mechanical Lagrangian. In the so--called
generalized Lagrange geometry one considers Sasaki type metrics (\ref{slm})
with certain general coefficients both for the d--metric and N--connection \
and canonical d--connection, i.e. when $^{(L)}g_{ij}\rightarrow \left[
g_{ij}(x,y),h_{ab}(x,y)\right] $, and $^{(L)}N_{j}^{i}\rightarrow
N_{j}^{i}(x,y)$. We shall use the term (generalized) Lagrange--Finsler
geometry for all such geometries modeled on tangent bundles or on arbitrary
N--anholonomic manifold $\mathbf{V}$. In original form, such spaces were
called generalized Lagrange spaces and denoted $GL^{n}=(M,g_{ij}(x,y))$, see %
\cite{ma1,ma2}.

\subsection{Analogous Models and Lagrange--Finsler geometry}

For a regular Lagrangian $\mathsf{L}(x,y)$ and corresponding Euler--Lagrange
equations it is possible to construct canonically a generalized Finsler
geometry on tangent bundle \cite{kern,ma1,ma2}. In our approach, such
constructions follow from the Theorem \ref{teleq} if instead of $g_{%
\underline{\alpha }\underline{\beta }}(x)$ there are considered the metric
components
\begin{equation}
g_{\underline{i}\underline{j}}(x,y)=\frac{1}{2}\frac{\partial ^{2}\mathsf{L}%
(x,y)}{\partial y^{\underline{i}}\partial y^{\underline{j}}},  \label{ilm}
\end{equation}%
introduced in (\ref{auxm}) for $e_{\alpha }^{~\underline{\alpha }%
}(x,y)=\delta _{\alpha }^{~\underline{\alpha }}$. For arbitrary given $g_{%
\underline{i}\underline{j}}(x,y)$ and $e_{\alpha }^{~\underline{\alpha }%
}(x,y)$ on $TM$, defining a d--metric (\ref{slme}), we can compute the
corresponding N--connection (\ref{cnlce}) and canonical d--connection
coefficients (\ref{candcon}). In general, the coefficients of d--curvatures (%
\ref{dcurvtb}) are not constant and we are not able to generate solitonic
hierarchies. In order to solve the problem we have to choose $e_{\alpha }^{~%
\underline{\alpha }}(x,y)$ so that the equations
\begin{equation*}
\frac{\partial ^{2}\mathsf{L}(x,y)}{\partial y^{\underline{a}}\partial y^{%
\underline{b}}}=2\mathring{g}_{ab}~e_{\underline{a}}^{~a}(x,y)~e_{\underline{%
b}}^{~b}(x,y)
\end{equation*}%
are satisfied (similarly to (\ref{aux4}), for given $\mathsf{L}$ and $%
\mathring{g}_{ab})$, resulting in constant curvature coefficients with
respect to N--adapted bases. This allows us to define $\mathbf{D}_{h\mathbf{X%
}}^{\mathsf{L}}$ and $\mathbf{D}_{v\mathbf{X}}^{\mathsf{L}}$ from formulas (%
\ref{trmap}), (\ref{1map}) or (\ref{-1map}), as pairs of h- and v--operators
with ``nonholonomic mixture'' (cf. Theorem \ref{mt}).

\begin{corollary}
Any regular Lagrangian $\mathsf{L}(x,y)$ induces a hierarchy of
bi-Hamiltonian N--adapted flows of curves described by geometric
nonholonomic (solitonic) map equations.
\end{corollary}

We can describe equivalently any regular Lagrange mechanics in terms of
solitonic equations, their solutions and symmetries.

\begin{remark}
Considering $\mathsf{L}(x,y)=F^{2}(x,y)$, where the homogeneous \newline
$F(x,\lambda y)=\lambda F(x,y)$ is a fundamental Finsler metric function
(see, for instance, \cite{ma2}), we can encode in solitonic constructions
all data for a Finsler geometry.
\end{remark}

Here it should be noted that there were elaborated alternative approaches
when gravitational effects are modeled in continuous and discrete media, see
review \cite{blv}, but they do not allow one to generate the Einstein equations
starting from a Lagrangian in a mechanical models like in \cite{kern,ma1,ma2}%
, or from a distribution on a nonholonomic manifold \cite{vr1,vr2,bejf,vsgg}%
. In our approach, for a d--metric (\ref{slme}) induced by (\ref{ilm})
(introduced in (\ref{auxm})), we can prove a geometric mechanical analog of
Theorem \ref{tisrg}:

\begin{theorem}
Any regular Lagrangian $\mathsf{L}(x,y)$ and frame structure $e_{\alpha }^{~%
\underline{\alpha }}(x,y)$ define a nonholonomic (semi) Riemannian geometry
on $TM$.
\end{theorem}

This way we can model various gravitational effects by certain mechanical
configurations.

\begin{corollary}
\label{corneq}A regular Lagrangian $\mathsf{L}(x,y)$ and frame structure $%
e_{\alpha }^{~\underline{\alpha }}(x,y)$ on $TM$ model a vacuum gravity
configuration with effective metric $\tilde{g}_{\alpha \beta }\ =[\tilde{g}%
_{ij},\tilde{g}_{ij}]$ (\ref{slme}) if the corresponding canonical
d--connection (\ref{candcon}) has vanishing Ricci d--tensor (\ref{dricci}).
\end{corollary}

We conclude that for a fixed Lagrangian $\mathsf{L}(x,y)$ we can define
frame structures $e_{\alpha }^{~\underline{\alpha }}(x,y)$ on $TM$ inducing
Einstein spaces or curved spaces with constant curvature coefficients
(computed with respect to certain N--adapted basis).

\begin{definition}
A geometric model defined by data $\mathbf{E}^{n+m}=[g_{ij}(x^{k},y^{c})$,
\newline
$g_{ab}(x^{k},y^{c}),N_{i}^{a}(x^{k},y^{c})]\ $ is N--anholonomically
equivalent to another one given by data $\widetilde{\mathbf{E}}^{n+m}=[%
\widetilde{g}_{ij}(x^{k},y^{c}),\widetilde{g}_{ab}(x^{k},y^{c}),\widetilde{N}%
_{i}^{a}(x^{k},y^{c})]$ if for the same splitting $n+m$ with $\widetilde{%
\mathbf{e}}_{\alpha }\rightarrow \mathbf{e}_{\alpha }$ there are nontrivial
polarizations $[\eta _{ij}(x^{k},y^{c}),\eta _{ab}(x^{k},y^{c})$, $\eta
_{i}^{a}(x^{k},y^{c})]$ for which $g_{\alpha \beta }=\eta _{\alpha \beta }%
\widetilde{g}_{\alpha \beta }$ and $N_{i}^{a}=\eta _{i}^{a}\widetilde{N}%
_{i}^{a}$.
\end{definition}

In general, the physical and geometric properties of two such
N--anholono\-mically related spaces $\mathbf{E}^{n+m}$ and $\widetilde{%
\mathbf{E}}^{n+m}$ are very different. Nevertheless, we are able to compute
the coefficients of geometrical and physical objects, define symmetries,
conservation laws and fundamental equations on a space from similar values
of another one if the polarizations are stated in explicit form. In some
particular cases, we can consider that $\mathbf{E}^{n+m}$ is defined by the
data with constant curvature coefficients with respect to an N--adapted
basis (or an exact solution of the Einstein equations) but $\widetilde{%
\mathbf{E}}^{n+m}$ is related to a regular Lagrange/ Finsler geometry.

\subsection{Modeling field interactions in Lagrange--Finsler ge\-ometry}

Let us consider a regular Lagrangian%
\begin{equation}
\mathsf{L}(x,y)=m_{0}\underline{a}_{ij}(x)y^{i}y^{j}+e_{0}\underline{A}%
_{i}(x)y^{i},  \label{lagr1}
\end{equation}%
where $m_{0},e_{0}$ are constants, $\underline{A}_{i}(x)$ is a vector field
and $\underline{a}_{ij}(x)=a_{\underline{i}\underline{j}}(x)$ is a second
rank symmetric tensor. The metric coefficients (\ref{ilm}) are $g_{%
\underline{i}\underline{j}}=m_{0}~a_{\underline{i}\underline{j}}(x)$ and
their frame transforms\ (\ref{auxm}) are given by
\begin{equation}
\widetilde{g}_{ab}(x,y)=e_{a}^{~\underline{a}}(x,y)~e_{b}^{~\underline{b}%
}(x,y)~g_{\underline{a}\underline{b}}(x)=m_{0}~a_{ab}(x,y),  \label{nonhtrm}
\end{equation}%
where $a_{ab}=e_{a}^{~\underline{a}}~e_{b}^{~\underline{b}}~a_{\underline{a}%
\underline{b}}$. For simplicity, we consider frame transforms to local bases
$e_{\alpha }=(e_{i},e_{a})$ when $e_{a}=e_{a}^{~\underline{a}}~\partial
/\partial y^{\underline{a}}$ are holonomic vectors but $e_{i}=e_{i}^{~%
\underline{i}}~\partial /\partial x^{\underline{i}}$ will be defined by a
N--connection like in (\ref{dder}). We compute (see Theorem \ref{teleq} and
formula (\ref{cnlce}))
\begin{equation*}
\widetilde{G}^{i}=\frac{1}{2}~^{g}\Gamma _{~jk}^{i}y^{j}y^{k}+\frac{1}{m_{0}}%
a^{ij}~^{g}F_{jk}~y^{k},
\end{equation*}%
where $^{g}\Gamma _{~jk}^{i}$ are Christoffel symbols of the tensor $%
\widetilde{g}_{ab}$ constructed by using derivatives $\partial /\partial
x^{i}$, $~^{g}F_{jk}=\frac{e_{0}}{4}(~^{g}D_{k}A_{j}-~^{g }D_{j}A_{k})$ with
the covariant derivative $~^{g}D_{j}$ defined by $^{g}\Gamma _{~jk}^{i}$,
and
\begin{equation}
\tilde{N}_{\ j}^{i}(x,y)=~^{g}\Gamma _{~jk}^{i}y^{k}-~^{g}F_{jk}.
\label{nceml}
\end{equation}%
The N--connection curvature (\ref{ncurv}) is
\begin{equation*}
\Omega _{ik}^{a}=y^{b}~^{g}R_{~bjk}^{a}-2~^{g}D_{[k}~^{g}F_{~i]}^{a} ,
\end{equation*}%
where $~^{g}R_{~bjk}^{a}$ is the curvature tensor of the Levi Civita
connection for $~g_{\underline{a}\underline{b}}(x)$, computed with respect
to a usual coordinate base and then nonholonomically transformed by $e_{a}^{~%
\underline{a}}(x,y)$.

For $m_{0}=mc$ and $e_{0}=2e/m$, where $m$ and $e$ are respectively the mass
and electric charge of a point particle and $c$ is the light speed, the
Lagrangian (\ref{lagr1}) describes the dynamics of a point electrically
charged particle in a curved background space $M$ with metric $\underline{a}%
_{ij}(x)$. The formula (\ref{nonhtrm}) states a class of nonholonomic
deformations of the metric on $M$ to $TM$. We can treat $~^{g}F_{jk}$ as an
effective electro-magnetic tensor field on $TM$. The auto-parallel curves $%
u^{\alpha }(\tau )=[x^{i}(\tau ),y^{i}(\tau )=dx^{i}/\tau ]$, parametrized
by a scalar variable $\tau $, adapted to $\tilde{N}_{\ j}^{i}(x,y)$ (\ref%
{nceml}), are
\begin{equation*}
\frac{dy^{a}}{d\tau }+~^{g}\Gamma _{~bc}^{a}(x,y)~y^{b}y^{c}=~^{g
}F_{~b}^{a}(x,y)~y^{b}.
\end{equation*}%
The nonholonomic Riemannian mechanical model of $\mathsf{L}(x,y)$ is
completely defined by a d--metric (\ref{slme}) with coefficients (\ref%
{nonhtrm}),
\begin{equation*}
\mathbf{\tilde{g}}=m_{0}~a_{ij}(x,y)\ e^{i}\otimes e^{j}+\
m_{0}~a_{ij}(x,y)\ \mathbf{\tilde{e}}^{i}\otimes \mathbf{\tilde{e}}^{j},
\end{equation*}%
where $\mathbf{\tilde{e}}^{i}$ are elongated by $\tilde{N}_{\ j}^{i}$ from (%
\ref{nceml}). From Corollary \ref{corneq}, we have:

\begin{conclusion}
Any solution of the Einstein--Maxwell equations given by gravitational field
$\underline{a}_{ij}(x)$ and electromagnetic potential $\underline{A}_{i}(x)$
is N--anholo\-no\-mically equivalent to a mechanical model with regular
Lagrangian (\ref{lagr1}) and associated nonholonomic geometry on $TM$ with
induced d--metric $\mathbf{\tilde{g}}$ and N--connection structure $\tilde{N}%
_{\ j}^{i}$.
\end{conclusion}

If the frame coefficients $e_{a}^{~\underline{a}}$ are chosen to yield
constant d--curvatures coefficients, we get a particular statement of the
main Theorem \ref{mt}:

\begin{corollary}
The Einstein--Maxwell gravity equations contain solitonic solutions that can
be modeled as a hierarchy of bi-Hamiltonian N--adapted flows of curves.
\end{corollary}

In a similar form, we can derive spinor--solitonic nonholonomic hierarchies
for Einstein--Dirac equations if we consider unitary groups and
distinguished Clifford structures \cite{vsgg}.

\section{Conclusion}

In this paper, it has been shown that the geometry of regular Lagrange
mechanics and generalized Lagrange--Finsler spaces can be encoded in
nonholonomic hierarchies of bi-Hamiltonian structures and related solitonic
equations derived for curve flows. Although the constructions are performed
in explicit form for a special class of generalized Lagrange spaces with
constant matrix curvature coefficients computed with respect to canonical
nonholonomic frames (induced by Lagrange or Finsler metric fundamental
functions), the importance of the resulting geometric nonlinear analysis of
such physical systems is beyond doubt for further applications in classical
and quantum field theory. This remarkable generality appears naturally for all
types of models of gravitational, gauge and spinor interactions, geometrized
in terms of vector/ spinor bundles, in supersymmetric and/or (non)
commutative variants when nonholonomic frames and generalized linear and
nonlinear connections are introduced into consideration.

If the gauge group structure of vector bundles and nonholonomic manifolds
is defined by an orthogonal group $\mathbf{H}=SO(n)\oplus SO(m)$ acting on
the base/ horizontal and typical fiber /vertical subspaces
with, say, respective dimensions of $n$ and $m$, then these subspaces
will be Riemannian symmetric spaces $\mathbf{G/H}$ whose geometric properties
are determined by the Lie groups
$\mathbf{G}=SO(n)\oplus SO(m)$ and $\mathbf{G}=SU(n)\oplus SU(m)$.
This structure can be formulated equivalently in terms of geometric objects
defined on pairs of Klein geometries.
The bi-Hamiltonian solitonic hierarchies are generated naturally by
recursion operators associated with the horizontal and vertical flows of
curves on such spaces, with these flows (of SG and mKdV type) being
geometrically described by wave maps and mKdV analogs of Schr\"odinger maps.

In the case of holonomic manifolds, the construction of bi-Hamiltonian curve
flows, soliton equations, and geometric PDE maps has recently been
generalized in Ref. \cite{ancjgp} to all Riemannian symmetric manifolds
$M=G/H$ including all compact semisimple Lie groups $K=G/H$ as given by
$G=K\times K$ and $H=$diag$G$. This generalization makes it possible to
study more general field models, such as quaternionic and octonian systems.

Considering arbitrary curved spaces and mechanical or field systems, it is not
clear if any bi-Hamiltonian structures and solitonic equations can be
derived from curve flows. The answer seems to be negative because arbitrary
curved spaces do not possess constant matrix curvatures with respect to
certain orthonormalized frames which is crucial for encoding recursion
operators and associated solitonic hierarchies.\footnote{a series of recent works \cite{vghrm,vrfsolw,vrf2,vrfsol,vdqfl,vdqes} provide a number of generalizations and examples when solitonic hierarchies can be derived for arbitrary Einstein, Riemann--Cartan and Larange--Finsler spaces, their Ricci flow evolutions and/or
quantum deformations} Nevertheless,
the results proved in our work provide a new geometric method of solitonic
encoding of data for quite general types of curved spaces and nonlinear
physical theories. This follows from the fact that we can always
nonholonomically deform a curved space having nonconstant curvature given by
the distinguished linear connection into a similar space for which this
curvature becomes constant.

Our approach employs certain methods elaborated in Finsler and Lagrange
geometry when the geometric objects are adapted to nonholonomic
distributions on vector/ tangent bundles, or (in general) on nonholonomic
manifolds. This accounts for existence of the canonical nonlinear connection
(defined as non-integral distributions into conventional horizontal and
vertical subspaces, and associated nonholonomic frames), metric and linear
connection structures all derived from a regular Lagrange (in particular,
Finsler) metric function and/ or by moving frames and generically off--diagonal
metrics in gravity theories. Of course, such spaces are not generally
defined to have constant linear connection curvatures and vanishing torsions
to which the curve-flow solitonic generation techniques can not applied.

We proved that for a very large class of regular Lagrangians and so--called
generalized Lagrange space metrics, there can be nonholonomic deformations
of geometric objects to equivalent ones on generalized Lagrange
configurations, for instance, with constant Hessian for the so--called
absolute energy function. The curvature matrix, with respect to the
correspondingly adapted (to the canonical connection structure) frames, can
be defined with constant, even vanishing, coefficients. For such
configurations, we can apply the former methods elaborated for symmetric
Riemannian spaces in order to generate curve-flow solitonic hierarchies.
Here should be noted that such special classes of nonholonomic manifolds are
equipped with generic off--diagonal metrics and possess nontrivial curvature
for the Levi Civita connection. They are characterized additionally by
nontrivial nonlinear connection curvature and nonzero anholonomy
coefficients for preferred frame structures, and anholonomically induced
torsion even their curvature of the canonical distinguished connection is
zero.

The utility of the method of anholonomic frames with associated nonlinear
connection structures is that we can work with respect to such frames and
correspondingly canonical metrics as defined on horizontal/ vertical pairs
of symmetric Riemannian spaces whose metric and linear connection structure
is soldered from an underlying Klein geometry. In this regard, the
bi-Hamiltonian solitonic hierarchies are generated as nonholonomic
distributions of horizontal and vertical moving equations and conservation
laws, containing all information for a regular Lagrangian and corresponding
Euler--Lagrange equations.

We note that curve-flow solitonic hierarchies can be constructed in a similar manner, for instance, for Einstein--Yang--Mills--Dirac equations, derived following the anholonomic frame method, in noncommutative generalizations of gravity and geometry and possible quantum models based on
nonholonomic Lagrange--Fedosov manifolds. During a long term review of this paper, a series of new important curve--flow solitonic results were obtained. For instance, the results of 
\cite{vhfrm} were generalized in \cite{vghrm} for arbitrary Einstein and (pseudo) Riemannian metrics for which alternative d--connections (to the Levi Civita one) were defined. It was proven that such metric compatible d--connections with constant curvature and Ricci d--tensor coefficients, with respect to certain N--adapted basis, can be constructed. All constructions can be redefined equivalently for the Levi Civita connection (because all connections under consideration are completely defined by a metric structure).

Finally, we conclude that solitonic hierarchies and the bi--Hamilon and N--connection formalisms happen to very efficient in the theory of nonholonomic Ricci flows and evolution of physically valuable nonlinear gravitational wave solutions \cite{vrfsolw,vrf2,vrfsol} and Fedosov quantization of Einstein and Lagrange--Finsler spaces \cite{vdqfl,vdqes}. So, the results of this paper can be naturally generalized for arbitrary classical and quantum (commutative and noncommutative of type \cite{vncg}) nonholonomic gravitational interactions,
generalized Lagrange--Finsler systems and their Ricci flow evolutions (see also review \cite{vrfg}). In general, we can encode the information on such
field/mechanical/evolution models into corresponding solitonic hierarchies and, inversely, to extract certain nonlinear interaction and/or evolution models from systems of solitonic equations.

\vskip 5pt

\textbf{Acknowledgement: } The second author (S.V.) is grateful to A. Bejancu for very important references on the geometry of nonholonomic manifolds and to T. Wolf for hosting a visit to the Mathematics Department at Brock University. He  also
thanks M. Anastasiei for kind support.

\appendix

\section{Some Local Formulas}

This Appendix outlines some local results from geometry of nonlinear
connections (see Refs. \cite{ma1,ma2,vncg,vsgg} for proofs and details).
There are two types of preferred linear connections uniquely determined by a
generic off--diagonal metric structure with $n+m$ splitting, see $\mathbf{g}%
=g\oplus _{N}h$ (\ref{m1}):

\begin{enumerate}
\item The Levi Civita connection $\nabla =\{\Gamma _{\beta \gamma }^{\alpha
}\}$ is by definition torsionless, $\mathcal{T}=0$, and satisfies the metric
compatibility condition$,\nabla \mathbf{g}=0$.

\item The canonical d--connection $\widehat{\mathbf{\Gamma }}_{\ \alpha
\beta }^{\gamma }=\left( \widehat{L}_{jk}^{i},\widehat{L}_{bk}^{a},\widehat{C%
}_{jc}^{i},\widehat{C}_{bc}^{a}\right) $ is also metric compatible, i.e. $%
\widehat{\mathbf{D}}\mathbf{g}=0$, but the torsion vanishes only on h-- and
v--subspaces, i.e. $\widehat{T}_{jk}^{i}=0$ and $\widehat{T}_{bc}^{a}=0$,
for certain nontrivial values of $\widehat{T}_{ja}^{i},\widehat{T}_{bi}^{a},%
\widehat{T}_{ji}^{a}$. For simplicity, we omit hats on symbols and write $%
L_{jk}^{i}$ instead of $\widehat{L}_{jk}^{i}$, $T_{ja}^{i}$ instead of $%
\widehat{T}_{ja}^{i}$ and so on, but preserve the general symbols $\widehat{%
\mathbf{D}}$ and $\widehat{\mathbf{\Gamma }}_{\ \alpha \beta }^{\gamma }$.
\end{enumerate}

By a straightforward calculus with respect to N--adapted frames (\ref{dder})
and (\ref{ddif}), one can verify that the requested properties for $\widehat{%
\mathbf{D}}$ on $\mathbf{E}$ are satisfied if
\begin{eqnarray}
L_{jk}^{i} &=&\frac{1}{2}g^{ir}\left( \mathbf{e}_{k}g_{jr}+\mathbf{e}%
_{j}g_{kr}-\mathbf{e}_{r}g_{jk}\right) ,  \label{candcon} \\
L_{bk}^{a} &=&e_{b}(N_{k}^{a})+\frac{1}{2}h^{ac}\left( \mathbf{e}%
_{k}h_{bc}-h_{dc}\ e_{b}N_{k}^{d}-h_{db}\ e_{c}N_{k}^{d}\right) ,  \notag \\
C_{jc}^{i} &=&\frac{1}{2}g^{ik}e_{c}g_{jk},\ C_{bc}^{a}=\frac{1}{2}%
h^{ad}\left( e_{c}h_{bd}+e_{c}h_{cd}-e_{d}h_{bc}\right) .  \notag
\end{eqnarray}%
For $\mathbf{E}=TM$, the canonical d--connection $\ \mathbf{\tilde{D}}=(h%
\tilde{D},v\tilde{D})$ can be defined in a torsionless form\footnote{%
Namely where the d--connection has the same coefficients as the Levi Civita
connection with respect to N--elongated bases (\ref{dder}) and (\ref{ddif})}
with the coefficients $\Gamma _{\ \beta \gamma }^{\alpha }=(L_{\
jk}^{i},L_{bc}^{a})$,
\begin{eqnarray}
L_{\ jk}^{i} &=&\frac{1}{2}g^{ih}(\mathbf{e}_{k}g_{jh}+\mathbf{e}_{j}g_{kh}-%
\mathbf{e}_{h}g_{jk}),  \label{candcontm} \\
C_{\ bc}^{a} &=&\frac{1}{2}h^{ae}(e_{c}h_{be}+e_{b}h_{ce}-e_{e}h_{bc}).
\notag
\end{eqnarray}

The curvature of a d--connection $\mathbf{D,}$
\begin{equation}
\mathcal{R}_{~\beta }^{\alpha }\doteqdot \mathbf{D\Gamma }_{\ \beta
}^{\alpha }=d\mathbf{\Gamma }_{\ \beta }^{\alpha }-\mathbf{\Gamma }_{\ \beta
}^{\gamma }\wedge \mathbf{\Gamma }_{\ \gamma }^{\alpha },  \label{curv}
\end{equation}%
splits into six types of N--adapted components with respect to (\ref{dder})
and (\ref{ddif}),
\begin{equation*}
\mathbf{R}_{~\beta \gamma \delta }^{\alpha }=\left(
R_{~hjk}^{i},R_{~bjk}^{a},P_{~hja}^{i},P_{~bja}^{c},S_{~jbc}^{i},S_{~bdc}^{a}\right) ,
\end{equation*}%
\begin{eqnarray}
R_{\ hjk}^{i} &=&\mathbf{e}_{k}L_{\ hj}^{i}-\mathbf{e}_{j}L_{\ hk}^{i}+L_{\
hj}^{m}L_{\ mk}^{i}-L_{\ hk}^{m}L_{\ mj}^{i}-C_{\ ha}^{i}\Omega _{\ kj}^{a},
\label{dcurv} \\
R_{\ bjk}^{a} &=&\mathbf{e}_{k}L_{\ bj}^{a}-\mathbf{e}_{j}L_{\ bk}^{a}+L_{\
bj}^{c}L_{\ ck}^{a}-L_{\ bk}^{c}L_{\ cj}^{a}-C_{\ bc}^{a}\Omega _{\ kj}^{c},
\notag \\
P_{\ jka}^{i} &=&e_{a}L_{\ jk}^{i}-D_{k}C_{\ ja}^{i}+C_{\ jb}^{i}T_{\
ka}^{b},~P_{\ bka}^{c}=e_{a}L_{\ bk}^{c}-D_{k}C_{\ ba}^{c}+C_{\ bd}^{c}T_{\
ka}^{c},  \notag \\
S_{\ jbc}^{i} &=&e_{c}C_{\ jb}^{i}-e_{b}C_{\ jc}^{i}+C_{\ jb}^{h}C_{\
hc}^{i}-C_{\ jc}^{h}C_{\ hb}^{i},  \notag \\
S_{\ bcd}^{a} &=&e_{d}C_{\ bc}^{a}-e_{c}C_{\ bd}^{a}+C_{\ bc}^{e}C_{\
ed}^{a}-C_{\ bd}^{e}C_{\ ec}^{a}.  \notag
\end{eqnarray}

Contracting respectively the components, $\mathbf{R}_{\alpha \beta
}\doteqdot \mathbf{R}_{\ \alpha \beta \tau }^{\tau }$, one computes the h-
v--components of the Ricci d--tensor (there are four N--adapted components)
\begin{equation}
R_{ij}\doteqdot R_{\ ijk}^{k},\ \ R_{ia}\doteqdot -P_{\ ika}^{k},\
R_{ai}\doteqdot P_{\ aib}^{b},\ S_{ab}\doteqdot S_{\ abc}^{c}.
\label{dricci}
\end{equation}%
The scalar curvature is defined by contracting the Ricci d--tensor with the
inverse metric $\mathbf{g}^{\alpha \beta }$,
\begin{equation}
\overleftrightarrow{\mathbf{R}}\doteqdot \mathbf{g}^{\alpha \beta }\mathbf{R}%
_{\alpha \beta }=g^{ij}R_{ij}+h^{ab}S_{ab}=\overrightarrow{R}+\overleftarrow{%
S}.  \label{sdccurv}
\end{equation}

If $\mathbf{E=}TM$, there are only three classes of d--curvatures,%
\begin{eqnarray}
R_{\ hjk}^{i} &=&\mathbf{e}_{k}L_{\ hj}^{i}-\mathbf{e}_{j}L_{\ hk}^{i}+L_{\
hj}^{m}L_{\ mk}^{i}-L_{\ hk}^{m}L_{\ mj}^{i}-C_{\ ha}^{i}\Omega _{\ kj}^{a},
\label{dcurvtb} \\
P_{\ jka}^{i} &=&e_{a}L_{\ jk}^{i}-\mathbf{D}_{k}C_{\ ja}^{i}+C_{\
jb}^{i}T_{\ ka}^{b},  \notag \\
S_{\ bcd}^{a}&=&e_{d}C_{\ bc}^{a}-e_{c}C_{\ bd}^{a}+C_{\ bc}^{e}C_{\
ed}^{a}-C_{\ bd}^{e}C_{\ ec}^{a},  \notag
\end{eqnarray}%
where all indices $a,b,\ldots,i,j,\ldots$ run over the same values and, for
instance, $C_{\ bc}^{e}\to $ $C_{\ jk}^{i}$, etc.


\begin{thebibliography}{99}
\bibitem{kern} Kern J., Lagrange geometry, Arch. Math. \textbf{25} (1974)
438--443

\bibitem{ma1} Miron R. and Anastasiei M., Vector Bundles and Lagrange Spaces
with Applications to Relativity (Geometry Balkan Press, Bukharest, 1997);
translation from Romanian of (Editura Academiei Romane, 1984)

\bibitem{ma2} Miron R. and Anastasiei M., The Geometry of Lagrange paces:\
Theory and Applications, FTPH no. \textbf{59} (Kluwer Academic Publishers,
Dordrecht, Boston, London, 1994)

\bibitem{deleon1} de Le\'{o}n M. and Rodrigues P. R., Methods of
Differential Geometry in Analytical Mechanics, Ser. 152 (North-Holland Math. Amsterdam, 1989)

\bibitem{deleon2} de Le\'{o}n M., Martin de Diego D. and Santamaria--Merino
A., Symmetries in classical field theory, International Journal of Geometric
Methods in Modern Physics, \textbf{1}\ (2004) 651--710

\bibitem{lam} Lamb G.L. Jr., Solitons on moving space curves, J. Math. Phys. \textbf{18}\ (1977) 1654--1661

\bibitem{gol} Goldstein R.E. and Petrich D.M., The Korteweg-de Vries hierarchy as dynamics of closed curves in the plane, Phys. Rev. Lett. \textbf{67}\ (1991) 3203-3206

\bibitem{nak} Nakayama K., Segur H., and Wadati M., Integrability and the motion of curves, Phys. Rev. Lett. \textbf{69}\ (1992) 2603--2606

\bibitem{lanper} Langer J. and Perline R., Curve motion inducing modified Korteweg-de Vries systems, Phys. Lett. A \textbf{239}\ (1998) 36--40

\bibitem{chou1} Chou K. -S. and Qu C., The KdV equation and motion of plane curves, J. Phys. Soc. Japan \textbf{70}\  (2001) 1912--1916

\bibitem{chou2} Chou K. -S. and Qu C., Integrable equations 
arising from motions of plane curves, Phys. D, \textbf{162}\ (2002) 9--33;\
  Integrable equations arising from motions of plane curves II, 
  J. Nonlin. Sci. \textbf{13}\ (2003) 487--517

\bibitem{mbsw} Mari Beffa G., Sanders J., Wang J. -P., Integrable systems in three--dimensional Riemannian geometry, J. Nonlinear. Sci., \textbf{12}\ (2002) 143--167

\bibitem{saw} Sanders J. and Wang J. -P., Integrable systems in $n$ dimensional Riemannian geometry, Mosc. Math. J. \textbf{ 3 }\ (2003) 1369--1393

\bibitem{anc1} Anco S. C., bi-Hamiltonian operators, integrable flows of curves using moving frames, and geometric map equations, J. Phys. A: Math. Gen. \textbf{39}\ (2006) 2043--2072

\bibitem{ath1} Athorne, C. Some integrable equations associated with symmetric spaces. Nonlinear evolutions [Balaruc-les-Bains, 198], (World Sci. Publ., Teaneck, NJ, 1988), pp. 191--198

\bibitem{ath2} Athorne C., Local Hamiltonian structures of multicomponent KdV equations, J. Phys. A: Math. Gen. \textbf{21}\ (1988) 4549--4556

%\bibitem{fours} Foursov M. V., Classification of certain integrable coupled
%potential KdV and modified KdV--type equations, J. Math. Phys., \textbf{41}
%(2000) 6173--6185

\bibitem{anc2} Anco S. C., Hamiltonian flows of curves in $G/SO(n)$ and
vector soliton equations of mKdV and sine--Gordon Type, Symmetry,
Integrability and Geometry: Methods and Applications, \textbf{2} (2006) 044

\bibitem{wang} Wang J.-P., Generalized Hasimoto transformation and vector
sine--Gordon equation, in SPT 2002: Symmetry and Perturbation Theory (Cala
Gonone), Editors  Abenda S., Gaeta G.  and  Walcher S., River Edge (WorldScientific, 2002), pp. 276--283

\bibitem{helag} Helagson S., Differential Geometry, Lie Groups, and
Symmetric Spaces (Providence, Amer. Math. Soc., 2001)

\bibitem{sw} Sokolov V. V. and Wolf T., Classification of integrable vector
polynomial evolution equation, J. Phys. A: Math. Gen. \textbf{34}\ (2001) 11139--11148

\bibitem{aw} Anco S. C. and Wolf T., Some symmetry classifications of hyperbolic vector evolution equations, J. Nonlinear Math. Phys., \textbf{12}\ 
(2005), suppl. 1, 13--31;\ Erratum, J. Nonlinear Math. Phys. \textbf{12}\ (2005) 607--608

\bibitem{vhfrm} Vacaru S., Curve flows and solitonic hierarchies generated by (semi) Riemannian metrics, math-ph/0608024

\bibitem{bej} Bejancu A., Finsler Geometry and Applications (Ellis Horwood, Chichester, England, 1990)

\bibitem{bejf} Bejancu A. and Farran H. R., Foliations and Geometric Structures (Springer, 2005)

\bibitem{vncg} Vacaru S., Exact solutions with noncommutative symmetries in
Einstein and gauge Gravity, J. Math. Phys. \textbf{46}\ (2005) 042503

\bibitem{vsgg} Clifford and Riemann- Finsler Structures in Geometric Mechanics and Gravity. Selected Works, by  Vacaru S., Stavrinos P., Gaburov E.  and Gon\c{t}a D.;\  Differential Geometry -- Dynamical Systems, Monograph 7 (Geometry Balkan Press, Bucharest, 2006);\ www.mathem.pub.ro/dgds/mono/va-t.pdf and gr-qc/0508023

\bibitem{yano} Yano K. and Ishihara S., Tangent and Cotangent Bundles (M. Dekker, Inc. New York, 1978)

\bibitem{kob} Kobayashi S. and Nomizu K., Foundations of Differential Geometry, Vols. I and II (Wiley, 1969)

\bibitem{sharpe} Sharpe R. W., Differential Geometry (New York, Springer--Verlag, 1997)

\bibitem{vr1} Vranceanu G., \ Sur les espaces non holonomes. C. R. Acad. Paris \textbf{103}\ (1926) 852--854

\bibitem{vr2} Vranceanu G., Le\c cons de Geometrie Differentielle, Vol II
(Edition de l'Academie de la Republique Populaire de Roumanie, 1957)

\bibitem{bcs} Bao D., Chern S. -S., and Shen Z., An Introduction to
Riemann--Finsler Geometry. Graduate Texts in Math., 200 (Springer--Verlag,
2000)

\bibitem{ancima} Anco S. C., Hamiltonian curve flows in Lie groups
G$\subset$U(N) and vector NLS, mKdV, sine-Gordon soliton equations,
IMA Volumes in Mathematics and its Applications, Vol. 144,
Symmetries and Overdetermined Systems of Partial Differential Equations (AMS 2007), 223--250

\bibitem{ancjgp} Anco S. C., Group-invariant soliton equations and
bi-Hamiltonian geometric curve flows in Riemannian symmetric spaces, J. Geom. Phys. \textbf{ 58 }\ (2008), 1--37

\bibitem{serg} Sergyeyev A., Why nonlocal recursion operators produce local
symmetries: new results and applications, J. Phys. A: Math. Gen. \textbf{38}\ (2005) 3397--3407

\bibitem{blv} Barcelo C., Liberati S. and Visser M., Analogue Gravity, Living Rev. Rel. \textbf{8}\  (2005) 12

\bibitem{mh} Miron R., Hrimiuc D., Shimada S. and Sabau V. S., The Geometry
of Hamilton and Lagrange Spaces (Kluwer Academic Publishers, Dordrecht, Boston, London, 2000)

\bibitem{fcartan} Cartan E., Les Espaces de Finsler\ (Paris, Hermann, 1935)

\bibitem{matsumoto} Matsumoto M., Foundations of Finsler Geometry and Special Finsler Spaces (Kaisisha: Shigaken, 1986)

\bibitem{vmon1} Vacaru S., Interactions, Strings and Isotopies in Higher
Order An\-iso\-tropic Superspaces (Hadronic Press, Palm Harbor, FL, USA, 1998), math--ph/ 0112065

\bibitem{esv} Etayo F., Santamari\'{a} R. and Vacaru S., Lagrange--Fedosov nonholonomic manifolds, J. Math. Phys. \textbf{46}\ (2005) 032901

\bibitem{vghrm} Vacaru S., Curve flows and solitonic hierarchies generated by Einstein metrics, arXiv: 0810.0707 [math-ph]
    
\bibitem{vrfsolw}Vacaru S., Ricci flows and solitonic pp--waves, Int. J. Mod. Phys. A \textbf{  21 }\ (2006) 4899-4912

\bibitem{vrf2} Vacaru S., Nonholonomic Ricci flows: II. Evolution equations and dynamics,  J. Math. Phys. \textbf{ 49 }\ (2008)  043504

\bibitem{vrfsol} Vacaru S., Nonholonomic Ricci flows: III. Curve flows and solitonic hierarchies, arXiv:  0704.2062  [math.DG]
    
\bibitem{vdqfl}  Vacaru S., Deformation quantization of almost K\" ahler models and Lagrange--Finsler spaces, J. Math. Phys. \textbf{ 48 }\ (2007) 123509
    
\bibitem{vdqes}  Vacaru S., Deformation quantization of nonholonomic almost K\" ahler models and Einstein gravity, Phys. Lett. A \textbf{ 372 }\ (2008) 2949-2955
    
\bibitem{vrfg} Vacaru S.,  Finsler and Lagrange geometries in Einstein and string gravity,  Int. J. Geom. Methods  Mod. Phys. (IJGMMP) \textbf{ 5 }\ (2008) 473-511
\end{thebibliography}
\end{document}